\pdfoutput=1
\documentclass[12pt]{report} 
\usepackage[sort&compress,numbers]{natbib}
\usepackage{lscape,utdiss2-05,amsfonts,makeidx,bm,eucal,tensor,mathrsfs,color,hyperref}

\hypersetup{hyperindex,backref,pdfpagemode=FullScreen,colorlinks=flase} 

\def\bra#1{\mathinner{\langle{#1}|}} 
\def\ket#1{\mathinner{|{#1}\rangle}} 
\def\braket#1{\mathinner{\langle{#1}\rangle}} 
\def\Bra#1{\left\langle#1\right|} 
\def\Ket#1{\left|#1\right \rangle} {\catcode`\|=\active \gdef\Braket#1{
\begingroup \mathcode`\|32768\let|\BraVert\left<{#1}\right>
\endgroup}} 
\def\BraVert{\egroup\,\mid\,\bgroup} 
\def\Brak#1#2#3{\bra{#1}#2\ket{#3}}

\author{Kavan Kishore Modi}
\address{4711 Duval St.\\
Austin TX, 78751}

\supervisor {E. C. George Sudarshan}
\committeemembers[Austin M. Gleeson][Allan H. MacDonald][John T. Markert]{Robert E. Wyatt} 
\title{A theoretical analysis of\\ experimental open quantum dynamics}
\previousdegrees{B.Sc., M.A.}
\graduationmonth{August} 
\graduationyear{2008} 
\degree{Doctorate of Philosophy} \degreeabbr{Ph.D.}
\oneandonehalfspacequote
\newcommand{\latexe}{{\LaTeX\kern.125em2
\lower.5ex\hbox{$\varepsilon$}}}

\chardef\bslash=`\\ 

\makeatletter 
\def\square{\RIfM@\bgroup\else$\bgroup\aftergroup$\fi \vcenter{\hrule\hbox{\vrule\@height.6em\kern.6em\vrule}
\hrule}\egroup} 
\makeatother 

\makeindex 

\begin{document} 

\copyrightpage
\newpage

%
%
%

\commcertpage 
\titlepage

%
\begin{dedication}
\index{Dedication @\emph{Dedication}}
To Mummy and Pappa
\end{dedication}

\begin{acknowledgments}
\index{Acknowledgments @\emph{Acknowledgments}}

I am in great debt to my teacher and my advisor Prof. E. C. George Sudarshan.  His insistence on examining each problem at its core has taught me to do science properly.  Under his supervision I worked on problems in open quantum systems, unstable quantum systems, and quantum optics.  I am grateful for the variety of physics that he has taught me.

I thank my group mates C\'esar Rodr\'iguez-Rosario, Antonia Chimonidou, and Kuldeep Dixit for many wonderful discussions and their friendships.  I thank Dr. Aik-meng Kuah; as much of the work in this dissertation was done is close collaboration with him.  I owe much to Dr. Anil Shaji, who patiently helped me with a great deal of physics that I know today.  A special thanks to Dr. Todd Tilma, who has given me numerous helpful suggestions on the topics discussed here.

I also learnt a great deal of physics and mathematics from Prof. Austin Gleeson and Prof. Luis Boya.  I thank them for teaching me and for their friendships.  Prof. Austin Gleeson helped at the lowest points of my graduate career; without him I would not have finished this dissertation. I would also like to thank my colleagues Ali Rezakhani, Masoud Mohseni, Tim Coffey, Animesh Datta, Dylan Miracle, Kevin Lee, and many others, whose names I cannot remember in the moment, for sharing many enlightening conversations.

\newpage


I express my deepest gratitude to Pappa, Mummy, and Chintan, who never lost the confidence in my abilities even when I did.  I am grateful to my girlfriend, Laura, for never getting jealous of Lady physics even during the busiest weeks;  her support has been invaluable to my success here.  I thank Laura's parents David and Beth for their support and interest in my work.  And to all of my friends, who made the last seven years enlightening and fun.

Lastly, I thank Dylan Miracle, Laura Speck, and James Zabel for proof reading this manuscript and suggesting corrections.
\end{acknowledgments}

\begin{preface}
Quantum process tomography is one of the most fundamental tool for the experimental study of open dynamics of a quantum system. Moreover, quantum process tomography is an important practical tool for implementing quantum information processing devices.  By the very nature of quantum mechanics, quantum information processing cannot be achieved in isolation.  The processing device must interact with a measuring apparatus to yield information.  Furthermore, the processing device, in general, is also affected by its surroundings (an environment) in an uncontrollable manner.  This results in the necessity of treating the dynamics of such devices with the quantum theory of open systems.  

The basic building blocks of quantum processing devices have only been experimentally realized in the last ten years; hence quantum process tomography is a relatively new procedure.  On the contrary, the theoretical analog of quantum process tomography is the dynamical map formalism, introduced almost fifty years ago by Sudarshan, Mathews, and Rau.  During the last five decades this formalism has matured significantly.  In some of the recent studies of the dynamical maps, a significant emphasis is placed on the reduced dynamics of initially correlated states.  While for quantum process tomography it is usually assumed that at the beginning of the experiment the state of the system is uncorrelated with its surroundings.  Based on some of the theoretical results, we study quantum process tomography for initially correlated systems.  

Since quantum process tomography is an experimental procedure, we are forced to deal with the idea of the preparation of initial states.  In fact, one must analyze the role of the preparation procedures for any quantum experiment that interacts with its surroundings non-trivially.  There are two different ways by which the preparation procedure can play a significant role in an open quantum experiment.  First is due to the initial correlations between the states of the system and the environment.  While the second deals the consistency of the preparation procedures; even when there are no initial correlations between the system and the environment.

The starting point of this dissertation is a review of the dynamical map formalism and several quantum process tomography procedures.  At this point, the mathematical structure of preparation procedures is discussed in detail.  This allows us to investigate the role of preparation in quantum process tomography without any assumptions.  

The study of quantum process tomography with preparation procedure leads to another process tomography method that is independent of the preparation procedure. Furthermore, this procedure leads to a surprising result; the map arising from this procedure (we call $\mathcal{M}$-map) leads a quantitative measure for the non-Markovian memory effect on the system due to the initial correlations between the system and the environment.  A measure for the non-Markovian memory effect can play a crucial role in a coherence control scheme.

Many of the results dealing with negativity and non-linearity in quantum process tomography are presented by concrete examples, rather than by a general mathematical analysis.  Based on the results of these examples we try to generalize the role of preparation in quantum process tomography.  On the other hand, the theory of preparation procedure, development of  $\mathcal{M}$-map and the memory matrix are worked out by general mathematical analysis.\\
\\
Kavan Modi, Austin Texas.

\end{preface}

\utabstract 
\index{Abstract @\emph{Abstract}} \indent 
In recent years there has been a significant development of the dynamical map formalism for initially correlated states of a system and its environment.  Based on some of these results, we study quantum process tomography for initially correlated states of the system and the environment.  This is beyond the usual assumption that the state of the system and the environment are initially uncorrelated. Since quantum process tomography is an experimental procedure, we wind up having to study the role of preparation of input states for open quantum experiments.  We work out a theory for the general preparation procedure, and study two preparation procedures in detail. In specific, we study the \emph{stochastic preparation procedure} and the \emph{ projective preparation procedure} and apply them to quantum process tomography. The two preparation procedures describe the ways to uncorrelate the state of the system and the environment.  However the specifics of how this is implemented plays a role on the outcomes of the experiment. 

When the stochastic preparation procedure is applied properly, quantum process tomography yields a linear process maps. We point out what it means to apply the stochastic preparation procedure properly by constructing several simple examples where inconsistencies in preparations leads to errors. When the projective preparation procedure is applied, quantum process tomography leads to a non-linear process map. We show that these processes can only be consistently described by a general dynamical map, which we call $\mathcal{M}$-map. The $\mathcal{M}$-map contains all of the dynamical information for the state of the system without the affects of a preparation procedure.  By carefully extracting some of this dynamical information, we construct a quantitative measure for the memory effect due to the initial correlations with the environment.

\tableofcontents 
\listoffigures     



\chapter{Introduction}\label{chapintro}
\index{Introduction @\emph{Introduction}}

Quantum information processing promises powerful computational methods that surpass the methods of classical information processing \cite{Nielsen00a,  shannon48a}. These methods rely on taking advantage of quantum parallelism by using quantum superposition and quantum entanglement as resources.  In order to implement such a device one must have precise control over the system, and isolate it from the surrounding environment to preserve coherence.  For delicate systems of this sort, it is nearly impossible to isolate the system of interest completely from its surroundings, while having a great deal of control.

With the rising interest in quantum computation and quantum information processing, quantum coherence experiments are performed readily these days, though with relatively small systems.  One of the major problem with these experiment is the loss of coherence due to the interaction between the system of interest and the unknown environmental states. The methods for studying the interaction between the system and the environment are given by the quantum theory of open systems. 

The quantum theory of open systems got its start in almost fifty years ago with the introduction of dynamical maps \cite{SudarshanMatthewsRau61, SudarshanJordan61} due to Sudarshan, Mathews, and Rau.  Fifty years after its conception, the dynamical map formalism is finally being tested in the laboratory setting.  The experimental determination of a dynamical map is achieved by a procedure called \emph{quantum process tomography} \footnote{Tomography, in the traditional sense, means to determine the internal structure of an object.  Here the process is thought to be a physical object that interacts with the system for an instant, hence process tomography means the action of the object onto the system.}.

Any experiment, including quantum process tomography experiments, requires a method to prepare the initial states of the system at the beginning of the experiment \cite{kuah:042113}.  We study the affects of the preparation procedure on quantum systems that interact with an environment \footnote{An environment is any degree of freedom that develops in time with the state of the system.}.  The act of preparation has been neglected from the theory of quantum process tomography (and for all quantum experiments that interact with a non-trivial environment).  We investigate this issue for quantum process tomography in detail in this dissertation. We present several simple examples to motivate our arguments.  Based these arguments, we analyze two recent quantum process tomography experiments and show their results to be consistent with our analysis.

In the course of our studies of the role of preparation procedures in quantum process tomography, we derive a new powerful method of quantum process tomography that is not affected by the preparation procedure.  An experimental recipe to carry out this procedure is given at the end.

The added advantage to this new procedure is an expression quantifying the memory due the initial correlations in the dynamics of the state of the system.  Quantifying the memory is the crucial step to develop a coherence control scheme.  If one can separate the purely dissipative terms from the terms that periodically recur, then a scheme can be developed to make use of the coherence periodicity, in battling decoherence.  

\section{Organization of this dissertation}\label{organization}

We start with a brief review of closed dynamics of quantum systems in Chap. \ref{chapopendyn}.  We then motivate the necessity for studying open quantum dynamics. In this dissertation we work in the dynamical map formalism to study the open dynamics of quantum systems\footnote{See \cite{davies76a,breuer02a} for other approaches to quantum theory of open systems.}.  Next we define the positivity classes associated with the dynamical maps and present an example of a not-completely positive dynamical map.  The not-completely positive nature is attributed to the initial correlations between the system and the environment. We use this example later on as an inspiration in analyzing quantum process tomography experiments.

In Chap. \ref{chapqpt} we review several existing methods of quantum process tomography.  We discuss the reasons for the different methods of quantum process tomography by analyzing the pluses and the minuses associated with each.  We  point out the central assumption in all of these procedure; that, the state of the system and the environment is uncorrelated at the beginning of the experiment.  When this assumption does not hold, a quantum process tomography experiment can yield a nonsensical process map for some cases.

In Chap. \ref{chapprp} we present one of the central points of this dissertation.  We discuss the  general theory for preparing input states  in terms of stochastic maps.  We discuss  in detail two of the most common preparation procedures practiced and compare them for closed and open quantum systems. For open systems, if care in not take in implementing a preparation procedure, then the state of the environment will pickup a non-trivial dependence on either the prepared state or the preparation procedure itself.  This dependence can lead to nonsensical experimental results.

In Chap. \ref{chapqptex} we revisit the quantum process tomography procedures armed with the preparation techniques.  We analyze the role that preparation procedure plays in a quantum process tomography experiment.  We show that for certain types of preparation procedures, the quantum process tomography will fail to obtain a process map that correctly describes the physical process.  We present three detailed examples dealing with these difficulties.

In Chap. \ref{appdvp}, we analyze the case when the preparation procedure itself is poor.  We present several more examples with different scenarios of poor preparation procedures.  The causes for the errors in this chapter are different from the causes discussed in Chap. \ref{chapexp}. We compile a list of operations that can lead to errors for a quantum process tomography experiment.

We offer a resolution by introducing a new process map that we call dynamical $\mathcal{M}$-map in Chap. \ref{chapgMmap}.  To obtain this map we extract the preparation procedure from the dynamics.  In this fashion the $\mathcal{M}$-map contains all of the dynamical information for the system unaffected by the preparation procedure.  We further show that, the $\mathcal{M}$-map contains information about the dynamics of the system when it is correlated and when it is uncorrelated with the environment.  This allows us isolate the reduced dynamics of the initial correlations between the system and the environment known as the memory due to the correlations and construct a quantitative measure for the memory.  Finally we give an example of $\mathcal{M}$-map, and from it calculate some of the underlying dynamics.

In Chap. \ref{chapexp}, we analyze two experiments where the preparation procedures were not carried out consistently.  The negative results obtained in these experiments are foreseen by our theory. We give the concluding remarks and potential future directions in Chap. \ref{chapcon}.

In App. \ref{appqst}, we briefly discuss quantum state tomography.  In App. \ref{mqpt} we lay out a procedure to experimentally determine the $\mathcal{M}$-map, and lay out a recipe for an experiment for a qubit system in App \ref{mrec}.  Finally in App. \ref{appnot} we lay out all of the notation used throughout this dissertation.
 
\chapter{Quantum theory of open system}\label{chapopendyn}
\index{Quantum theory of open system @\emph{Quantum theory of open system}}

Pure states in quantum mechanics are generally represented by rays in the 
\index{density matrix}
Hilbert space.  Rays  adequately explain the phenomena of quantum superposition, but they cannon represent classical mixtures of quantum states.  For instance, consider a beam of particles with 70\% in state $\ket{1}$ and 30\% in state $\ket{0}$; this situation is not the same as quantum superposition. We can represent the state of the beam by a density matrix as $\rho=0.7\ket{1}\bra{1}+0.3\ket{0}\bra{0}$. Here 0.7 and 0.3 represent the classical probabilities of finding the system in state $\ket{1}$ or $\ket{0}$. Any state of a quantum system can be written in terms of a density matrix\footnote{We only consider finite dimensional systems throughout this dissertation.}, be it a pure quantum state or a classically mixed one.

Since density matrices describe classical mixtures of quantum states, they must have the following three properties.
\begin{center}
\begin{tabular}{l l}
$\mbox{Tr}[\rho]=1$ \hspace{4cm} & Normalization, \cr
$\rho=\rho^\dag$ \hspace{4cm} & Hermitian, \cr
$x^*_r\rho_{rs}x_s\geq 0$ \hspace{4cm} & Positivity. \cr
\end{tabular}
\end{center}
The eigenvalues of the density matrix represent classical probabilities for each pure state given by it's corresponding eigen-projectors.  Thus, the trace condition gives us the conservation of probabilities, Hermiticity of $\rho$ guarantees that the eigenvalues are real, and finally with the last condition we demand that the eigenvalues are positive.

Density matrices form a convex set; if $\{\rho^{(j)}\}$ are a set of 
\index{density matrix! convexity}
density matrices then
\begin{eqnarray}
\rho=\sum_{j}p_j\rho^{(j)}
\end{eqnarray}
is also a density matrix as long as $p_j$ are real, positive, and $\sum_{j}p_j=1$.  Any density matrix that cannot be written in terms of a convex sum is called extremal. Extremal states are also pure states. An important decomposition is the eigen-decomposition, 
\begin{eqnarray}
\rho=\sum_{j}\lambda_j\ket{\psi_j}\bra{\psi_j},
\end{eqnarray}
where the $\lambda_j$ and $\ket{\psi_j}\bra{\psi_j}$ are the eigenvalues and eigenprojectors of $\rho$ respectively. 

For a qubit (a two-level system), we can write the density matrix as the 
\index{density matrix! of one qubit}\index{qubit} 
following
$$
\rho=\frac{1}{2}\{\mathbb{I}+a_1\sigma_1+a_2\sigma_2+a_3\sigma_3\},
$$
where $\mathbb{I}$ is the $2\times 2$ identity matrix, $\sigma_j$ are the 
\index{Pauli spin matrices}
Pauli spin matrices given by 
$$
\sigma_1=\left(\matrix{0&1\cr1&0}\right),\;\;
\sigma_2=\left(\matrix{0&-i\cr i&0}\right),\;\;
\sigma_3=\left(\matrix{1&0\cr0&-1}\right),
$$
and $(a_1,a_2,a_3)$ are the components of the Bloch vector \cite{PhysRev.70.460} satisfying $|{\vec{a}}|\leq 1$.  The Bloch vector $\vec{a}$ has a nice geometric interpretation for a qubit; it can be thought of as a vector inside of the unit sphere.  The magnitude of $\vec{a}$ represents the polarization of the density matrix.  For instance, if $|\vec{a}|=1$ then $\rho$ is a pure state and if $|\vec{a}|=0$ then 
\index{Bloch vector} \index{Bloch sphere} \index{polarization}
$\rho$ is a completely mixed state, with everything else in-between. 

In matrix form the density matrix takes the form
$$
\rho=\frac{1}{2}\left(\matrix{1+a_3 & a_1-ia_2\cr a_1+ia_2 & 1-a_3}\right),
$$
with eigenvalues $\frac{1\pm|\vec{a}|}{2}$.

When a state of the system evolves with an environment, it exchanges physical quantities such as polarization or phase with the environment.  It also becomes correlated with the state of the environment.  Such dynamics cannot be described adequately by a pure state; therefore density matrices are employed to study open dynamics.

\section{Closed dynamics}
\index{closed evolution|see{ unitary evolution}}
The dynamics of a density matrix is governed by the von-Neumann equation
\begin{eqnarray}\label{vonnuemann}
i\hbar\frac{\partial{\rho}}{\partial t}=[H(t),\rho],
\end{eqnarray}
where $H(t)$ is the time dependent Hamiltonian\footnote{We take $\hbar=1$ from here on.}.  
\index{von Neumann equation}
By integrating the above equation, we find that the time evolution is also be expressed by a unitary operator, $U$, as follows
\begin{eqnarray}
\rho(t)=\mathcal{U}(\rho(t_0))
=U(t,t_0)
\rho(t_0)
U^\dag(t,t_0).
\end{eqnarray}
The unitary operator, $U(t,t_0)$, takes the density matrix from time $t_0$ 
\index{unitary evolution}
to $t$ in a linear fashion
\begin{eqnarray}
U\left(\sum_jp_j\rho^{(j)}(t_0)\right)U^\dag
=\sum_jp_j\rho^{(j)}(t),
\end{eqnarray}
where $\sum_jp_j=1$. That is, the unitary operations preserve the mixture.
An unitary operator in general can be written as
\begin{eqnarray}
U=\mathscr{T}\;\left\{exp\left(-i \int_{t_0}^{t}H(t') dt'\right)\right\},
\end{eqnarray}
where $\mathscr{T}$ represents time ordering and $H(t')$ is the time dependent Hamiltonian.  In this dissertation we will only work with time independent Hamiltonians for simplicity, though all results \index{Hamiltonian evolution|see{ unitary evolution}}
apply to time dependent cases as well. 

In quantum information theory, the information is processed by passing quantum states through various quantum logic gates, which are represented by unitary operations.  Then we are interested only in the finite time evolution of a quantum state.  Therefore, it is far more convenient to describe the evolution of quantum states with unitary operators rather than with the von Neumann differential equation. 

In the next section we investigate the characteristics of an open quantum system.  Our goal is to describe the most general time evolution of the system of interest.  The most general evolution includes non-dissipative evolution, dissipative evolution, and the combination of the two.

\section{Open dynamics}
\index{open evolution}
In general, a state going through a quantum logic gate is not only subject to the gate, but also to the surrounding environment.  Such unwanted influences lead to noisy information and processing errors.  This noise should be minimized for optimal performance.  To optimize a gate we must first understand the additional influences from the environment on the state.  The presence of an interacting  environment leads the system to experience ``open dynamics", where physical quantities such as polarization and phase are affected by the state of the environment.

More specifically, a quantum state in dynamics can experience changes in the relative phase between the orthogonal components as well as changes in its polarization.  We can recognize that the magnitude of the polarization of the system density matrix alone does not change under unitary transformations; that is eigenvalues of the density matrix are invariant under unitary transformations but not the eigenvectors.  

For a qubit, the Bloch vector under a unitary transformation goes through a rotation in the Bloch sphere, $\vec{a}(t_0)\rightarrow\vec{a}'(t)$. While the magnitude of the Bloch vector remains unchanged, $|\vec{a}(t_0)|=|\vec{a}'(t)|$.  In open evolution, the Bloch vector will experience rotations as before; additionally its magnitude may change in time, $|\vec{a}(t_0)|\neq|\vec{a}'(t)|$ (see Fig \ref{opendyn} for a graphical explanation).

\begin{figure}[!ht]
\begin{center}
\resizebox{10 cm}{9.6 cm}
{\includegraphics{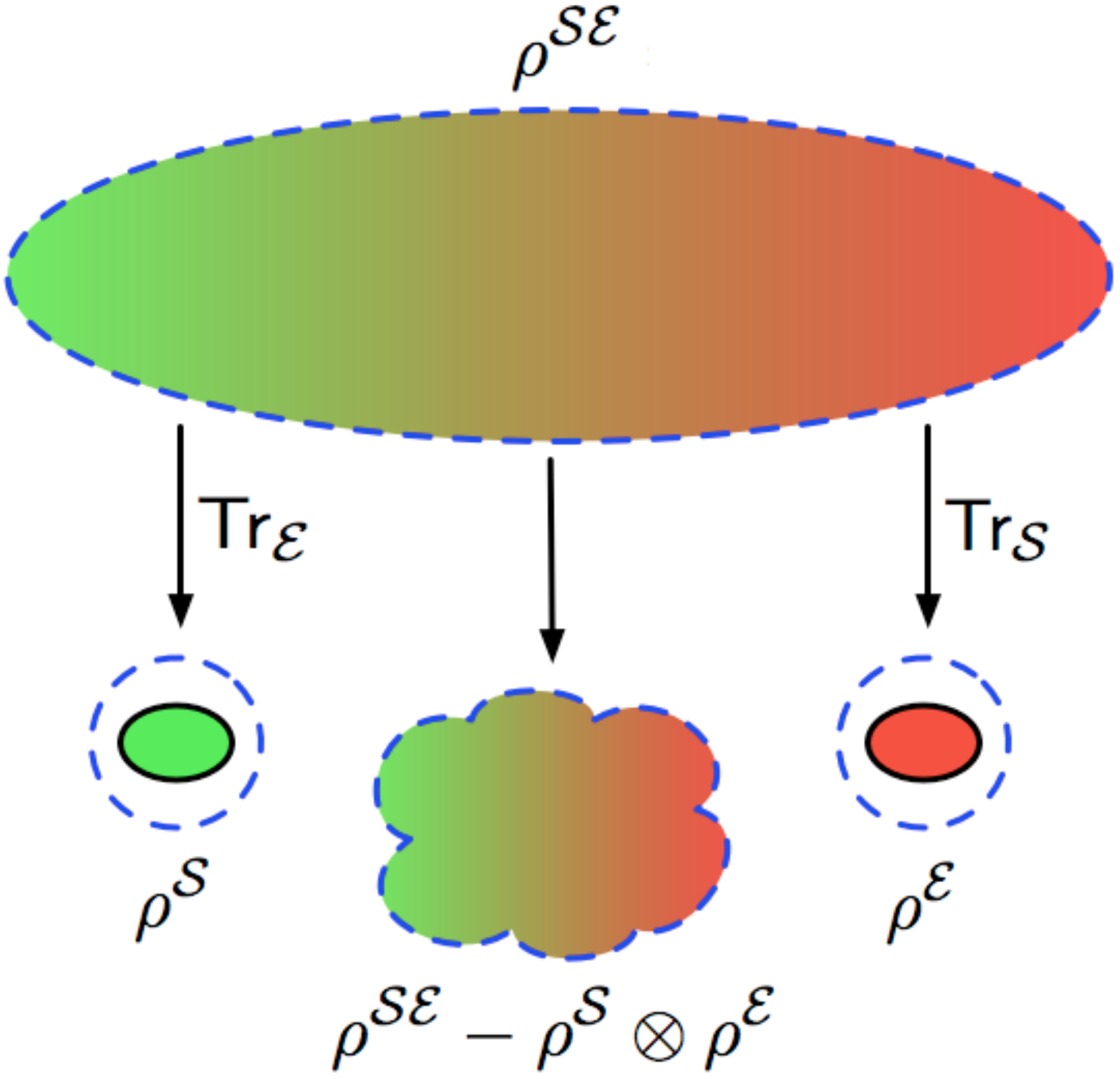}}
\caption{\label{bipartite}
The total state of the system and environment ($\rho^\mathcal{SE}$), with the state of the system ($\rho^\mathcal{S}$) is represented by color green and the state of the environment ($\rho^\mathcal{E}$) is represented by color red.  The fuzzy part in-between represent the correlations between the system and the environment ($\rho^\mathcal{SE}- \rho^\mathcal{S} \otimes \rho^\mathcal{E}$).  The dotted blue lines represent the set all physical states in that space.  Due to the correlations not all possible states of the system and the environment are allowed. Only the states that are compatible with the correlations are allowed, represented by the green and red ellipses on the bottom .}
\end{center}
\end{figure}

We can write down the interaction between the system and environment easily by treating the combined evolution in the closed from. Consider a bipartite state $\rho^{\mathcal{SE}}$ of the system (labeled by $\mathcal{S}$) and the environment (labeled by $\mathcal{E}$). The total unitary evolution is as follows
\begin{eqnarray}
\rho^{\mathcal{SE}}(t)=U\rho^{\mathcal{SE}}(t_0)U^\dag.
\end{eqnarray}
For generality, the state of the system and the environment is assumed to be correlated initially (see Fig. \ref{bipartite}). We also assume that we cannot observe the state of the environment and do not know what the interaction unitary transformations are.  If we had the knowledge of the state of the environment and the unitary transformations, by calculating the closed evolution we would know what the state of system will be at any point. Our only knowledge is of the initial and final states of the system.  The initial and final states can be obtained by averaging over the environment in the above equation. 
\index{partial trace} 
Mathematically this corresponds to a partial trace operation
\begin{eqnarray}
\rho^{\mathcal{S}}(\cdot)=\mbox{Tr}_{\mathcal E}[\rho^{\mathcal{SE}}(\cdot)].
\end{eqnarray}
See Fig \ref{opendyn} for a graphical explanation.
\begin{figure}[!ht]
\begin{center}
\resizebox{10 cm}{3.33 cm}
{\includegraphics{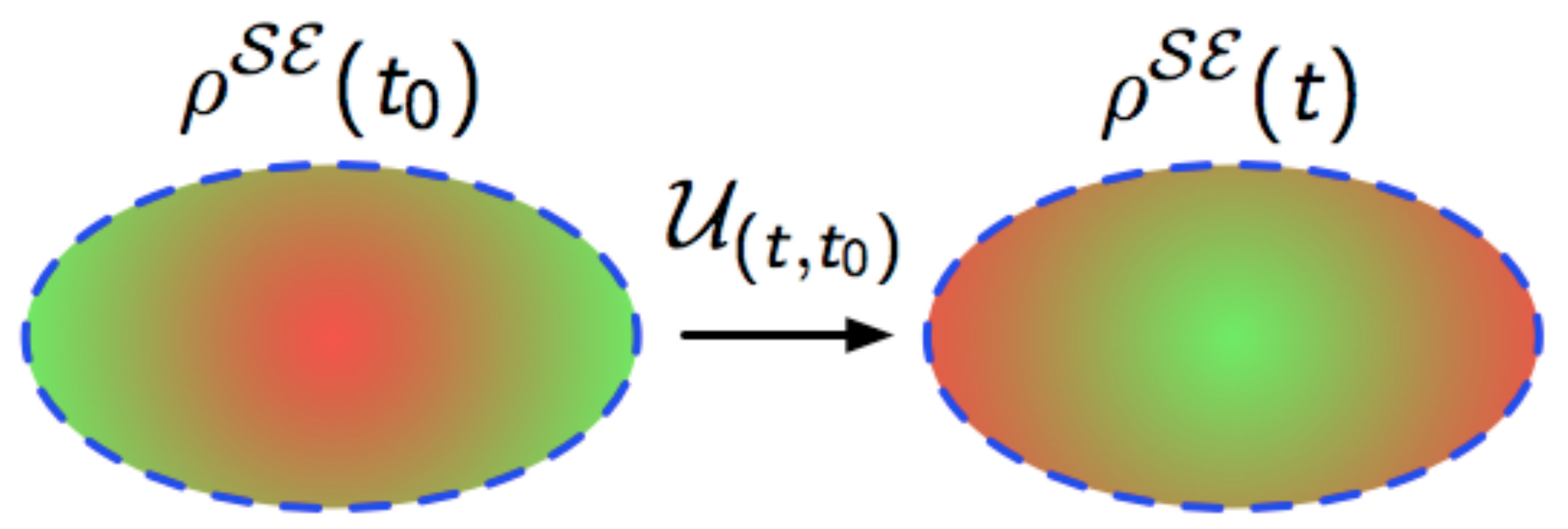}}
\resizebox{10 cm}{4.67 cm}
{\includegraphics{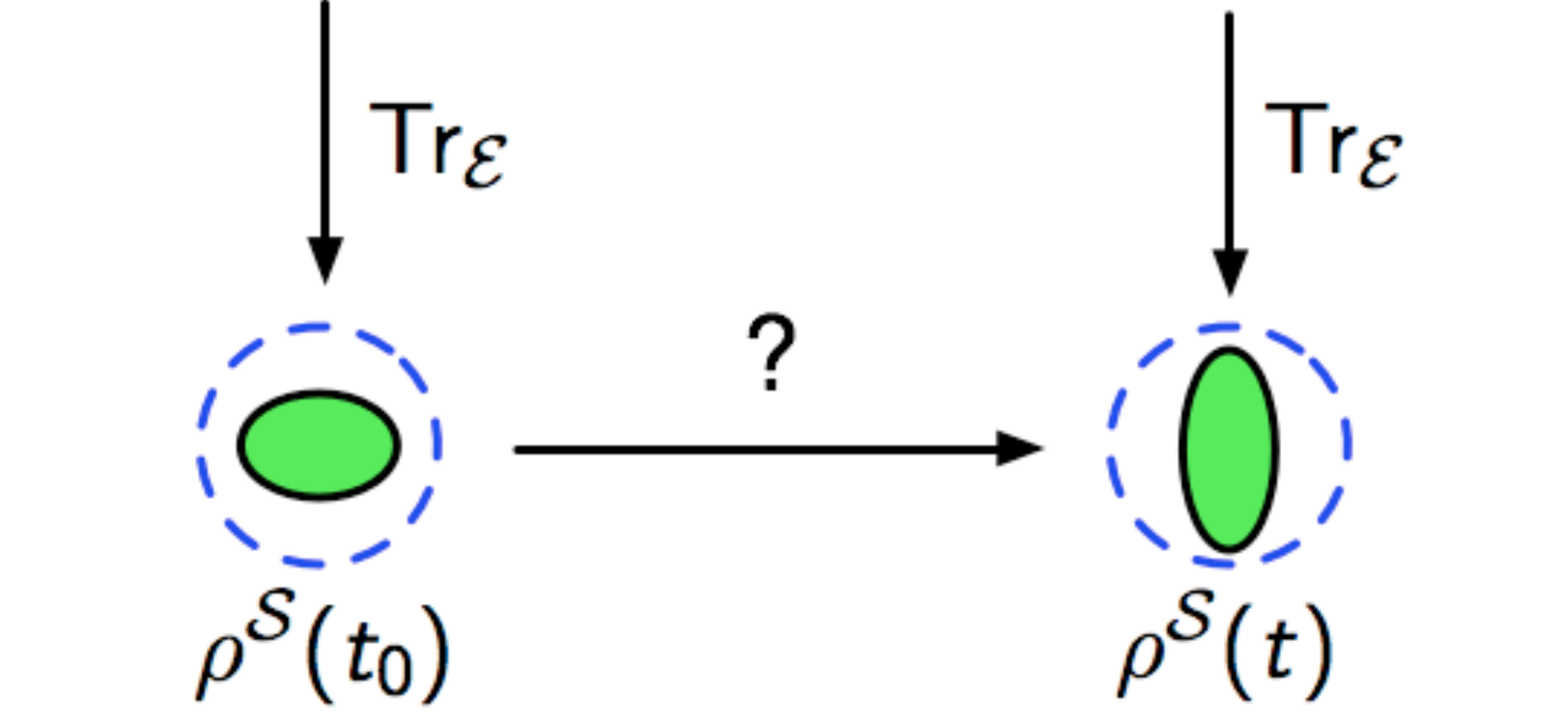}}
\caption{\label{opendyn}
[Top] The total state $\rho^\mathcal{SE}$ evolves unitarily, which changes the polarization of the system, the environment, and correlations between them (note the color changes).
[Bottom]
While the reduced state of the system does not evolve unitarily.  Note that the shape and the size of the set of states of the system (green ellipses) change from the initial time to the final time. We want to be able to describe this transformation.}
\end{center}
\end{figure}

There is no general equation in the differential form (like Eq. \ref{vonnuemann}) that governs the dynamics of the state of the system from $t_0$ to $t$\footnote{If the process is Markovian in nature then the Kossakowski-Lindblad equation can be applied \cite{kossakowski,lindblad,sudarshangorini,spohn80a}. A Markovian process is when the state of the system only depends on the state in the previous step.}.  It is easy to see that the mixture of the states of the system initially will not be the final mixture.  For instance, $\rho^{\mathcal{S}}(t_0)$ could begin as a pure state, and evolve into a mixed state, or vice versa.  This is due the exchanges of physical quantities, such as polarization, between the system and the environment. They may also become correlated by loosing polarization to correlations.  This exchange can be periodic\footnote{Systems that have a memory are called non-Markovian. In this case the memory effect is due to the correlations with the environment \cite{cesarnonmarkov}.}, semi-periodic, or purely relaxing.  For periodic and semi-periodic cases, the system may start pure, become mixed at some intermediary step, and then become pure or almost pure once again.

\section{Dynamical map formalism}
\index{dynamical map} \index{open evolution}

Suppose we are not interested in the dynamics of the system for all times 
\index{open quantum theory}
but rather in the transformation of the state from time $t_0$ to time $t$.  Then we need to define a mapping from density matrices to density matrices such that all allowed initial states of the system are mapped to the corresponding final states in a linear fashion.  We closely follow the arguments originally put forth by Sudarshan et al. \cite{SudarshanMatthewsRau61}.

\subsection{$\mathcal{A}$-form}
\index{dynamical map! $\mathcal{A}$-form}

Consider an operator acting on the density matrix mapping the state to another density matrix linearly
\begin{eqnarray}
\rho^{\mathcal{S}}_{r's'}(t_0)\rightarrow\mathcal{A}_{rs;r's'}\rho^{\mathcal{S}}_{r's'}(t_0)=\rho^{\mathcal{S}}_{rs}(t).
\end{eqnarray}
Above $\rho^{\mathcal{S}}$ is labeled by two indices and the operator $\mathcal{A}$ is labeled by four, meaning if $\rho^{\mathcal{S}}$ is a $d\times d$ matrix then $\mathcal{A}$ is $d^2\times d^2$ matrix.  We can think of the above equation as a super matrix $\mathcal{A}$ acting on a column vector $\rho^{\mathcal{S}}$. This is very similar to a classical stochastic process, where a stochastic matrix maps a probability vector to another probability vector. Matrix is $\mathcal{A}$ called the stochastic map \cite{SudarshanMatthewsRau61,Davies70}.

The only restriction we need to place on the stochastic map is that it map a density matrix to another density matrix.  This implies that it must preserve trace, Hermiticity, and positivity of the density matrix.  These restriction translate into the following properties for $\mathcal{A}$
\begin{center}
\begin{tabular}{l l}
$\mathcal{A}_{nn,r's'}=\delta_{r's'}$ \hspace{1cm} 
& Trace preservation, \cr
$\mathcal{A}_{rs,r's'}=\left(\mathcal{A}_{sr,s'r'}\right)^*$ \hspace{1cm} &
Hermiticity preservation, \cr
$x^*_r x_s\mathcal{A}_{rs,r's'}y_{r'}y^*_{s'}\geq 0$ \hspace{1cm} &
Positivity. \cr
\end{tabular}
\end{center}

\subsection{$\mathcal{B}$-form}
\index{dynamical map! $\mathcal{B}$-form}

Following Sudarshan et al. \cite{SudarshanMatthewsRau61} again, let us define a dynamical map from the stochastic map by
\begin{eqnarray}
\mathcal{B}_{rr';ss'}=\mathcal{A}_{rs;r's'}.
\end{eqnarray}
For a $4\times 4$ map, matrix $\mathcal{A}$ tranforms as follows
\begin{eqnarray}\label{atobform}
\left(
\begin{array}{cccc}
\mathcal{A}_{11} &\mathcal{A}_{12} &\mathcal{A}_{13} &\mathcal{A}_{14} \\
\mathcal{A}_{21} &\mathcal{A}_{22} &\mathcal{A}_{23} &\mathcal{A}_{24} \\
\mathcal{A}_{31} &\mathcal{A}_{32} &\mathcal{A}_{33} &\mathcal{A}_{34} \\
\mathcal{A}_{41} &\mathcal{A}_{42} &\mathcal{A}_{43} &\mathcal{A}_{44} \\
\end{array}
\right)
\rightarrow
\left(
\begin{array}{cccc}
\mathcal{A}_{11} &\mathcal{A}_{12} &\mathcal{A}_{21} &\mathcal{A}_{22} \\
\mathcal{A}_{13} &\mathcal{A}_{14} &\mathcal{A}_{23} &\mathcal{A}_{24} \\
\mathcal{A}_{31} &\mathcal{A}_{32} &\mathcal{A}_{41} &\mathcal{A}_{42} \\
\mathcal{A}_{33} &\mathcal{A}_{34} &\mathcal{A}_{43} &\mathcal{A}_{44} \\
\end{array}
\right).
\end{eqnarray}

The properties of the stochastic map translate in the following manner for the dynamical map \cite{SudarshanMatthewsRau61}.
\begin{center}
\begin{tabular}{l l}
$\mathcal{B}_{nr',ns'}=\delta_{r's'}$ \hspace{1cm} 
& Trace preservation, \cr
$\mathcal{B}_{rr',ss'}=\left(\mathcal{B}_{ss',rr'}\right)^*$ \hspace{1cm} &
Hermiticity preservation, \cr
$x^*_r y_{r'}\mathcal{B}_{rr',ss'}x_{s}y^*_{s'}\geq 0$ \hspace{1cm} &
Positivity. \cr
\end{tabular}
\end{center}

The advantage of writing the dynamical map $\mathcal{B}$ is that it has nicer properties than the stochastic map $\mathcal{A}$. The dynamical map is Hermitian and its diagonal $d\times d$ block elements have unit trace.  For these nicer properties we have sacrificed the simplicity of the composition of the stochastic map on the state.  The stochastic map acts as a matrix on a state that is a column vector, while the dynamical map acts on the state in the following manner
\begin{eqnarray}
\mathcal{B}_{rr';ss'}\rho^{\mathcal{S}}_{r's'}(t_0)
=\rho^{\mathcal{S}}_{rs}(t).
\end{eqnarray}
The consequence of sacrificing the simple action of the stochastic map on density matrices is that we have sacrificed the semi-group property that comes with it.  We will discuss this matter in the next section.

First, we should remark that since the dynamical map can transform all possible states to all possible states, it describes the most general evolution of a quantum state for finite time.  We need nothing more to to describe any time evolution of a state. Furthermore, the set of density matrices form a convex set, then the set of trace preserving positive maps that act on density matrices also form a convex set.  Then any map written as
\begin{eqnarray}
\mathcal{B}=\sum_j q_{j} B_{j}\;\;\;;\;\;\;0<q_j\;\;\;;\;\;\;\sum_j q_j=1
\end{eqnarray}
is a valid trace preserving positive map, given that all $\mathcal{B}_{j}$ positive and preserve trace.  Any positive maps that cannot be written as a 
\index{dynamical map! convexity of}\index{extremal maps}
convex sum of other positive map is called an extremal map.

\subsection{Semi-group property}
\index{semi-group property}
Suppose map $\mathcal{A}^{(1)}(t_1,t_0)$ takes a state from $t_0$ to $t$ and map $\mathcal{A}^{(2)}(t_2,t)$ takes the state from $t_1$ to $t_2$.  Then the map that takes the state from $t_0$ to $t_2$ is simply
\begin{eqnarray}
\rho^{\mathcal{S}}_{rs}(t_2)
&=&\mathcal{A}^{(2)}_{rs;r's'}(t_2,t_1)
\rho^{\mathcal{S}}_{r's'}(t_1)\\
&=&\mathcal{A}_{rs;r's'}^{(2)}(t_2,t_1)
\mathcal{A}^{(1)}_{r's';r''s''}(t_1,t_0)
\rho^{\mathcal{S}}_{r''s''}(t_0)\\
&=&\mathcal{A}^{(3)}_{rs;r''s''}(t_2,t_0)
\rho^{\mathcal{S}}_{r''s''}(t_0),
\end{eqnarray}
where $\mathcal{A}^{(3)}=\mathcal{A}^{(2)}\mathcal{A}^{(1)}$ simple by matrix multiplication.

While for the corresponding dynamical maps we have
\begin{eqnarray}
\rho^{\mathcal{S}}_{rs}(t_2)
&=&\mathcal{B}^{(2)}_{rr';ss'}(t_2,t_1)
\rho^{\mathcal{S}}_{r's'}(t_1)\\
&=&\mathcal{B}^{(2)}_{rr';ss'}(t_2,t_1)\circ\mathcal{B}^{(1)}_{r'r'';s's''}(t_1,t_0)
\rho^{\mathcal{S}}_{r''s''}(t_0)\\
&=&\mathcal{B}^{(3)}_{rr'';ss''}(t_2,t_0)\rho^{\mathcal{S}}_{r''s''}(t_0),
\end{eqnarray}
where $\circ$ represents composition of two maps, and not simple matrix multiplication.  The same simple composition property of the stochastic maps does not hold for the dynamical maps, i.e. $\mathcal{B}^{(3)} \neq\mathcal{B}^{(2)}\mathcal{B}^{(1)}$. In general, knowing the mapping from $t_{0}$ to $t$ does not tell us very much about the dynamics from $t$ to $t_2$.  This is due to the fact, whatever the correlations that the system and environment build up during their interaction from time $t_0$ to $t$ can comeback and play a role in the dynamics of the system from time $t$ to $t_2$.  This memory effect in the dynamics results in  non-Markovian dynamics \cite{sudarshangorini,cesarnonmarkov}.

A special but important case is when the dynamical map has the semi-group property. The resulting dynamics is Markovian. A Markovian system has no memory, meaning the state of the environment and the correlations with the environment do not affect the dynamics of the system.  In that case we can simply write down the state of the system at $t_n$ as
\begin{eqnarray}
\rho^{\mathcal{S}}(t_n)&=&\mathcal{B}(t_n,t_{n-1})\circ\cdots\circ\mathcal{B}(t_2,t)\circ\mathcal{B}(t_1,t_0)\rho^{\mathcal{S}}(t_0)\\
&=&\left(\mathcal{B}(t_1,t_0)\right)^{n}\rho^{\mathcal{S}}(t_0).
\end{eqnarray}
This case leads to the well known Kossakowski-Lindblad master equation \cite{kossakowski,lindblad}.

\subsection{Positivity classes}
\index{positivity classes of maps}
One advantage of the dynamical map is that it is Hermitian.  This means that it has a real spectrum, and we may talk about its positivity.  The positivity constraint placed on the dynamical map earlier turns out to be too strong \cite{pechukas94a,PhysRevLett.75.3020}.  This in itself is a rich topic, and still a highly debated issue in the community \cite{CarteretTernoZyczkowski05, CesarEtal07}. We will not go into the details of that discussion here; though it will be fruitful to define positivity classes associated with maps.  There are three classes associated with the positivity of  dynamical maps:

\begin{itemize}
\item[]{\emph{Completely positive}: A map is called completely positive if all of its eigenvalues are positive semi-definite. Such maps always map all positive states to positive states \cite{choi72a,choi75}. Thus, the valid domain for a completely positive map is the set of all states.  Unitary maps are an example of this class.}
\item[]{\emph{Positive}: A map is called positive if not all of its eigenvalues are positive, but it maps all positive states to positive states. The transpose map for a qubit has this quality. It has one negative eigenvalue, but the transpose of any density matrix is another valid density matrix \cite{Sudarshan86}.}
\item[]{\emph{Negative}: A map is called negative if it maps any positive state to a negative state. These maps have at least one negative eigenvalue. Maps of this kind are still physically valid for a certain set of states \cite{jordan06a}.  This set is called the compatibility domain \cite{shaji05dis}.}
\end{itemize}
The set of positive and negative maps are often put in the same category known as \emph{not-completely positive maps}.  For our purpose, we will assume that the dynamical map need not always be completely positive. In fact the negative eigenvalues will come in handy later on in chapter \ref{chapqptex}.

\subsection{Choi representation}\label{osr}
\index{Choi representation}
\index{operator sum representation|see{ Choi representation}}
Because the dynamical map has a real spectrum, we can write its action in terms of its eigenmatrices $\{\zeta^{(m)}\}$ and eigenvalues $\{\lambda_m\}$ as
\begin{equation}\label{oper1}
\mathcal{B}_{rr';ss'}\rho^{\mathcal S}_{r's'}
=\sum_{m}\lambda_{m}\;
\zeta^{(m)}_{rr'}\rho^{\mathcal S}_{r's'}\;{\zeta^{(m)}_{ss'}}^{*},
\end{equation}
where $1\leq m\leq d^2$ (because $\mathcal{B}$ is a $d^2\times d^2$ matrix, it has $d^2$ eigenmatrices).  If the eigenvalues are positive then we can absorb them into the eigenmatrices
$C^{(m)}\equiv\sqrt{\lambda_{m}}\zeta^{(m)}$ to get
\begin{equation}
\mathcal{B}\rho^{\mathcal S}=\sum_{m}C^{(m)}\rho^{\mathcal S}\; {C^{(m)}}^{\dagger},\label{oper2}
\end{equation}
with
\begin{eqnarray}
\sum_{m}{C^{(m)}}^{\dagger}C^{(m)}=1.
\end{eqnarray}
Eq. \ref{oper2} is know as the \emph{Choi representation}\footnote{ The $C$-matrices are not unique; they can be unitarily rotated, but the dynamical map $\mathcal{B}$ is unique.}.  This representation was first pointed out by Sudarshan et al. \cite{SudarshanMatthewsRau61}, however Choi made use of this representation to study the positivity classes for maps  \footnote{This representation is also known as the \emph{operator sum representation}, the \emph{Stinespring form}, and the \emph{Kraus canonical form}.}.  Any map that can be written in the Choi form is completely 
\index{$C$-matrices}\index{complete positivity}
positive \cite{choi72a,choi75, Sudarshan86, SudChaos}, which is a convenient definition for complete positivity. 

The advantage of writing the dynamical map in the Choi representation is that now the dynamics is governed by a set of operators in the space of the system.  Unitary evolution is a special case in this representation
$$
\rho^\mathcal{S}\rightarrow U\rho^{\mathcal S} U\;\;\;
\mbox{with}\;\;\;C^{(1)}\equiv U,\;\;\; C^{(m)}=0 \;\;\;(\mbox{for all } m>1).
$$

\section{Dynamical maps from contraction}
\index{dynamical map! from contraction}

Let us now consider dynamical maps coming from the contraction of Hamiltonian  evolution.  If we know the initial state of system 
\begin{eqnarray}
\rho^{\mathcal S}(t_0)=\mbox{Tr}_{\mathcal E}[\rho^{\mathcal{SE}}(t_0)]
\end{eqnarray}
and the final state
\begin{eqnarray}
\rho^{\mathcal S}(t)=\mbox{Tr}_{\mathcal E}[\rho^{\mathcal{SE}}(t)],
\end{eqnarray}
can we find the dynamical map? 

To make our examples physical we consider a toy model by taking a combined state of the system and the environment, and letting it develop in time unitarily. Then we can find the map by considering the evolution of the reduced states of the system
\begin{eqnarray}
\rho^{\mathcal S}(t)&=&\mathcal{B}\rho^{\mathcal S}(t_0)\\
&=&\mbox{Tr}_{\mathcal E}[U\rho^{\mathcal{SE}}(t_0)U^\dag],
\end{eqnarray}
where the initial and final states are obtained by contracting the environmental degrees of freedom.  This is all we are allowed to know in order find the dynamical map. We will workout two examples in two different manners to obtain the dynamical map in each case.

\subsection{Initially product states}\label{simpsep}
\index{example of! dynamical map! initially product state}

Let us now construct the dynamical map from physical motivations.  Let us start by looking at the special case of an initially  product state \index{simply separable|see{ product state}}\index{product state}
(simply separable) $\rho^{\mathcal{SE}}=\rho^{\mathcal{S}} \otimes \rho^{\mathcal{E}}$.  We can write the action of the dynamical map in terms of matrix indices as follows
\begin{eqnarray}
\mathcal{B}_{rr';ss'}\rho^\mathcal{S}{r's'}
=\sum_{\alpha\beta\epsilon}
U_{r\epsilon ;r'\alpha}
\rho^{\mathcal{S}}_{r's'} \rho^{\mathcal{E}}_{\alpha\beta} 
U^*_{s\epsilon ;s'\beta}.
\end{eqnarray}
Then we can construct the dynamical map right away by removing $ \rho^{\mathcal{S}}$ from both sides.  In the $\mathcal{B}$ form the map is
\begin{eqnarray}\label{envsize}
\mathcal{B}_{rr';ss'}=\sum_{\epsilon\alpha\beta}U_{r\epsilon ;r'\alpha}
\rho^{\mathcal{E}}_{\alpha\beta} U^*_{s\epsilon ;s'\beta}.
\end{eqnarray}
The final state of the system is given by
\begin{eqnarray}\label{finstatedyn}
\rho^{\mathcal S}_{rs}(t)
&=&\mathcal{B}_{rr';ss'}\rho^{\mathcal{S}}_{r's'}(t_0)\nonumber\\
&=&\left(\sum_{\epsilon\alpha\beta}
U_{r\epsilon;r'\alpha}
\rho^{\mathcal E}_{\alpha\beta} 
U^*_{s\epsilon;s'\beta}\right)\rho^{\mathcal S}_{r's'}(t_0).
\end{eqnarray}
We can see in the last equation that the dynamical map acts linearly on the state of the system, i.e. it does not depend on any of the parameters of the initial state of the system. The linearity here is consequence of the linearity of quantum mechanics.    The map only connects the initial state of system to the final state of the system.

Since $\rho^{\mathcal{E}}$ is a positive matrix, we can take its square root and distribute it.  This way we can obtain the Choi representation
\begin{eqnarray}
\mathcal{B}_{rr';ss'}
&=&\sum_{\epsilon\gamma}
\left(\sum_{\alpha}U_{r\epsilon;r'\alpha}
\left[\sqrt{ \rho^{\mathcal{E}}}\right]_{\alpha\gamma} \right)
\left(\sum_{\beta}U_{s\epsilon;s'\beta}
\left[\sqrt{\rho^{\mathcal{E}}}\right]_{\beta\gamma}\right)^*\\
&=&\sum_{\epsilon\gamma}
C^{(\epsilon\gamma)}_{rr'}{C^{(\epsilon\gamma)}_{ss'}}^*.
\end{eqnarray}
Finally we have
\begin{eqnarray}
\mathcal{B}_{rr';ss'}\rho^{\mathcal S}_{r's'}
=\sum_m C^{(m)}_{rr'}
\rho^{\mathcal S}_{r's'}{C^{(m)}_{ss'}}^*.
\end{eqnarray}
This last equation is in the same form as Eq. \ref{oper2}, which proves that maps arising from initially uncorrelated states are always completely positive.  Conversely, any completely positive map can be thought as coming from a unitary evolution of the system and the environment in the product form.

\subsection{Initially correlated states}\label{inicorr}
\index{example of! dynamical map! initially correlated state}

Let us now look at an example of a dynamical map for a total state where the system and the environment are initially correlated. Let us work with the simplest open system, a state of two qubits.  We will treat the first qubit as the system and the second as the environment.  The density matrix for two qubits in general is
\begin{eqnarray}
\rho^{\mathcal{SE}} =  \frac{1}{4} \{\mathbb{I}\otimes\mathbb{I}
+a_j \sigma_j\otimes\mathbb{I}
+b_k \mathbb{I}\otimes\sigma_k
+c_{jk}\sigma_j\otimes\sigma_k\},
\end{eqnarray}
where $\sigma_j$ are the Pauli spin matrices, $a_j$ and $b_k$ are the Bloch 
\index{density matrix! of two qubits}
vector components of the system and the environment respectively, and $c_{jk}$ represent the correlations between the system and the environment.

For our example we set $b_k=c_{jk}=0$ except $c_{23}$ to simplify the calculations.  Then we have the following density matrix for our two qubit system
\begin{eqnarray}\label{sep1}
\rho^{\mathcal{SE}}(t_0) =  \frac{1}{4} \{\mathbb{I}\otimes\mathbb{I}+a_j
\sigma_j\otimes\mathbb{I}+c_{23}\sigma_2\otimes\sigma_3\},
\end{eqnarray}
This state is separable but it is not a product state.  Let us evolve the total state with the unitary operator
\begin{eqnarray}\label{unitarys}
U = e^{-iHt} 
= \prod_{j=1,2,3}\left\{\cos\left(\omega t\right)\mathbb{I}\otimes\mathbb{I} -i\sin\left(\omega t\right)\sigma_j\otimes\sigma_j\right\},
\end{eqnarray}
where 
\begin{eqnarray}
H =  \omega\sum_j \sigma_j \otimes \sigma_j.
\end{eqnarray}
The evolved state $U\rho^{\mathcal SE}(t_0)U^\dag$ is
\begin{eqnarray}
\rho^{\mathcal{SE}}(t)
&=&\frac{1}{4}
\left\{\mathbb{I}\otimes\mathbb{I}
+a_j \cos^2(2\omega t)\;\sigma_j\otimes\mathbb{I}
-c_{23} \cos(2\omega t)\sin(2\omega t) \;\sigma_1\otimes\mathbb{I}\right.\nonumber\\
&&\;\;\;\;\;+a_j\sin^2(2\omega t)\;\mathbb{I}\otimes\sigma_j
+ c_{23} \cos(2\omega t)\sin(2\omega t)\;\mathbb{I}\otimes\sigma_1\nonumber\\
&&\;\;\;\;\;+a_l \cos(2\omega t)\sin(2\omega t) \;(\sigma_m\otimes\sigma_n-\sigma_n\otimes\sigma_m)
\nonumber\\
&&\;\;\;\;\;\left.+c_{23}\;\cos^2(2 \omega t)\;\sigma_2\otimes\sigma_3
+c_{23}\;\sin^2(2 \omega t)\;\sigma_3\otimes\sigma_2\nonumber
\right\},
\end{eqnarray}
where $j$ runs from 1 to 3 and $l,m,n$ are distinct and cyclic.  

We are now in the position to find the dynamical map for the system going from $t_0$ to $t$. For that we need the states of the system at those times.  These are obtained by tracing the environment out of the total state at these two times.  The initial state is
\begin{eqnarray}\label{initialstate}
\rho^{\mathcal{S}}(t_0)=\mbox{Tr}_\mathcal{E}[\rho^{\mathcal{SE}}(t_0)] =\frac{1}{2}\{ \mathbb{I}+a_j \sigma_j \},
\end{eqnarray}
and the the state of the system at time $t$ is given by
\begin{eqnarray}\label{finalstate}
\rho^{\mathcal S}(t)
&=&\mbox{Tr}_\mathcal{E}[U \rho^{\mathcal{SE}}(t_0) U^\dagger]\nonumber\\
&=&\frac{1}{2}\{ \mathbb{I} + \cos^2\left(2\omega t\right) a_j
\sigma_j -c_{23}\cos\left(2\omega t\right) \sin\left(2\omega
t\right)\sigma_1 \}.
\end{eqnarray}

We can see how the linearly independent elements, that span the space of the system, are mapped from $t_0$ to $t$ \footnote{A more general prescription of one qubit system and one qubit environment is given in \cite{rodr,jordan06a}.}. 
\begin{eqnarray}
\mathcal{A}:\;\;\;
\frac{1}{2}\left(\begin{array}{c}
1+a_3 \\ a_1-i a_2  \\ a_1+ia_2 \\ 1-a_3
\end{array}\right)\longrightarrow
\frac{1}{2}\left(\begin{array}{c}
1+ C^2 a_3 \\ C^2 (a_1-i a_2) -c_{23} CS \\ C^2 (a_1+ia_2)-c_{23} CS \\ 1-C^2 a_3
\end{array}\right),
\end{eqnarray}
where $C=\cos(2\omega t)$ and $S=\sin(2\omega t)$. Since the stochastic map $\mathcal{A}$ acts on the state as a matrix on a vector, we can construct it by inspection as follows
\begin{eqnarray}
\mathcal{A}=\frac{1}{2}
\left(\begin{array}{cccc}
1+C^2 & 0 & 0 & 1-C^2 \\
-c_{23} CS & 2 C^2 & 0 & -c_{23} CS\\ 
-c_{23} CS & 0 & 2 C^2 & -c_{23} CS\\ 
1-C^2 & 0 & 0 & 1+C^2
\end{array}\right).
\end{eqnarray}
Notice that $\mathcal{A}$ is not Hermitian, and the element of the first and the last rows add to unity.

To construct the dynamical map, we simply rearrange each row of $\mathcal{A}$ into the corresponding block matrix (see Eq. \ref{atobform}),
\begin{eqnarray}\label{dynamicalmap}
\mathcal{B}=\frac{1}{2}\left(
\begin{array}{cccc}
1+C^2 & 0 & -c_{23} CS & 2 C^2 \cr
0 & 1-C^2 & 0 &-c_{23} CS \cr
-c_{23} CS & 0 & 1-C^2  & 0 \cr
2C^2& -c_{23} CS & 0 &1+C^2 \cr 
\end{array}
\right).
\end{eqnarray}
Note that $\mathcal B$ is Hermitian, the $2\times 2$ block diagonal elements add to unity, and is an affine transformation \cite{jordan:034101}
that squeezes the Bloch sphere of the qubit into a sphere of radius $\cos^2 (2\omega t)$ and shifts its center by $c_{23}\cos(2\omega t)\sin(2 \omega t)$ in the $\sigma_1$ direction.

\begin{center}
\begin{figure}[!ht]
\resizebox{12.8 cm}{7.8 cm}{\includegraphics{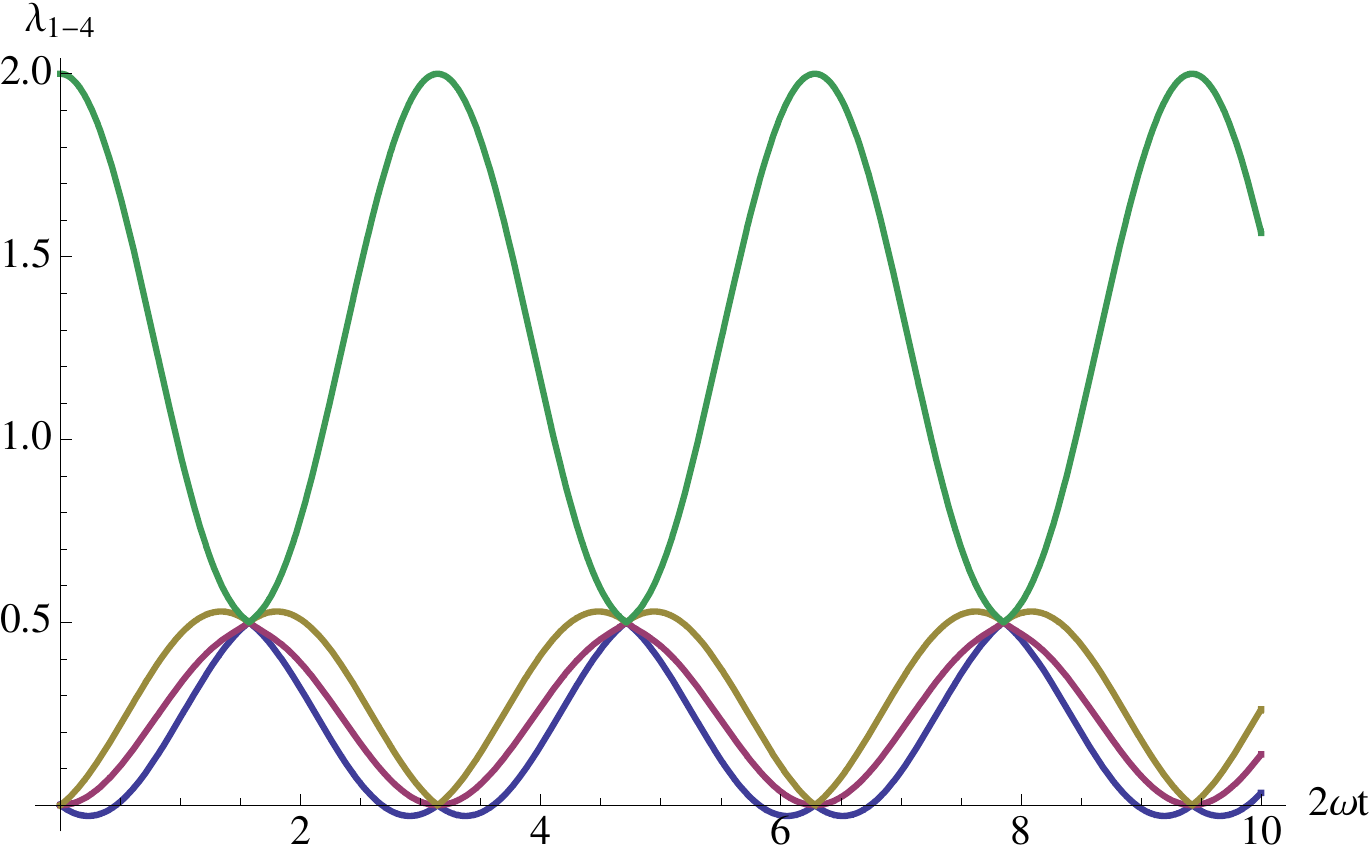}}
\caption{\label{fig2_1}
The eigenvalues of the dynamical map in Eq \ref{dynamicalmap} are plotted as function of $2 \omega t$.  One of the eigenvalue is negative for certain 
\index{negative maps! due to correlations}
values of $\omega t$; we have taken $c_{23}=0.5$.  The negative eigenvalue is due to the initial correlations between the system and the environment.}
\end{figure}
\end{center}

The eigenvalues of the map are
\begin{eqnarray*} 
\lambda_{1,2}&=&\frac{1}{2}\big[1-\cos^2 (2\omega t)\pm c_{23}\cos (2 \omega
t)\sin (2 \omega t ) \big], \\
\lambda_{3,4}&=&
\frac{1}{2}\left[1+ \cos^2 (2\omega t ) \nonumber
\pm \cos (2\omega t )\sqrt{ 4\cos^2 ( 2\omega
t )+c_{23}^{2} \sin^2 ( 2\omega t )} \right].
\end{eqnarray*}
It is easily seen that $\lambda_{3,4}$ are always positive. While, for
$\lambda_{1,2}$ to be positive, we need $\sin^2 (2\omega t ) \geq \pm c_{23}\cos
(2\omega t)\sin (2\omega t)$. We can choose $c_{23}$ such that this condition will be violated for some values of $\omega t$ making the map $\mathcal{B}$ not-completely positive.  The eigenvalues are plotted in Fig. \ref{fig2_1}. 

It has been previously shown that not-completely positive maps come from initial entanglement \cite{jordan:052110}. This example shows that initially correlated states can lead to not-completely positive maps. A similar example has been worked out in \cite{CarteretTernoZyczkowski05, CesarEtal07}. The map $\mathcal{B}$ has physical interpretations as long as it is applied to initial states $\rho^{\mathcal{S}}(t_0)$ that are {\em compatible} with the total state $\rho^{\mathcal{SE}}(t_0)$ \cite{shaji05dis}. Let us discuss 
\index{compatibility domain}\index{positivity domain}
these ideas in more detail.

\begin{figure}[!ht]
\begin{center}
\resizebox{9.33 cm}{8.4 cm}{\includegraphics{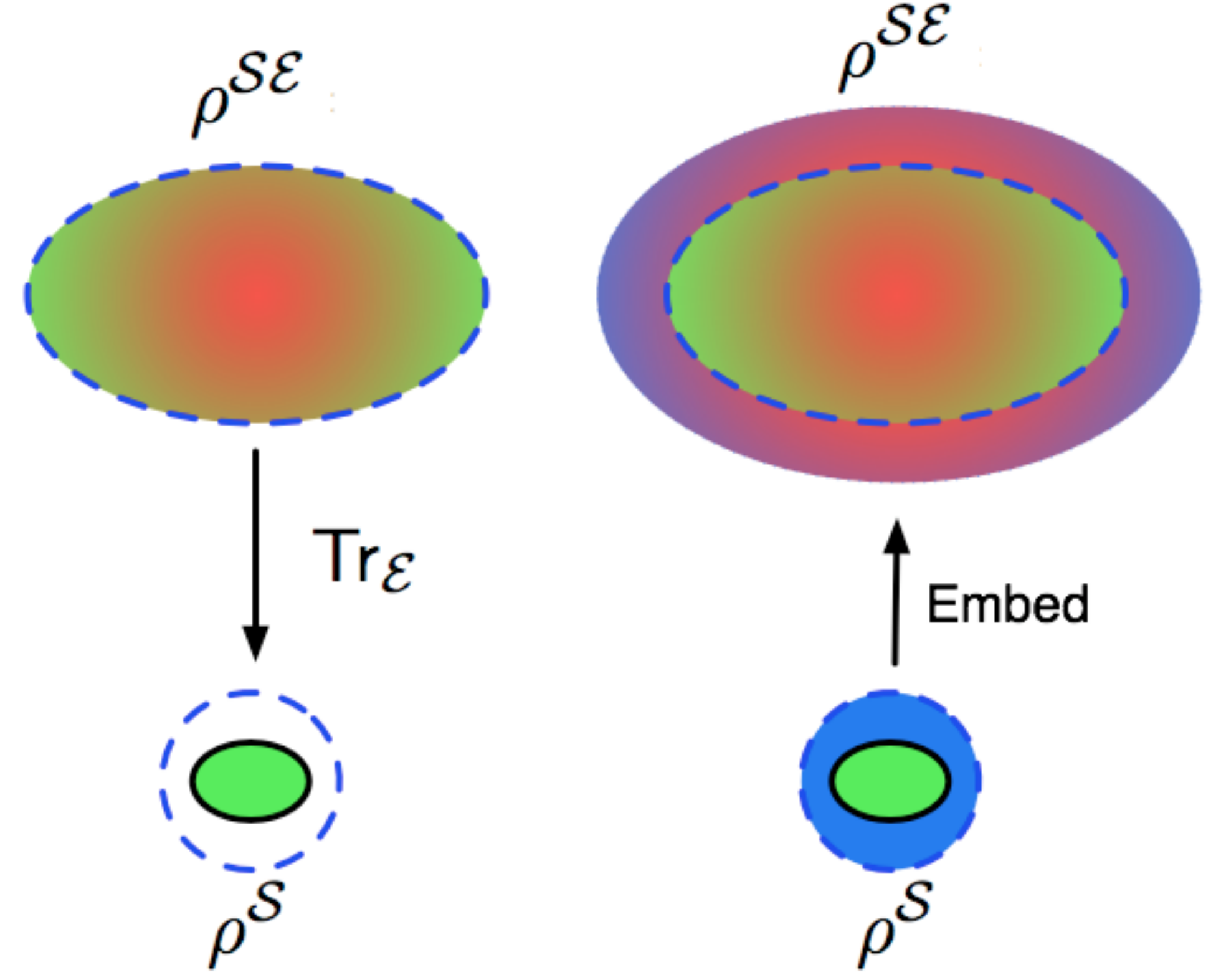}}
\caption{\label{embed} [Left]  The total state of the system and the environment has correlations, therefore the set states that are physically available for the system part (green ellipse on bottom) are less than set of all state (dotted blue circle). This is the compatibility domain. [Right] Conversely, if one demands to embed set of all system states into the total state (compatible are in the green ellipse and incompatible are represented by the blue part of the circle on the bottom), then the total state may become unphysical (fuzzy blue and red area above).}
\end{center}
\end{figure}

The state $\rho^{\mathcal{S}}(t_0)$ depends on the parameters ${a_j}$.  For $\rho^{\mathcal{S}}(t_0)$ to be positive the Bloch vector must be smaller than unity in magnitude, $|\vec{a}|^2\leq 1$. Additionally, the Bloch vector is constrained by the positivity condition of the \emph{total} state $\rho^{\mathcal{SE}}(t_0)$.  This constraint arises from the presence of the correlation term $c_{23}$. Because of this term, $\rho^{\mathcal S}(t_0)$ cannot be any state, for instance a pure state
\index{embedding}
(see Fig. \ref{embed} for a graphical explanation).  

\begin{figure}[!ht]
\begin{center}
\resizebox{10.27 cm}{8.4 cm}{\includegraphics{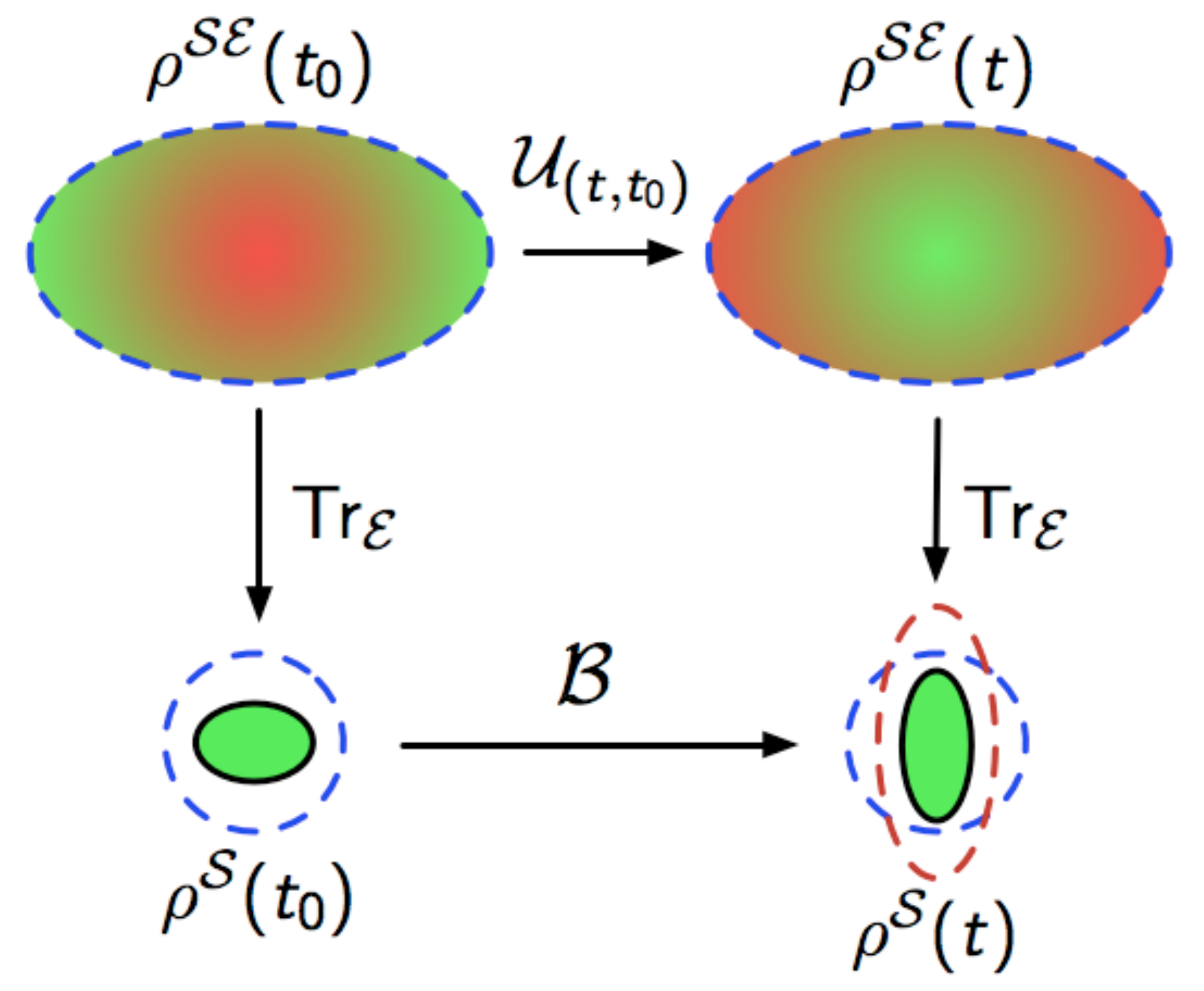}}
\caption{\label{NCPMapfig} 
[Top] The total state evolves unitarily, which changes the polarization of the system, the environment, and the correlations between them (note the color difference).
[Bottom] The dynamical map acting on the compatibility domain results in physical states (green ellipses from left to right).  But its action on set of all states (dotted blue circle on left) can take physical states to non-physical states (dotted red ellipse on right).}
\end{center}
\end{figure}

The value of the correlation can range from $-1\leq c_{23}\leq 1$.  If we fix $c_{23}$ in that range, then for the eigenvalues for $\rho^{\mathcal{SE}}(t_0)$ to be positive we get the following condition
\begin{eqnarray}
1\geq |a_1|^2 + (|a_2|+|c_{23}|)^2 +|a_3|^2.
\end{eqnarray}
According to the inequality, when $|c_{23}|>0$, $\rho^{\mathcal S}(t_0)$ cannot be a pure state; otherwise $\rho^{\mathcal{SE}}$ will not be positive. The volume enclosed by this inequality is the valid domain of the map, and the action of the map on this set of states has a physical interpretation. This domain is known as the \emph{compatibility domain} \cite{jordan:052110}.  The map acting outside this domain may transform a physical state to a non-physical state as shown in Fig \ref{NCPMapfig}. 

There is another domain associated with dynamical maps.  This domain comes from the unitary operators, and it called the \emph{positivity domain}. The positivity domain is the set of positive states that get mapped to a positive state by the dynamical map. The compatibility domain is smaller than the positivity domain, and it is a subset of the positivity domain. We will not show an example of the positivity domain here; it may be found in \cite{shaji05dis}.


\section{Size of the environment}\label{secenvsize}
\index{size of the environment}

The Choi representation shown in Sec. \ref{osr} tells us that only a finite number of operators govern the most general dynamics of the system.  Then we may ask what is the largest dimension of the environment that is necessary to simulate the most general dynamics?  It is often assumed that the dimensions of the environment is very large compared to the system.  We show here that this is not the case.

In Sec. \ref{simpsep} we showed that any completely positive map can be thought of as a contraction of initial product states evolving.  Let us re-examine Eq. \ref{envsize}
\begin{eqnarray}
\mathcal{B}_{rr';ss'}=\sum_{\epsilon\alpha\beta}U_{r\epsilon;r'\alpha}
\rho^{\mathcal{E}}_{\alpha\beta} U^*_{s\epsilon;s'\beta}.
\end{eqnarray}
Let us consider the state of the environment to be a pure state
\begin{eqnarray}
\mathcal{B}_{rr';ss'}&=&\sum_{\epsilon\alpha\beta}
U_{r\epsilon;r'\alpha}
\ket{\phi}_{\alpha}\bra{\phi}_{\beta}
U^*_{s\epsilon;s'\beta}\\
&=&\left(\sum_{\epsilon\alpha}
U_{r\epsilon;r'\alpha}
\ket{\phi}_{\alpha}\right)
\left(\sum_{\epsilon\beta}
U_{s\epsilon;s'\beta}
\ket{\phi}_{\beta}\right)^*\\
&=&\sum_{\epsilon}
C^{(\epsilon)}_{rr'}{C^{(\epsilon)}_{ss'}}^*.
\end{eqnarray}
If $\mathcal{B}$ is defined on a $d$ dimensional system then $\mathcal{B}$ is a $d^2\times d^2$ matrix.  Hence it has only $d^2$ $C$-matrices, therefore the index $\epsilon$ only needs to run to $d^2$.  The index $\epsilon$ also represents the dimensionality of the environment.  

Conversely, any completely positive map can be thought of as a contraction of a pure state of the environment of dimension $d^2$ and unitary $U$ (see \cite{Sudarshan86} for more details).  The unitary operator can be constructed from the $C$-matrices as
\begin{eqnarray}
C^{\epsilon\alpha}_{rr'}=U_{r\epsilon;r'\alpha}.
\end{eqnarray}
The value of $\alpha$ is fixed by the choice of the state of the environment.  Different $C$-matrices are determined by the value of $\epsilon$. All other elements of the unitary operator can be arbitrary, as long as they satisfy the unitarity condition for $U$.

\section{Discussion}

Let us review the most essential statements made in this chapter.  We have shown that dynamical maps govern the most general finite time evolution of a quantum state.  The not-completely positive maps are due to the initial correlations between the system and the environment.  These maps have a physical interpretation as long as their action is restricted to the compatibility domain.

We also showed that the size of the environment need not be very large for the most general evolution.  It has to be at most $d^2$, where $d$ is the size of the system.  With this in mind we will justify the validity of the simple examples that will be studied through out this dissertation.

It should now be clear that the dynamical map formalism is a very powerful theoretical tool.  One may ask, are dynamical maps experimentally realizable?  The answer is yes, quantum process tomography is an experimental procedure that allows one determine the open evolution of a system.  We will review several quantum process tomography procedures in the next chapter.
 

\chapter{Quantum process tomography}\label{chapqpt}
\index{Quantum process tomography @\emph{Quantum process tomography}}%

There has been significant experimental interest in the dynamics of quantum correlations, entanglement, and coherence in the context of quantum information theory \cite{Nielsen00a,bennett02a,bennett99b,StelmachovicBuzek01,shabani06a}. All of these categories require the study of multi-partite quantum systems.  Experimentally, for all of these cases, the external influences due to the environment are not uncommon. In the last chapter we introduced the dynamical map formalism to describe the open dynamics of a quantum system.  We now look at the experimental analog to dynamical maps, which we call 
\index{process map}
\emph{process maps}.

Quantum process tomography \cite{JModOpt.44.2455, PhysRevLett.78.390} is the experimental tool that determines the open evolution of a system that interacts with the surrounding environment. It is the tool that allows an experimenter to determine the unwanted action of a quantum process
\footnote{Sometimes the process is known, but we are only interested in characterizing the effects due to an unknown process.} on the quantum bits going through it. It is an important tool for quantum information processing. A state going through a quantum gate or a quantum channel will experience some interactions with the surrounding environment.  Quantum process tomography allows the experimentalist to distinguish the differences between the ideal process and the process found experimentally.  Therefore it is an important tool in quantum control design and battling decoherence (loss of polarization).

Today there are many variations of the original quantum process tomography procedure, namely \emph{ancilla (entanglement) assisted process tomography} \cite{PhysRevLett.86.4195, PhysRevA.67.062307, PhysRevLett.90.193601, PhysRevLett.91.047902}, \emph{direct characterization of quantum dynamics} \cite{mohseni:170501,mohseni:062331}, \emph{selective efficient quantum process tomography}\cite{paz}, and \emph{symmetrized characterization of noisy quantum processes}\cite{laflamme}. Some of these procedures have been experimentally tested \cite{Nielsen:1998py, PhysRevA.64.012314, PhysRevLett.91.120402, Wein:121.13, orien:080502, Howard06,Howard05, myrskog:013615}.

\section{Quantum process tomography: general description}
\index{quantum process tomography! general description}

The objective of quantum process tomography is to determine how a quantum process acts on different states of the system.  In very basic terms, a quantum process connects different quantum input states to different output states:
\begin{equation}
\mbox{input states} \rightarrow \mbox{process} \rightarrow \mbox{output states}.
\end{equation}
The complete behavior of the quantum process is known if the output state 
\index{input state}\index{output state}
for any given input state can be predicted.

A quantum process can be anything from controlled unitary evolution to dissipative open evolution (or in most cases some combination of the two) experienced by the state of the system.  The quantum process, thus can be described by a map just like the dynamical map from the last chapter.  The only difference being, this is now to be done experimentally.

The tomography aspect of quantum process tomography is to use a finite number of input states, instead of all possible states, to determine the quantum process.  
For instance, to determine the dynamical map we only need to know the mapping of the elements of the density matrix from an initial time to a final time,
\begin{eqnarray}
\mathcal{B}:\rho_{ij}(t_0)\rightarrow\rho_{ij}(t).
\end{eqnarray}
The elements of the density matrix linearly span the whole state space.  

Experimentally, we do not have the access to the individual elements of the density matrix; we can only prepare physical states.  Thus, a set of physical states that linearly span the state space will be sufficient for the experiment.  
A state space of dimension $d$ requires $d^2$ states to span the space.  Once the evolution of each these input states is known, by linearity the 
\index{linearly independent set}
evolution of any input state is known (see \cite{Nielsen00a} for detailed discussion).\footnote{The linearity of the quantum process is an assumption here.}

For example, the following four projections as input states are necessary 
to linearly span the whole state space of a qubit:
\begin{eqnarray}\label{prj}
P^{(1,-)} = \frac{1}{2}(\mathbb{I} - \sigma_1),&&
P^{(1,+)} = \frac{1}{2}(\mathbb{I} + \sigma_1 ),\nonumber\\
P^{(2,+)} = \frac{1}{2}(\mathbb{I} + \sigma_2), &&
P^{(3,+)} = \frac{1}{2}(\mathbb{I} + \sigma_3).
\end{eqnarray}
Any state of a qubit can be written as a unique linear combination of these four projections\footnote{The linear combination will not always be convex.  For example $P^{(2,-)}=P^{(1,+)}+P^{(1,-)}-P^{(2,+)}$.  Also notice that these four states form a linearly independent set, but they are not orthogonal to each other.}

Using the set linearly independent input states $P^{(m)}$, and measuring the corresponding output states $Q^{(m)}$\footnote{This requires performing \emph{quantum state tomography}, see App. \ref{appqst}.}, the evolution of an arbitrary input state can be determined. Let $\Lambda$ be the map describing the process, which we call \emph{process map}, and an arbitrary input state be expressed (uniquely) as a linear combination $\sum_j p_m P^{(m)}$.  The action of the map in terms of the matrix elements is as follows:
\begin{eqnarray*}
\sum_{r's'}\Lambda_{rr';ss'}\left(\sum_j p_j P_{r's'}^{(j)}\right) 
&=& \sum_j p_j Q_{rs}^{(j)}.
\end{eqnarray*}

\section{Standard quantum process tomography}
\index{standard quantum process tomography|see{ quantum process tomography}}
\index{quantum process tomography! standard}

In quantum process tomography, it is often assumed that at the beginning  of the experiment the state of the system and environment are uncorrelated.  At that point they evolve in a closed form under some unitary transformation.  Under this assumption we write down the dynamical equation in terms of matrix indices
\begin{eqnarray}
Q^{(m)}_{rs}
&=&\sum_{\epsilon,\alpha,\beta}
U_{r\epsilon;r'\alpha}
P^{(m)}_{r's'}
\rho^{\mathcal{E}}_{\alpha\beta}
U^*_{s\epsilon;s'\beta}\\
&=&\left(\sum_{\epsilon,\alpha,\beta}\label{processmapsto}
U_{r\epsilon;r'\alpha}
\rho^{\mathcal{E}}_{\alpha\beta}
U^*_{s\epsilon;s'\beta}\right)
P^{(m)}_{r's'}\\
&=&\Lambda_{rr';ss'}P^{(m)}_{r's'}.\label{processeq}
\end{eqnarray}
We call Eq. \ref{processeq} the \emph{linear process equation}. An equation 
\index{process equation! linear}
that relates the input states to output states is called a process equation. Notice that the linear process equation looks very much like Eq. \ref{finstatedyn} which describes the dynamical map for the initially product states in Sec \ref{simpsep}.  

Experimentally, we do not have any information about the global unitary operators \footnote{Unitary operators that acts on a bipartite state are called global, while unitary operators that acts only on the subpart are called local.} 
\index{local unitary operator} \index{global unitary operator}
or the state of the environment.  But if we know the output states corresponding to the necessary input state, then we find the map by the following expression.
\begin{eqnarray}\label{sqptlin}
\Lambda_{rr';ss'} = \frac{1}{d^2}\sum_n Q_{rs}^{(m)} {\tensor*{\tilde{P}}{*^{(m)}_{r's'}}}^*,
\end{eqnarray}
where $d$ is the size of the system and $\tilde{P}^{(n)}$ are the duals of 
\index{process map! linear}
the input states satisfying the scalar product 
\begin{eqnarray*}
{\tensor{\tilde{P}}{^{(m)}}}^{\dag}P^{(n)}=\sum_{rs}{\tensor*{\tilde{P}}{*^{(m)}_{rs}}}^*{P^{(n)}_{rs}} = \delta_{mn}.
\end{eqnarray*}
The duals for the projections 
\index{dual set! for a qubit}
in Eq. \ref{prj}  are
\begin{eqnarray}\label{dual}
\tilde{P}^{(1,-)} = \frac{1}{2}(1 - \sigma_1- \sigma_2- \sigma_3), &&
\tilde{P}^{(1,+)} = \frac{1}{2}(1 + \sigma_1-\sigma_2-\sigma_3 ), \nonumber\\
\tilde{P}^{(2,+)} =\sigma_2,&&
\tilde{P}^{(3,+)} =\sigma_3. 
\end{eqnarray}
This procedure is known as \emph{standard process tomography}.

The map in Eq. (\ref{sqptlin}) determines the open evolution of any state sent through that process.  This is a powerful statement, yet this procedure has two serious downsides to it.  First, the procedure assumes that a set of linearly independent states can be prepared for any experimental setup.  This may not be the case in reality.  In fact for certain situation the experimentalist may only care about what happens to certain elements of the density matrix and may not care about what happens to the rest of the state space.  In that case, determining the whole process map is an over kill. For one qubit this may not seem like a serious issue, because performing additional two or three experiments is generally not difficult.   But consider an experiment with $n$ qubits.  In that case, to determine the whole map requires preparing $2^{4n}$ input states.  This may not be practical for all experimental setups (see for a detailed analysis \cite{mohseni:032322}).   The second issue is precisely the number of required inputs. This number grows exponentially, and so do the number of necessary experiments with it. There are elegant methods that assist in overcoming both of the issues.  We tackle the first issue first with ancilla assisted process tomography in the next section.

\section{Ancilla assisted quantum process tomography}
\index{ancilla assisted process tomography|see{ quantum process tomography}}
\index{entanglement assisted process tomography|see{ quantum process tomography}}
\index{quantum process tomography! ancilla assisted}
\index{quantum process tomography! entanglement assisted}

Let us consider a situation where the experimentalist has an ancillary system available in addition to the system of interest.  The ancillary system can be used to help overcome the limitations of preparing a set of linear independent states.  In fact, we will only need to prepare one state with this procedure, known as \emph{ancilla assisted process tomography} \cite{PhysRevLett.91.047902, PhysRevLett.90.193601}. Let us once again assume that the initial state of the system, environment, and the ancilla are uncorrelated at the beginning of the experiment.  We start by entangling the ancilla with the system with unitary $V$ that acts on the combined state.
\begin{eqnarray}
\rho^{\mathcal{AS}}=V \rho^{\mathcal{A}}\otimes \rho^{\mathcal{S}} V^\dag,
\end{eqnarray}
where $\rho^{\mathcal{A}}$ is the initial ancilla state, $\rho^{\mathcal{S}}$ is the initial system state, and $R^{\mathcal{AS}}$ is 
\index{entanglement}
the maximally entangled state of the system and the ancilla.

At this point we send the system state through the quantum process. We then make a measurement $J^{(m)}$ on the ancilla after the system has gone through the complete quantum process, and analyze the output state of the system.  The linear process equation now looks as
\begin{eqnarray}
Q^{(m)}
&=&\frac{1}{\mbox{Tr}[J^{(m)}\rho^\mathcal{AS}]}
\mbox{Tr}_{\mathcal A}\left[
J^{(m)}
\mbox{Tr}_{\mathcal E}\left[
U \rho^{\mathcal{AS}}
\rho^{\mathcal{E}}
U^\dag \right]
J^{(m)} \right],
\end{eqnarray}
where we have normalized the final state by dividing the probability that the ancilla part of $\rho^{\mathcal{AS}}$ will collapse to $J^{(m)}$.  Since the system and the ancilla are maximally entangled, we are guaranteed the perfect knowledge of the state of the system before it goes through the quantum process by simply knowing the state of the ancilla.

Notice that $J^{(m)}$ commutes with the unitary $U$ and trace over the environment since only the system is going through the quantum process.  We can pull $J^{(m)}$ inside the trace over the environment
\begin{eqnarray}
Q^{(m)}_{rs}&=&
\frac{1}{\mbox{Tr}[J^{(m)}\rho^{\mathcal{AS}}]}
\sum_{\epsilon,\alpha,\beta,z}
U_{r\epsilon;r'\alpha}
J^{(m)}_{z x}
\rho^{\mathcal{AS}}_{xr';ys'}
J^{(m)}_{yz}
\rho^{\mathcal E}_{\alpha\beta}
U^*_{s\epsilon;s'\beta}\nonumber\\
&=&\sum_{\epsilon,\alpha,\beta,z}
U_{r\epsilon;r'\alpha}
J^{(m)}_{zz}P^{(m)}_{r's'}
\rho^{\mathcal{E}}_{\alpha\beta} 
U^*_{s\epsilon;s'\beta}\nonumber\\
&=&\left(\sum_{\epsilon,\alpha,\beta}
U_{r\epsilon;r'\alpha}
\rho^{\mathcal{E}}_{\alpha\beta} 
U^*_{s\epsilon;s'\beta}\right)
P^{(m)}_{r's'}\nonumber
=\Lambda_{rr';ss'}P^{(m)}_{r's'},
\end{eqnarray}
where $P^{(m)}$ is the state the system collapses to when the ancilla collapses to $J^{(m)}$. $P^{(m)}$ is also the desired input state.  Since acting with $J^{(m)}$ on the ancilla before the system goes through the process or after is the same, we can chose the any input state $P^{(m)}$ by making measurement on the ancilla after the system has long passed through the process.  Notice the process map $\Lambda$ has the same form as in Eq. \ref{processeq}, therefore the rest of this procedure is the same as the standard process tomography procedure from the last section.

By this method we have eliminated the problem of having an experimental setup equipped to prepare a variety of input states.  Now we need to prepare only one entangled state of the system and the ancilla. The correlations between the system and the ancilla do the rest of the work for us.  This is an example of quantum parallelism.  It is interesting to note that this method works even when the system and ancilla aren't entangled \cite{PhysRevLett.90.193601}, though not as well.  There are quantum correlation other than entanglement that can be used as resources as well. A measure of correlations due to Ollivier and Zurek and Hederson and Vedral called \emph{quantum discord} \cite{PhysRevLett.88.017901, henderson01a, PhysRevA.40.4277} gives some insight on how entanglement, quantum
\index{entanglement}\index{classical correlations}\index{quantum correlations}\index{quantum discord}
correlation, and classical correlations are different.

Though, in solving one problem we have added another. On the one hand, our ability to prepare a verity of input states has increased dramatically, on the other hand we now have to worry about keeping the ancilla completely isolated from the environment.  This is almost never practical.  We will not discuss this particular problem in this dissertation (See. VIII in \cite{kuah:042113} for a discussion).

\section{Direct characterization and other methods}
\index{direct characterization of quantum dynamics|see{ quantum process tomography}}
\index{quantum process tomography! direct characterization of quantum dynamics}

The trick with the ancilla assisted tomography method is to create a maximally entangled state to prepare all possible inputs simultaneously.  In a sense entanglement is being used as a resource in this method.  We can further utilize this resource to decrease the number of experiments necessary to find the process map using error correction techniques.  This is known as the \emph{direct characterization of quantum dynamics} \cite{mohseni:062331,mohseni:170501}.

The basic idea is to modify the ancilla assisted procedure by making a joint measurement on the system and the ancilla after the system has passes through the quantum process, instead of making a measurement on just the ancilla.  
\begin{eqnarray}
\Lambda_{mn}
=\mbox{Tr}_{\mathcal E}\left[
J^{(n)}
U \rho^{\mathcal{AS}^{(m)}}
\rho^{\mathcal{E}}
U^\dag
J^{(m)} \right],
\end{eqnarray}
above $\rho^{\mathcal{AS}^{(m)}}$ an entangled state (not always maximal) of the system and ancilla. $J^{(m)}$ now represents a joint measurement on the combined state of the ancilla and the system.  It turns out that what arises is simply an element of the process map $\Lambda$.  By choosing proper inputs and proper measurements we can find the map without every looking at the output state. The philosophy here is that determining the process map is important not the input and the output states.  Therefore in this method quantum state tomography is unnecessary.

This method gets its inspiration from \emph{quantum error correction codes}, which is beyond the scope of this dissertation. We will not go into the details of any of these methods, since they are of no consequence to our work here.  Though we would like to point out that these methods resolve the issue of preparing a large number input states.  In the direct characterization of quantum dynamics the number of inputs necessary to determine the process map grow polynomially with the size of the system rather than exponentially as in the standard quantum process tomography procedure.  This makes it possible to carry out quantum process tomography for large systems. There two other techniques along this line, \emph{selective efficient quantum process tomography} \cite{paz}, and \emph{symmetrized characterization of noisy quantum processes} \cite{laflamme}, that we will not review here.

\section{Discussion}
\index{quantum process tomography! weak coupling assumption}

In every quantum process tomography procedure above the input states are thought to be pure states.  There are two advantages of using pure states as inputs.  First, it is easier to span the space of the system with a set of pure state than it is with a set of mixed states. Second, pure states are always uncorrelated, which is one of the central assumption in every procedure above. This is often called the \emph{weak coupling assumption}.
\index{weak coupling assumption}
It is simply a matter of preparing necessary pure states to perform a quantum process tomography experiment.

In reality, the weak coupling assumption may not be true for some experiments.  In that case the initial state of the system will not be pure.  Thus, some preparation procedures must be applied on the system to prepare a pure state.  We relax the weak coupling assumption in the next chapter and analyze state preparation for open systems.  In chapter \ref{chapqptex} we will return to quantum process tomography armed with the knowledge of how preparation procedures work.


\chapter{Preparation of input states}\label{chapprp}
\index{Preparation of input states @\emph{Preparation of input states}}

In this chapter we study the effect of the preparation of the input states 
\index{preparation}
for a generic quantum experiment.  In quantum process tomography, a linearly independent set of states that span the system space have to be prepared.  But for a generic experiment, we may need to prepare a set that is larger than the linearly independent set.  In this section, we will not concern ourselves with quantum process tomography and keep the discussion focused on preparation only.

We can describe a quantum experiment in three general steps.  The experiment begins with an unknown state that has to be altered into known input state.  After the preparation, the prepared state is subjected to some quantum operation, and finally the outcome is analyzed.  In this chapter we will only concern ourselves with the first step of state preparation. 

Preparation procedures are very complicated in practice. Since we cannot describe each preparation procedure in detail, we will attempt to develop a general theory of preparation.  For clarity, we will first deal with the preparation of a general state $\rho$, and then embark onto the preparation of a state of a system that may be correlated and  may interact with an environment.  

\section{Preparations in quantum mechanics}
\label{prepinqm}

We can summarize a set of preparations by considering an experiment that 
\index{preparation! stochastic}
starts with a generic state of the system $\rho$, which is then altered into a set of desired inputs $\{P^{(m)}\}$. We now have to somehow connect the initial state to the input states.

As we saw in chapter \ref{chapopendyn}, the most general dynamics of a quantum state are described by a stochastic map.  We can denote the 
\index{stochastic process}
procedure for preparing the $m$th state with a map $\mathcal{P}^{(m)}$.  The only restriction we put on the preparation map is that it be completely positive.  This is because no matter what the initial state of the system is, we should be able to prepare the state.  If $\mathcal{P}$ is completely positive, then its action is defined on all possible states of the system.  

The action of $\mathcal{P}^{(m)}$ onto $\rho$ can be written as
\begin{eqnarray}
\mathcal{P}^{(m)}_{rr';ss'}\rho_{r's'}
=P^{(m)}_{rs}.
\end{eqnarray}

Before we look at the preparation procedure for an open system, we should discuss how a most general preparation can be carried out with the aid of an ancillary system.   As we saw in Sec. \ref{secenvsize}, to describe the most general completely positive dynamics of a quantum state of dimension $d$, we only need an environment of dimension $d^2$.  Our goal here is to be able to prepare the system, so we need control over the system.  Therefore, instead of an environment we place two ancillary systems of dimension $d$ to perform the most general input state.

\subsection{Trace preservation}
It turns out that not all preparation procedures are trace preserving.  The action of such maps is written as
\begin{eqnarray}
\frac{1}{r^{(m)}}
\mathcal{P}^{(m)}_{rr';ss'}\rho_{r's'}
=P^{(m)}_{rs},
\end{eqnarray}
where the normalizing factor 
\begin{eqnarray}
r^{(m)}=\mbox{Tr} [\mathcal{P}^{(m)}\rho]
\end{eqnarray}
is the probability with which $\rho$ will become $P^{(m)}$.  

However, there is a complete map that preserves the trace. Consider a 
\index{preparation! trace preservation}
completely positive preparation map, $\mathcal{P}$, acting on $\rho$.   We can rewrite this map in the Choi representation (Sec. \ref{osr}) as
\begin{eqnarray}
\mathcal{P}(\rho)=\sum_{m}\pi^{(m)}\rho\;{\pi^{(m)}}^\dag,
\end{eqnarray}
where $\pi^{(m)}$ are the eigenmatrices of $\mathcal{P}$ satisfying $\sum_m{\pi^{(m)}}^\dag\pi^{(m)}=\mathbb{I}$, which guarantees trace preservation.  We now combine, for each $m$, the eigenmatrices $\pi^{(m)}$ with its Hermitian adjoint ${\pi^{(m)}}^\dag$ to get
\begin{eqnarray}
\mathcal{P}(\rho)=\sum_{m}\mathcal{P}^{(m)}\;\rho.
\end{eqnarray}
It is now clear that the $m$th preparation is given by the action of $\mathcal{P}^{(m)}$, which does not preserve trace. But a set of preparations $\mathcal{P}^{(m)}$, belonging to $\mathcal{P}$, together preserve the trace.

This suggests that, to prepare $m$ input states, we will in general need $m$ preparation maps. In other words, an experiment will require $m$ preparation procedures.  Not all of these procedures will be completely unrelated, in fact, most will share some common features.

\section{Preparations in open quantum mechanics}

How does the situation above change when we consider open quantum systems?  
\index{preparation! in open quantum mechanics}
We simply follow the procedure laid out in section \ref{prepinqm}, but instead of using the generic initial state $\rho$, we must use a bipartite state of the system and the environment $\rho^\mathcal{SE}$.  The preparation map still only acts on the state of the system.

\begin{figure}[!ht]
\begin{center}
\resizebox{10 cm}{7.87 cm}
{\includegraphics{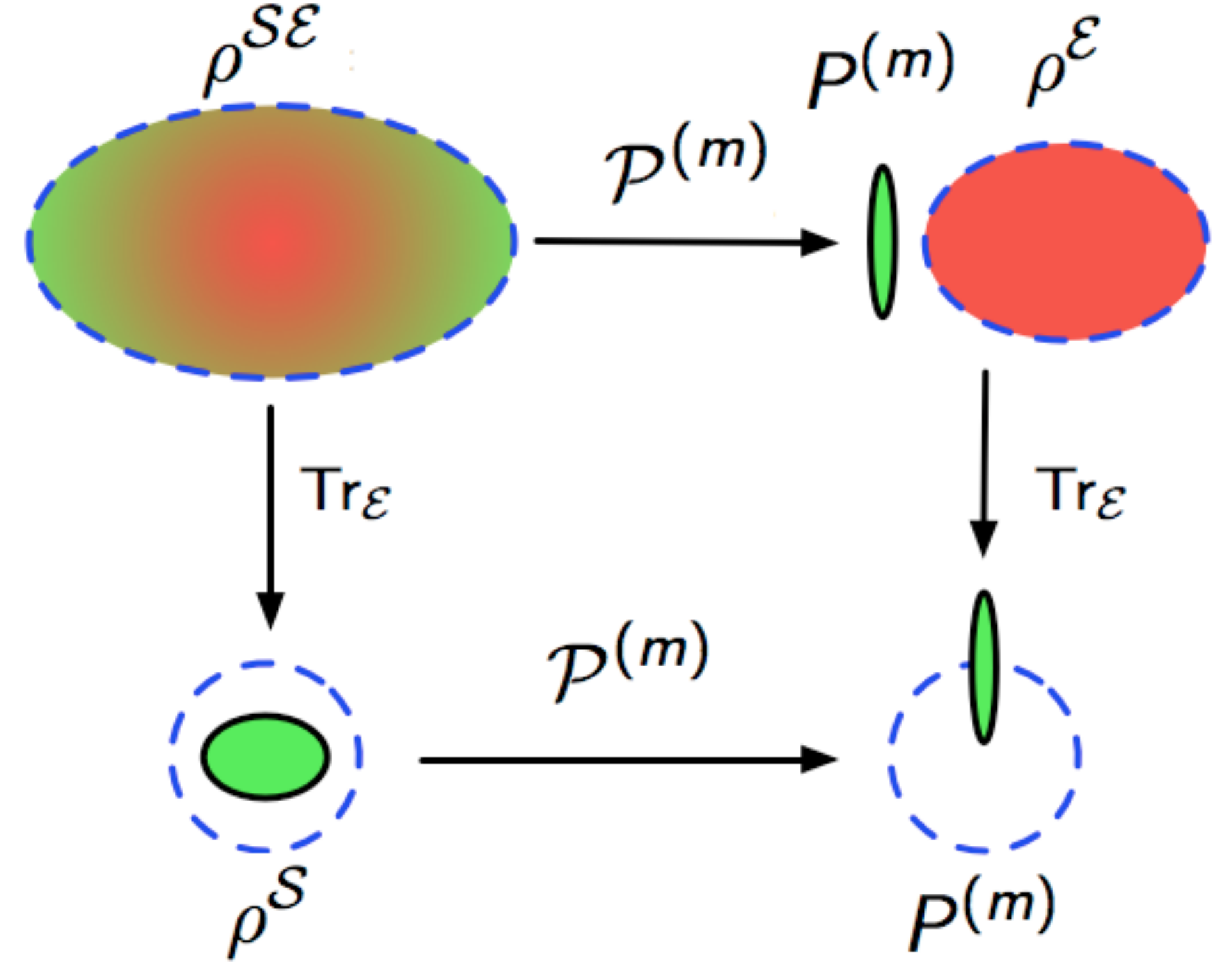}}
\caption{\label{prepfig}
At the beginning of the experiment an unknown state of the system and the environment is present.  The $m$th state of the system is then prepared, which diminishes the correlations between the system and the environment.}
\end{center}
\end{figure}

The preparation of a system belonging to a bipartite state of the system and the environment we get
\begin{eqnarray}
R_{r'\alpha ;s'\beta}^{\mathcal{SE}^{(m)}}
=\frac{1}{r^{(m)}}
\mathcal{P}^{(m)}_{r'r'';s's''} \mathcal{I}_{\alpha\alpha;\beta\beta}
\rho^{\mathcal{SE}}_{r''\alpha;s''\beta},
\end{eqnarray}
where indices $\alpha$ and $\beta$ belong to the state of the environment and $R^{\mathcal{SE}^{(m)}}$ is the bipartite state of the system and environment after the preparation.  Notice that the preparation map still acts on only the system part of the initial bipartite state. For completeness, in the equation above we have included an identity map $\mathcal{I}$ acting on the state of the environment, but for simplicity we will omit writing it from here on. Fig. \ref{prepfig} shows graphically the action of the preparation procedure.

Even though the preparation procedure only acts on the state of the system, the state of the environment in general will indirectly be affected by the procedure. Suppose our goal is to prepare an uncorrelated state. Simplifying the notation above  we get
\begin{eqnarray}
R^{\mathcal{SE},(m)}
=\mathcal{P}^{(m)}
\rho^{\mathcal{SE}}
=P^{(m)}\otimes\rho^{\mathcal{E},(m)}.
\end{eqnarray}

In the last equation, we see that the state of the environment has picked a superscript $m$, because if $\rho^{\mathcal{SE}}$ is initially correlated, then an action on the system part will necessarily effect the state of the environment.  However, if the weak coupling assumption is retained, then the preparation procedures for the closed and the open cases are the same.  Let us investigate this through two specific procedures.  See \cite{Kuah02} for a similar analysis for quantum measurements.

\section{Two common preparation procedures}

Uncorrelated states are often desired for experiments. Therefore we will only focus on preparing pure states. We describe two of the most common preparation procedures to prepare pure states (thus uncorrelated) used today. We first analyze each preparation procedure for closed quantum systems, and then analyze them for open quantum systems \cite{kuah:042113}.  In this manner, the differences between two will be apparent. More complicated procedures can be easily studied by writing down the corresponding preparation map.

\subsection{Stochastic preparation}
Many quantum experiments begin by initializing the system to a specific state. For instance, in the simplest case, the system can be prepared to the ground state by cooling it to near absolute zero temperature \cite{Wein:121.13, orien:080502, Howard06,Howard05, myrskog:013615}. Mathematically, these set of operations can be written as a pin map \cite{GoriniSudarshan}.  Consider the following map
\begin{eqnarray}
\Theta_{rr';ss'}
=\left[\ket{\Phi}\bra{\Phi}\right]_{rs}
\mathbb{I}_{r's'}.
\end{eqnarray}
The action of this map on the initial state of the system is
\begin{eqnarray}
\Theta\rho&=&
\sum_{r's'}
\left[
\ket{\Phi}\bra{\Phi}
\right]_{rs}
\mathbb{I}_{r's'}
\rho_{r's'}\nonumber\\
&=&\ket{\Phi}\bra{\Phi}
\mbox{Tr}[\rho]
=\ket{\Phi}\bra{\Phi}.
\end{eqnarray}
In the procedure above, no matter what the initial state of the system was, it is ``pinned" to the final state $\ket{\Phi}\bra{\Phi}$.  Along with the pin map, a set of unitary transformations on the system are sufficient to prepare the $m$th input state.  In that case we can write the preparation as a composition of the stochastic pin map and a set of local unitary transformations (acting only on the system) $\Omega^{(m)}$ as
\begin{eqnarray}
\mathcal{P}^{(m)}\rho
=\Omega^{(m)}\circ\Theta\rho
=P^{(m)}.
\end{eqnarray}
We call this procedure the \emph{stochastic preparation} 
\index{stochastic preparation procedure|see{ preparation procedure}}
\index{preparation procedure! stochastic! for close systems}
method.

Each preparation map is trace preserving, because the pin map and the unitary transformations that follow it are all trace preserving transformations.  Also, to prepare $m$ input states, we need $m$ unitary rotations, hence we have $m$ preparation procedures.  However as claimed before, each of these procedure has the pin map as a common step.  

For a closed system, we could just as well use $m$ pin maps to prepare $m$ input states.  The trouble only arises when the system is interacting with an environment.  Let us look at this in detail.

\subsubsection{For open systems}\label{stoprep}

Let us consider the action of the pin map $\Theta$ on a bipartite state of the system and the environment.
\begin{eqnarray}\label{pinmap}
\Theta \rho^\mathcal{SE}
&=&\left[\Ket{\Phi}\Bra{\Phi}\right]_{rs} \mathbb{I}_{r's'} 
\rho^\mathcal{SE}_{r'\alpha;s'\beta} \nonumber\\
&=&\Ket{\Phi}\Bra{\Phi}\otimes
\mbox{Tr}_{\mathcal{S}}[\rho^\mathcal{SE}]
=\Ket{\Phi}\Bra{\Phi}\otimes\rho^\mathcal{E}.
\end{eqnarray}
As we can see, not much happens differently when a stochastic map is applied to a bipartite state.  The only difference is the presence of the environmental partner $\rho^\mathcal{E}$.  The pin map fixes the system into a single pure state, which means that the state of the environment is 
\index{preparation procedure! stochastic! for open systems}
fixed into a single state as well.  The purpose of the pin map is to decouple the system from the environment, to eliminate any correlation 
\index{pin map}
between the system state and the environment state.

Once the pin map, $\Theta$, is applied, the system is prepared in the various different input states by applying local (only on the system part) unitary transformations as before.  The prepared state $R^\mathcal{SE}$ takes the following form.
\begin{eqnarray}\label{stocmap}
{R^\mathcal{SE}}^{(m)}
&=&\mathcal{P}^{(m)}\rho^\mathcal{SE}\nonumber
=\Omega^{(m)} \circ\Theta(\rho^\mathcal{SE})\\
&=&\Omega^{(m)}\left(\Ket{\Phi}\Bra{\Phi}\right)\otimes
\rho^\mathcal{E}\nonumber
= P^{(m)}\otimes\rho^\mathcal{E}.
\end{eqnarray}

Strictly speaking, in a real experiment the pin map is a complicated set of steps.  For instance, cooling the system so it relaxes to the ground state is very different from optically pumping a system to an excited state.  Two different pin maps would correspond to these two different methods.  Then, the state of the environment in each case can also be different.  We have assumed that an identity map acts on the environment in our mathematical formulation.  This may not be the case; we cannot analyze the state of environment before or after the pin map is applied.  Therefore, to be careful, we should label $\rho^\mathcal{E}$ as $\rho^\mathcal{E}_\Theta$ to clarify that the state of the environment may depend on certain steps involved in applying the pin map.
  
Then, it should be emphasized that the initial pin map is critical. It may be tempting to simply use a set of pin maps, $\Theta^{(m)}$, to prepare the various input states $P^{(m)}$.  The different input states will be
\begin{eqnarray}
R^{\mathcal{SE}^{(m)}}
&=&\mathcal{P}^{(m)}\rho^\mathcal{SE}
=\Theta^{(m)}\rho^\mathcal{SE}\nonumber\\
&=& P^{(m)}\otimes
\rho^{\mathcal{E}}_{\Theta^{(m)}}.
\end{eqnarray}
As seen in the last line of the equation above, the state of the environment  has taken the subscript $\Theta^{(m)}$ through different pin maps along with the input state.  What this means is that the state of the environment is affected by the preparation procedure. We show in the next chapter, when the preparation procedure is not carried out carefully in a quantum process tomography experiment, the process map can behave non-linear.

\subsection{Projective preparation}
Let us look at another preparation method that is also very common. Consider an optical experiments, where a beam of photons is sent through a polarizer \cite{Nielsen:1998py, PhysRevA.64.012314, PhysRevLett.91.120402}.  Polarizers project the polarization vector of the photons into a set 
\index{preparation procedure! projective! for close systems}
orthogonal directions.  By using a quarter wave plate, all necessary input states can be prepared.


We can write the action of $m$th preparation map acting on the initial state of the system as the following:
\begin{eqnarray}
\frac{\mathcal{P}^{(m)}\rho}
{\mbox{Tr}[\mathcal{P}^{(m)}\rho]}=
\frac{1}{r^{(m)}}P^{(m)}\rho{P^{(m)}}^\dag=P^{(m)},
\end{eqnarray}
where $\frac{1}{r^{(m)}}$ is the normalization factor with $r^{(m)}=\mbox{Tr}[\mathcal{P}^{(m)}\rho]=\mbox{Tr}[P^{(m)}\rho]$; this is the probability of obtaining that particular input state from a von-Neumann measurement \cite{vonNeu}.  Since projective transformation are not trace preserving, the preparation map does now preserve trace.  The normalization factor, $\frac{1}{r^{(m)}}$, is simply the inverse of the probability that the system will collapse to the $m$th state, which can be measured experimentally. In this fashion we can prepare any pure state as long as we properly normalize them. We call this method \emph{projective preparation}.
Note that to prepare $m$ input states $m$ different preparation procedures are necessary.

Though each projective preparation map does not preserve trace, there is an 
\index{preparation! trace preservation}
overall map that preserves trace.  Let us write a complete map in the operator sum representation, where we let the eigenmatrices of the map to be a set of orthogonal projections. Thus we have
\begin{eqnarray}
\mathcal{P}\rho=\sum_{m}P^{(m)}\rho{P^{(m)}}^\dag ,
\end{eqnarray}
where $\{P^{(m)}\}$ form an orthogonal set satisfying $P^{(m)}P^{(n)}=\delta_{mn}P^{(m)}$ and $\sum_m {P^{(m)}}^\dag P^{(m)}=\sum_m P^{(m)}=\mathbb{I}$.  Now we can define each map $\mathcal{P}^{(m)}$ for a single preparation as $P^{(m)}{P^{(m)}}^\dag$.  Now we see that for a non-trace preserving preparation there is a complete map that does preserve the trace. In that sense, the projections are the eigenmatrices of the map $\mathcal{P}^{(m)}$.

\subsubsection{For open systems}\label{mespreps}
\index{projective preparation procedure|see{ preparation procedure}}
\index{preparation procedure! projective! for open systems}
A projective preparation is given by a set of orthonormal projections. If a particular outcome is observed from the measurement, the state of the system collapses to that corresponding projection.  Therefore, the input states can be prepared for an experiment by suitably fixing our measurement basis. With the knowledge of the basis and the outcomes, the exact input state becomes known.

Suppose the $m$th input state is prepared, given by the projection $P^{(m)}$, by projecting the system into that state.  For the open system, the total state becomes
\begin{eqnarray}\label{mesprep}
R^{\mathcal{SE}^{(m)}}
&=&\frac{1}{r^{(m)}}\mathcal{P}^{(m)}\rho^{\mathcal{SE}}
=\frac{1}{r^{(m)}} 
P^{(m)} \rho^\mathcal{SE} P^{(m)}\nonumber\\
&=&P^{(m)}\otimes\rho^{\mathcal{E},(m)},
\end{eqnarray}
where once again $r^{(m)} = \mbox{Tr}[\mathcal{P}^{(m)}\rho^\mathcal{SE}] \mbox{Tr}[P^{(m)}\rho^\mathcal{SE}]$ is the normalization factor.  Notice that the state of the environment has picked the superscript $m$.  This is due to the fact that in general $\rho^\mathcal{SE}$ contains correlations between the system and the environment.  We should once again emphasize that the preparation procedure only acts on the system part and affects the state of the environment only indirectly.

The dependance that the state of the environment picks up is very easy to demonstrate.  Consider the following one qubit system and one qubit environment state:
$$
\rho^\mathcal{SE}=\frac{1}{4}(\mathbb{I}\otimes\mathbb{I}+\sigma_1\otimes\sigma_1-\sigma_2\otimes\sigma_2+\sigma_3\otimes\sigma_3).
$$
This is a maximally entangled state of two qubits.  It is easy to check that the projective preparation of this state with different projections $P^{(\pm j)}=\frac{1}{2}(\mathbb{I}\pm\sigma_j)$ for $\{j=1,2,3\}$ leads to different states of the environment in each case:
\begin{eqnarray}
\frac{1}{r^{(\pm j)}}\mathcal{P}^{(\pm j)}(\rho^{\mathcal{SE}})
&=&\frac{1}{r^{(\pm j)}} 
P^{(\pm j)} \rho^\mathcal{SE} P^{(\pm j)}\nonumber\\
&=&P^{(\pm j)}\otimes P^{(\pm j)}.
\end{eqnarray}
Once again, we see that the state of the environment is affected by the preparation of system state.  This dependance cannot be ignored, and should be accounted for whenever the system evolves openly. In the next chapter, we will show  that when the preparation procedure is ignored in quantum process tomography, the process maps obtained can behave non-linearly.

\begin{figure}[!ht]
\begin{center}
\resizebox{10.4 cm}{7.74cm}
{\includegraphics{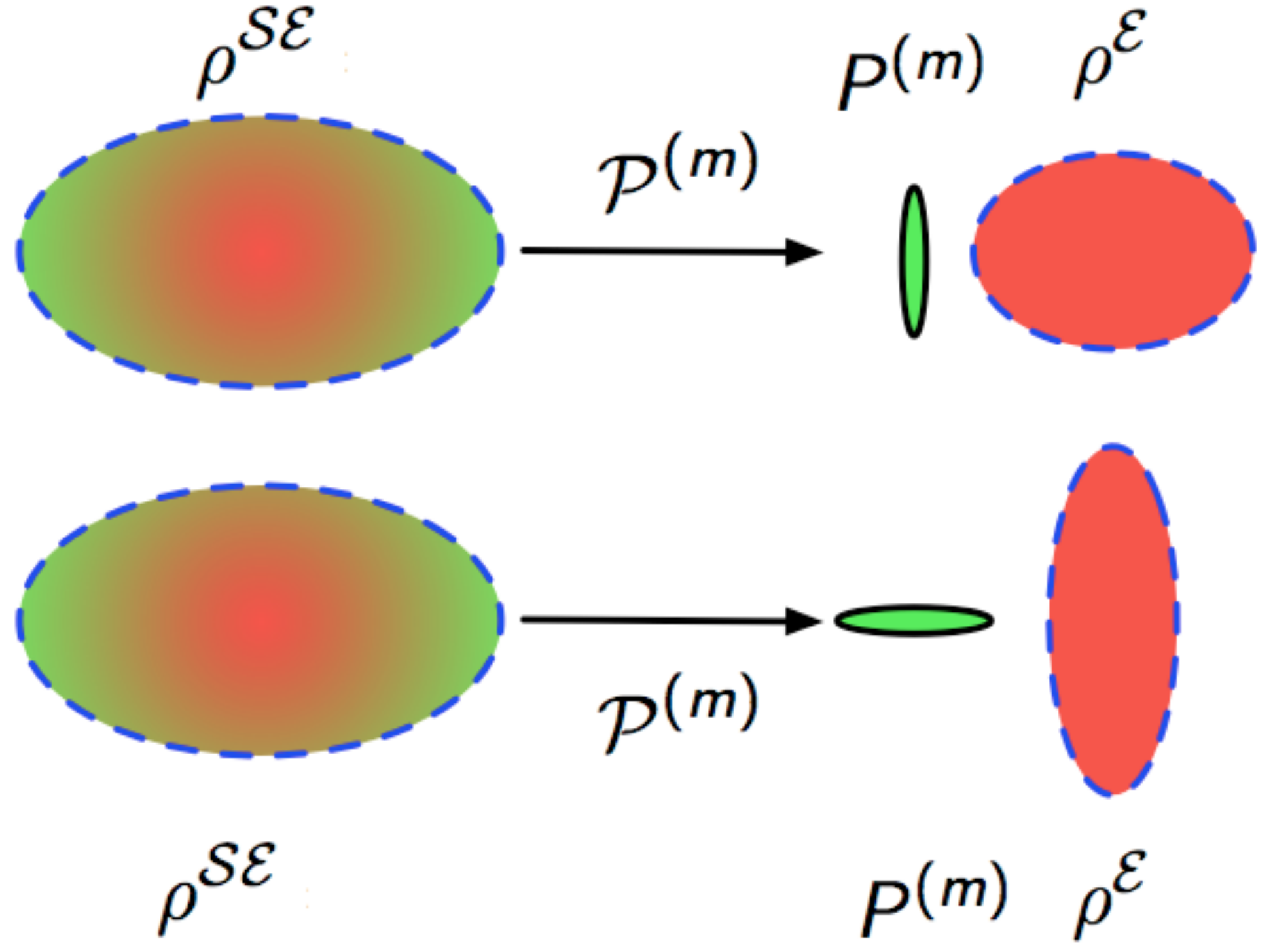}}
\caption{\label{prepfigE}
The figure above shows how that for two different projective preparations lead to the desired input states, but the corresponding states of the environment can be different.  In both cases the initial total state is the same.}
\end{center}
\end{figure}

\section{Discussion}

The examples above only aspire to prepare pure states.  However, much more can be obtained from the prescription spelled out above.  In principle, nothing is preventing us from preparing mixed states by using so called \emph{generalized preparations} (see \cite{kuah:042113}).
\index{preparation! generalized}
Furthermore, our discussion so far has largely focused on preparing uncorrelated states; the very same techniques can be applied in preparing entangled states as well.  

We should also add that any time an ancillary system is used in an open experiment, it is an unreasonable assumption that the ancillary is uncorrelated with the environment. Furthermore, the ancillary may even interact with the environment.  We will not discuss this topic in this dissertation, but see Sec. VIII in \cite{kuah:042113} for a discussion.

Note that, if the system evolves in a closed form then all preparation procedures are equivalent.  Trouble only arises when the preparation procedure indirectly affects the state of the environment which then interacts with the system. Additionally, even though we will only use the results here in analyzing quantum process tomography procedures, they apply to any quantum experiment that interacts with an environment. 
 
\chapter{Preparation and quantum process tomography}\label{chapqptex}
\index{Preparation and quantum process tomography
@\emph{Preparation and quantum process tomography}}

Each of the quantum process tomography procedure discussed in Chap. 
\index{preparation! in quantum process tomography}
\ref{chapqpt} assumes the system and the environment are initially uncorrelated.  What if we depart from this assumption? In chapter \ref{chapopendyn} we showed that initially correlated states can lead to not-completely positive dynamics for the system.  In many recent experiments, the process maps that characterize the quantum operations have been plagued with negative eigenvalues and occasional non-linear behavior (see references in \cite{kuah:042113}). We now examine the how the two cases are related.

The quantum process tomography procedures we reviewed require the input states be uncorrelated with the environment. Just before the experiment begins, in general the state of the system may be correlated with the environment. Thus, at the beginning of the experiment it is necessary to prepare the initially correlated total state into an uncorrelated state.  The difference between the process maps found from a quantum process tomography experiment and the dynamical maps calculated theoretically is precisely the act of preparation of input states.  Let us investigate this issue by analyzing the steps involved in a quantum process tomography experiment.

\section{Quantum process tomography experiment in steps}

The basic steps in a 
\index{quantum process tomography}
quantum process tomography experiment are broken down below:
\begin{itemize}
\item {Just before the experiment begins, the system and environment is in an unknown state $\rho^{\mathcal{SE}}$. The system and the environment in general are correlated; we are no longer making the weak coupling assumption.}
\item {The system is altered to a known input state by a preparation procedure.  The system and environment state after preparation is therefore given by 
$$\frac{1}{r^{(m)}}\mathcal{P}^{(m)} \rho^{\mathcal{SE}}.$$ 
The input state is given by taking trace with respect to environment $$P^{(m)}=\frac{1}{r^{(m)}}\mbox{Tr}_{\mathcal{E}}\left[\mathcal{P}^{(m)} \rho^{\mathcal{SE}}\right].$$}
\item {The system is then sent through a quantum process.  We consider the evolution to be a global unitary transformation in the space of the system \emph{and} the environment:
\begin{eqnarray*}
\frac{1}{r^{(m)}}
U \mathcal{P}^{(m)}
\rho^{\mathcal{SE}} 
U^\dag.
\end{eqnarray*}}
\item {Finally the output state is observed.  Mathematically it is the trace with respect to the environment
\begin{eqnarray}\label{output}
Q^{(m)}=\frac{1}{r^{(m)}}\mbox{Tr}_{\mathcal E}\left[U \mathcal{P}^{(m)}
\rho^{\mathcal{SE}} U^\dag \right],
\end{eqnarray}
we will call this the 
\index{process equation! generalized}
\emph{generalized process equation}.}
\item  {Finally using the input and the output states, we construct a map describing the process is constructed.}
\end{itemize}
The procedure above is identical to the procedure to find a dynamical map, except for the preparation procedure.  In the next two sections we will analyze quantum process tomography with the two preparation procedures discussed in last chapter.  The differences between what we find here and dynamical maps will be due to the preparation procedures.
\section{Stochastic preparation}\label{stochprep}
\index{quantum process tomography! stochastic preparation}
\index{preparation! stochastic! quantum process tomography}

In this section we analyze quantum process tomography procedures using the stochastic preparation procedure to generate input states.  The stochastic preparation map's action on a bipartite state is given by \ref{pinmap}.  We plug that into the generalized process equation \ref{output} to get
\begin{eqnarray}\label{stocproceq}
Q^{(m)} 
&=& \mbox{Tr}_\mathcal{E} \left[ 
U \left(\Omega^{(m)} \circ\Theta\right)
\rho^{\mathcal{SE}}
U^\dagger\right]\nonumber \\
& =& \mbox{Tr}_\mathcal{E}\left[U P^{(m)} \otimes \rho^{\mathcal{E}} U^\dagger\right].
\end{eqnarray}
The last equation is the same as the linear process equation 
\index{process equation! linear}
(Eq. \ref{processeq}) from Chap. \ref{chapqpt}. Therefore once the input states are prepared, the procedure for quantum process tomography is the same as the given in Chap. \ref{chapqpt}.  This is a generalization of linear quantum process tomography.  Here, we have assumed that only one pin map, $\Theta$, is used. Therefore we have not labeled the state of the environment with the pin map.  We will relax this assumption in the second example below.

\subsection{An example with stochastic preparation}
\index{example of! process map! stochastic preparation}

Consider the following two qubit state as the available state to the experimenter at $t=0_-$:
\begin{eqnarray}\label{totst}
\rho^{\mathcal{SE}}=\frac{1}{4}\left(\mathbb{I}\otimes\mathbb{I}+a_j\sigma_j\otimes\mathbb{I}+c_{23}\sigma_2\otimes\sigma_3\right).
\end{eqnarray}
This is the same state we used in Chap. \ref{chapopendyn} for the dynamical map example.

Let the pin map, $\Theta$, for our example be
\begin{eqnarray}\label{pin1}
\Theta=\ket{\phi}\bra{\phi}\otimes\mathbb{I},
\end{eqnarray}
where $\ket{\phi}\bra{\phi}$ is a pure state of the system. The preparation of $\rho^{\mathcal{SE}}$ with this pin map leads to
\begin{eqnarray}\label{inp1}
\Theta\rho^{\mathcal{SE}}=\ket{\phi}\bra{\phi}\otimes\frac{1}{2}\mathbb{I},
\end{eqnarray}
yielding the initial state $\ket{\phi}\bra{\phi}$ for the system qubit and a completely mixed state for the environment qubit. The next step is to create the rest of the input states using maps $\Omega^{(m)}$.  In this case, the fixed state $\ket{\phi}\bra{\phi}$ can be locally rotated to get the desired input state $P^{(m)}$ (given in Eq. \ref{prj})
\begin{eqnarray}\label{inp2}
\Omega^{(m)}\ket{\phi}\bra{\phi}\otimes\frac{1}{2}\mathbb{I}
&=& V^{(m)}\ket{\phi}\bra{\phi}{V^{(m)}}^{\dag}
\otimes\frac{1}{2}\mathbb{I}\nonumber\\
&=&P^{(m)}\otimes\frac{1}{2}\mathbb{I},
\end{eqnarray}
where $m=\{(1,-),(1,+),(2,+),(3,+)\}$ and $V^{(m)}$ are local unitary operators acting on the space of the system.  

Now each input state is sent through the quantum process. The output states can be calculated using Eq. \ref{stocproceq}. Once again let us choose the same unitary transformations as in Eq. \ref{unitarys}. The output states are
\begin{eqnarray}\label{stoout}
Q^{(1,-)}=\frac{1}{2}\{\mathbb{I}-\cos^2(2\omega t)\sigma_1\},&&
Q^{(1,+)}=\frac{1}{2}\{\mathbb{I}+\cos^2(2\omega t)\sigma_1\}\nonumber\\
Q^{(2,+)}=\frac{1}{2}\{\mathbb{I}+\cos^2(2\omega t)\sigma_2\},&&
Q^{(3,+)}=\frac{1}{2}\{\mathbb{I}+\cos^2(2\omega t)\sigma_3\}.
\label{stochoutputs}
\end{eqnarray}

\begin{center}
\begin{figure}[!htt]
\resizebox{12.8 cm}{7.8 cm}{\includegraphics{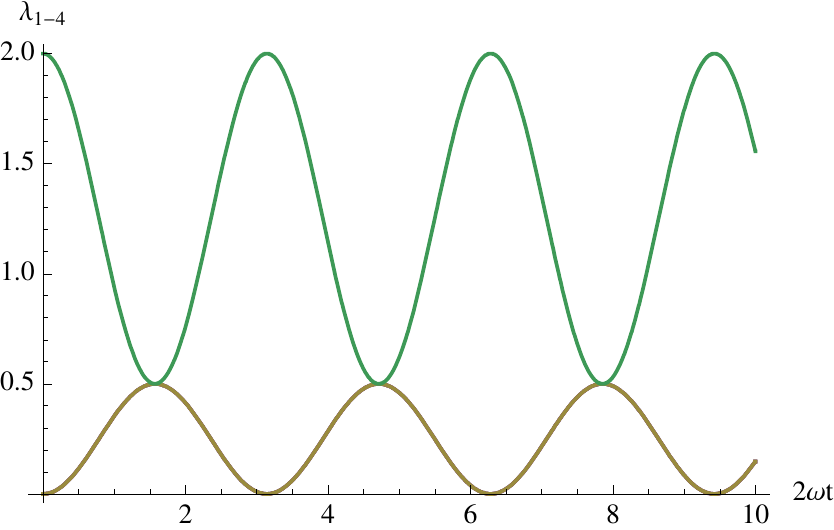}}
\caption{\label{fig5_1}
The eigenvalues of the process map in Eq. \ref{sact} are plotted as function of $2 \omega t$. As expected the eigenvalues are always positive. The stochastic preparation procedure allows the experimentalist to prepare any pure state for the system.  By convexity, then all possible states of the system are in the compatibility domain of the process map in Eq. \ref{sact}.}
\end{figure}
\end{center}

The linear process map is constructed using Eq. \ref{sqptlin}, the duals in Eq. \ref{dual}, and the output states:
\begin{eqnarray}\label{sact}
\Lambda_s=\frac{1}{2}
\left(\begin{array}{cccc}
1+C^2 &0&0&2C^2\vspace{.2cm}\\
0& 1-C^2&0 & 0 \vspace{.2 cm}\\
0 & 0& 1-C^2 &0 \vspace{.2 cm}\\
2C^2& 0&0& 1+C^2 \vspace{.2 cm}
\end{array}\right),
\end{eqnarray}
where $C=\cos(2\omega t)$.  The eigenvalues of the process map are plotted 
\index{process map! stochastic}
in Fig. \ref{fig5_1}

\subsection{An example with multiple stochastic preparations} \label{exmultstoch}
\index{example of! process map! multiple stochastic preparations}
\index{preparation! multiple stochastic}
The pin map used in stochastic preparation must be used consistently.  Let us show an example of what happens when two stochastic preparation procedures are used.  Consider a quantum process tomography experiment where the following linearly independent states are used to span the space of the system
\begin{eqnarray}
P^{\mathbb{I}}=\frac{1}{2}
\mathbb{I},\;\;P^{(1+)},\;\;P^{(2+)},\;\;P^{(3+)}.
\end{eqnarray}
These states form the linearly independent set $\{\mathbb{I},\sigma_j\}$; which is different from the set given in Eq. \ref{prj}.

Once again, we take the same initial state and unitary operators as in the last example.  Now suppose the pin map given in Eq. \ref{pin1} is used to prepare the state $\ket{\phi}\bra{\phi}$, and then by local transformations  $P^{(1,+)}$, $P^{(2,,+)}$, and $P^{(3,+)}$ are prepared. Finally the mixed state is prepared by letting $P^{(3+)}$ decohere.  This is a different pin map than the one in Eq. \ref{pin1}, so we are using two pin maps to prepare two sets of input states.  The unitary operator, we are using, is often called the \emph{swap gate}, because it swaps the states of two qubits with the period of $t=\frac{\pi}{4\omega}$.  Then at $t=\frac{\pi}{4\omega}$, the total state will be
\begin{eqnarray}
\rho^{\mathcal{SE}}\left(\frac{\pi}{4\omega}\right)
=\frac{1}{4}
\{\mathbb{I}\otimes\mathbb{I}
+a_j \mathbb{I}\otimes\sigma_j\}.
\end{eqnarray}
The state of the system has fully decohered.  

The corresponding output states for the input states above are found using Eq. \ref{stocproceq}. The state of environment in that equation for input $P^{\mathbb{I}}=\frac{1}{2}\mathbb{I}$ is $\rho^{\mathcal E}=\frac{1}{2} \{\mathbb{I} +a_j\sigma_j\}$, while for the other inputs the state of the environment $\rho^{\mathcal E}=\frac{1}{2}\mathbb{I}$.  Then the corresponding output states are
\begin{eqnarray}\label{multstocout}
Q^{(\mathbb{I})}=\frac{1}{2}\{\mathbb{I}+\sin^2(2\omega t)\sigma_3\},&&
Q^{(1,+)}=\frac{1}{2}\{\mathbb{I}+\cos^2(2\omega t)\sigma_1\},\\
Q^{(2,+)}=\frac{1}{2}\{\mathbb{I}+\cos^2(2\omega t)\sigma_2\}, && Q^{(3,+)}=\frac{1}{2}\{\mathbb{I}+\cos^2(2\omega t)\sigma_3\}.
\end{eqnarray}
The last three output states are the same as in the last example, but the fourth one is different.

Now suppose we calculate the output state corresponding to the input $P^{(-1)}$. By linearity we have $P^{(-1)}=\mathbb{I}-P^{(+1)}$.  If the process is linear, then the output state for this input state is given by
\begin{eqnarray}
Q^{(-1)}&=&2 Q^{(\mathbb{I})}-Q^{(+1)}\\
&=&\frac{1}{2}\{\mathbb{I}+2\sin^2(2\omega t)\sigma_3
-\cos^2(2\omega t)\sigma_1\}.
\end{eqnarray}
This state is not physical for certain times, and therefore the linearity and the positivity of the process is violated.  

\begin{center}
\begin{figure}[!htt]
\resizebox{12.8 cm}{7.8 cm}{\includegraphics{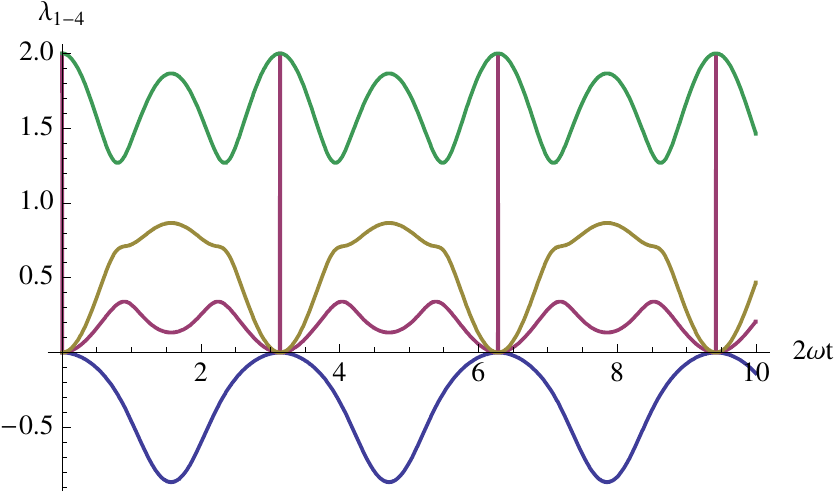}}
\caption{\label{fig5_2}
The eigenvalues of the process map in Eq. \ref{msact} plotted as function of $2 \omega t$. One of the eigenvalue is negative for certain values of $\omega t$.  The negativity is due to the inconsistency in the preparation procedure, and not due to the initial correlations with the environment.}
\end{figure}
\end{center}

The process has not changed from the last example, only the method of determining the process has.  Thus when the stochastic map is not used consistently, the process map can behave nonlinear.  For completeness we find the process map for this example,
\begin{eqnarray}\label{msact}
\Lambda_{ms}=\frac{1}{2}
\left(\begin{array}{cccc}
1+C^2 &-(1+i)S^2&0&2C^2\vspace{.2cm}\\
-(1-i)S^2& 1-C^2+2S^2&0 & 0 \vspace{.2 cm}\\
0 & 0& 1-C^2 &(1+i)S^2 \vspace{.2 cm}\\
2C^2& 0&(1-i)S^2& 1+C^2-2S^2 \vspace{.2 cm}
\end{array}\right),
\end{eqnarray}
where $C=\cos(2\omega t)$ and $S=\sin(2\omega t)$. The eigenvalues of this process map are negative, as seen in Fig. \ref{fig5_2}. The fact that the 
\index{process map! multiple stochastic}
state of the environment is not a constant for all of the input states leads to the negativity and non-linearity of the process map.

The origin for the negative eigenvalues here is completely different than 
\index{negative maps! multiple stochastic preparations}
the negative eigenvalues found for the dynamical maps in Sec. \ref{inicorr}.  The negative eigenvalues in the dynamical map were due to the initial correlations between the system and the environment. Since none of the inputs were correlated with the environment here, negativity arises from the inconsistencies in the preparation procedure.  Furthermore, the dynamical map in Sec. \ref{inicorr} is linear and has a physical interpretation within the compatibility domain.  No such interpretation is found for the process map in this example.

While this example may not seem realistic, the point regarding inconsistency arising from multiple pin maps still stands.  In realistic cases, the trouble may not be seen so easily, due to complicated interactions with the environment. We will discuss an experiment where multiple stochastic preparation procedures are implemented in Chap. \ref{chapexp}.

\section{Projective preparation}
\index{quantum process tomography! projective preparation}
\index{preparation! projective! quantum process tomography}
Let us now consider projective preparation procedure in quantum process tomography experiments. Combining the projective preparations given by Eq. \ref{mesprep} with the generalized process equation \ref{output}, we get 
\index{process equation! projective}
the projective process equation for the $m$th input state
\begin{eqnarray}\label{BasicProcessEquation}
Q^{(m)} = \frac{1}{r^{(m)}} \mbox{Tr}_\mathcal{E} \left[UP^{(m)} \rho^{\mathcal{SE}} P^{(m)}U^\dagger\right] .
\end{eqnarray}

Is this process given by a linear map?  Dynamically, the  evolution of the total state $\rho^{\mathcal{SE}}$ is  linear, because every operation on $\rho^{\mathcal{SE}}$ is linear.  However, for the purpose of tomography, the dynamics of the prepared input states $P^{(m)}$ is of interest.  But $P^{(m)}$ appears twice in the process equation, therefore the output states, $Q^{(m)}$, depend bilinearly on $P^{(m)}$. This bilinearity can also be seen from the dependence of the environment state on the state of the system.  To see this expand Eq. \ref{BasicProcessEquation} as:
\begin{eqnarray}
Q^{(m)} &=&  \mbox{Tr}_\mathbb{B} \left[U P^{(m)} \otimes \rho^{\mathcal{E},(m)} U^\dagger\right],\\
\mbox{with}\hspace{.5cm}&\\
\rho^{\mathcal{E},(m)}  &=& \frac{1}{r^{(m)}} \mbox{Tr}_\mathcal{S}\left[P^{(m)}\rho^{\mathcal{SE}} \right].
\end{eqnarray}
The last equation shows the environment state is in effect a function of $P^{(n)}$.  It is well known \cite{CarteretTernoZyczkowski05, RomeroEtal04, Ziman06} that if the initial state of the system is related to the state of the environment by some function $f$ as
\begin{equation}\label{nonlinev}
\rho^\mathcal{SE} = \sum_j \rho^\mathcal{S}_j \otimes f(\rho^\mathcal{S}_j),
\end{equation}
then the evolution of the reduced matrices, $\rho^\mathcal{S}$, cannot be consistently described by a single linear map.   Above $f(\rho^{\mathcal{S}})$ are the density matrices of the state of the environment with dependence on the state of the system. In this case, the function $f$ is of a specific form that gives us a bilinear dependence.

It is instructive to now look at a simple example to demonstrate that when input states are prepared by projective preparations, the results cannot be consistently described by a linear map. 

\subsection{An example with projective preparation}\label{prjex}
\index{example of! process map! projective preparation}

Suppose we are unaware of the dependence of the state of the environment on the state of the system due to projective preparation and assume that the process is given by a linear map.  We would prepare a set of linearly independent input states, then construct the linear process map from the output states and the duals of input states. 

We once again start with the total state given by Eq. \ref{sep1} and the input states given in Eq. \ref{prj}. The state of the system plus the environment after each projective preparation takes the following form:
\begin{eqnarray}\label{mesinputs1}
P^{(m)}\rho^\mathcal{SE}P^{(m)}\rightarrow P^{(m)}\otimes\frac{1}{2}\mathbb{I}\;\;(\mbox{for}\;m=\{(1,\pm),(3,+)\} )\nonumber\\
P^{(2,+)}\rho^\mathcal{SE}P^{(2,+)}\rightarrow P^{(2,+)}\otimes\frac{1}{2}\left(\mathbb{I}+c'_{23}\sigma_3\right),
\hspace{.5cm}
\end{eqnarray}
where $c'_{23}=\frac{c_{23}}{1+a_2}$.

Following the same recipe as before, we obtain the output states using Eq. \ref{BasicProcessEquation}. We will use the same global unitary from the last two examples given by Eq. \ref{unitarys}.  The output states are as follows:
\begin{eqnarray}
Q^{(1,-)}&=&\frac{1}{2}\left\{\mathbb{I}-\cos^2(2\omega t)\sigma_1\right\},\\
Q^{(1,+)}&=&\frac{1}{2}\left\{\mathbb{I}+\cos^2(2\omega t)\sigma_1\right\},\\
Q^{(2,+)}&=&\frac{1}{2}\left\{\mathbb{I}-c'_{23}\cos(2\omega t)
\sin(2\omega t)\sigma_1\right.\\
&&\hspace{.5cm}\left.+\cos^2(2\omega t)\sigma_2
+c'_{23}\sin^2(2\omega t)\sigma_3\right\},\nonumber\\
Q^{(3,+)}&=&
\frac{1}{2}\left\{\mathbb{I}+ \sin^2(2\omega t)\sigma_3\right\}.
\end{eqnarray}

The linear process map is constructed using Eq. \ref{sqptlin}, the duals in Eq. \ref{dual}, and the output states:
\begin{eqnarray}\label{prjmap}
\small{\Lambda_{p}=\frac{1}{2}\left(\begin{array}{cccc}
1+C^2 & ic'_{23} S^2 &0 & 2C^2-ic'_{23}C S  \vspace{.2 cm}\\
-ic'_{23}S^2 & 1-C^2 & ic'_{23}C S & 0 \vspace{.2 cm}\\
0 & -ic'_{23}C S &1-C^2 & -ic'_{23}S^2  \vspace{.2 cm}\\
 2C^2+ic'_{23}C S & 0& ic'_{23}S^2 & 1+C^2 \vspace{.2 cm}
\end{array}\right).}
\end{eqnarray}
\begin{center}
\begin{figure}[!htt]
\resizebox{12.8 cm}{7.8 cm}{\includegraphics{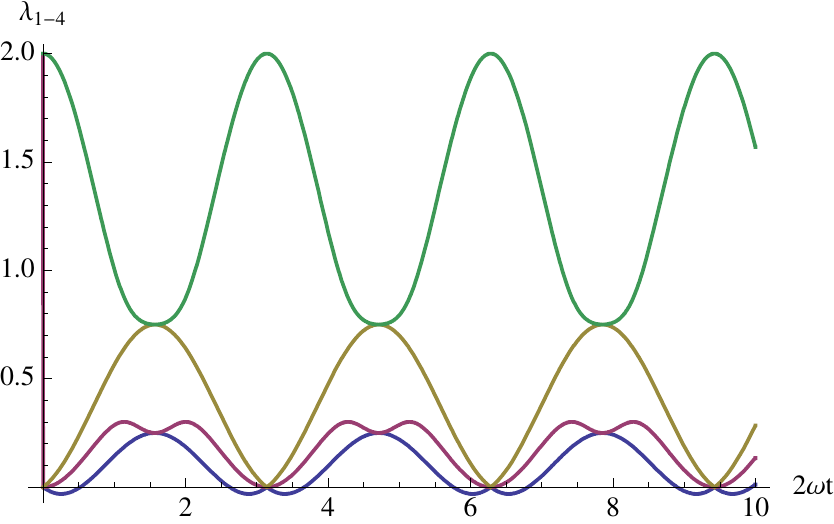}}
\caption{\label{fig5_3}
The eigenvalues of the process map in Eq. \ref{prjmap} are plotted as function of $2\omega t$. One of the eigenvalue is negative for certain values of $\omega t$; we have taken $c'_{23}=0.5$.  The negativity here is 
\index{process map! projective}
\index{negative maps! projective preparation}
due to the initial correlations and the inconsistence preparation procedure.}
\end{figure}
\end{center}

Now consider the action of $\Lambda_p$ on state $P^{(2,-)}=\frac{1}{2}(1-\sigma_2)$. $P^{(2,-)}=P^{(1,-)}+P^{(1,+)}-P^{(2,+)}$  is a linear combination of three of the input states used above.  If the action of $\Lambda_m$ is linear then the output state corresponding to $P^{(2,-)}$ should be 
\begin{eqnarray}\label{lact}
\Lambda_p\left(P^{(2,-)}\right)&=&\Lambda_p\left(P^{(1,-)}+P^{(1,+)}-P^{(2,+)}\right)\nonumber\\
&=&Q^{(1,-)}+Q^{(1,+)}-Q^{(2,+)}\\
&=&\frac{1}{2}\left(\mathbb{I}+c'_{23}
\cos(2\omega t)\sin(2\omega t)\sigma_1\right.\\
&&\hspace{.5cm}\left.-\cos^2(2\omega t)\sigma_2
-c'_{23}\sin^2(2\omega t)\sigma_3\right).\nonumber
\end{eqnarray}
Let us check if this the same as if the system was prepared in the state $P^{(2,-)}$ by projective preparation.  The output state for input $P^{(2,-)}$ can be calculated using Eq. \ref{BasicProcessEquation},
\begin{eqnarray}\label{realst}
Q^{(2,-)}&=&\frac{1}{r^{(2,-)}}\mbox{Tr}_{\mathbb B}\left[UP^{(2,-)}\rho^\mathcal{SE}P^{(2,-)}U^{\dag}\right]\nonumber\\
&=&\frac{1}{2}\left(\mathbb{I}-c''_{23} \cos(2\omega t)\sin(2\omega t)\sigma_1\right.\\
&&\hspace{.5cm}-\left.\cos^2(2\omega t)\sigma_2- c''_{23}\sin^2(2\omega t)\sigma_3\right),
\end{eqnarray}
where $c''_{23}=\frac{c_{23}}{1-a_2}$.  

The output state predicted by the linear process map in Eq. \ref{lact} is not the same as the real output state calculated dynamically in Eq. \ref{realst}, hence the linear process map $\Lambda_m$ does not describe the process correctly.  This is not surprising; observe the state of the environment in Eq. \ref{mesinputs1}. It depends on  $a_2$, and subsequently  the linear process map depends on $a_2$, thus on the initial state of the system.  The dependence of the environment on the system is the cause for the non-linear behavior of the process.

\section{Discussion}

The examples with multiple stochastic preparations (in Sec. \ref{exmultstoch}) and projective preparations (in Sec. \ref{prjex}) look very similar.  For multiple stochastic preparations, the state of the environment depends on the stochastic map.  If we think of each projective preparations as an independent stochastic preparation then the state of the environment, in the example in Sec. \ref{prjex}, depends on the preparation map.  Then are the two situations the same? Let us show below that the two situations are fundamentally different. 

In the weak coupling limit, the projective preparation procedure does not play an important role.  As we saw in the last chapter, when the system and the environment are initially uncorrelated, the projective preparation procedure will have no affect on the state of the environment. This is because the state of the environment is affected only indirectly due to the 
\index{quantum process tomography! weak coupling limit}
initial correlations between the system and the environment.  Thus the projective preparations, in the weak coupling limit, yield a linear process map.

This is not the case in the multiple stochastic preparation example. When multiple stochastic maps are used to prepare different input states, the inconsistencies do not stem from the initial correlations between the system and the environment.  In fact, in our example all input states are initially uncorrelated from the environment.  The inconsistencies arise from the preparation procedures themselves, leading to different states of the environment for different input states.  These inconsistencies will be absent in the case where the system develops in a closed form, since the system does not feel the presence of the environment during the quantum 
\index{quantum process tomography! weak interaction limit}
process.  Hence in the ``weak interaction limit'' multiple stochastic preparations yield a linear process map.

Lastly, note that each of the process map found in this chapter, given by Eqs. \ref{sact}, \ref{msact}, and \ref{lact}, are different from each other.  This clearly shows that the preparation procedures play a non-trivial role in open quantum system experiments.  The preparation procedure is the only thing that distinguished each case.  For the case where no preparation procedure is applied, i.e. the case of the dynamical map (Eq. \ref{dynamicalmap}), the situation is still different.  The dynamical map, which has negative eigenvalues, is linear and has a valid interpretation within the compatibility domain.  The process maps in Eqs. \ref{msact} and \ref{lact} do not have any consistent interpretation. 

When negative eigenvalues are found in a process map, they hint to some problem in the preparation procedure.  Although this is a bit premature to state at this point; we analyze the negative eigenvalues in process map in more detail in the next chapter. In Chap. \ref{chapgMmap} we show that a processes can be correctly described using a general dynamical map, which we call dynamical $\mathcal{M}$-map.  This map is independent of the preparation procedure.

\chapter{More on stochastic preparations}\label{appdvp}
\index{More on stochastic preparations@\emph{More on stochastic preparations}}%


Dynamical maps \cite{SudarshanMatthewsRau61, SudarshanJordan61} are a 
\index{dynamical map}
generalization of the unitary transformation much like density matrices are a generalization of the pure state rays.  Dynamical maps allow the description of 
\index{stochastic process}
stochastic processes as well as the evolution of open systems.  Dynamical maps used to describe the evolution of open systems are usually defined with a constant environment state $\rho^\mathcal{E}$:
\begin{eqnarray}
\mathcal{B}\rho^{\mathcal S} = \mbox{Tr}_\mathcal{E}\left[U \rho^\mathcal{S} \otimes \rho^\mathcal{E} U^\dagger\right].
\end{eqnarray}
Implicitly, the state of the environment, $\rho^\mathcal{E}$, is a parameter of the map $\mathcal{B}$.  Therefore, the linear dynamical map $\mathcal{B}$ would only consistently describe an experiment if different input states $\rho^\mathcal{S}$ can be prepared independently of the environment state $\rho^\mathcal{S}$.  The actual issue of how this can be executed is not addressed.

It is possible to consider a dynamical map where the environment is not fixed, such as the reduced dynamical evolution of a initially correlated state $\rho^\mathcal{SE}$ \cite{jordan:052110}:
\begin{eqnarray}\label{dmaps}
\mathcal{B}\mbox{Tr}_{\mathcal{E}} \left[\rho^\mathcal{SE}\right]
=\mbox{Tr}_{\mathcal E}\left[U\rho^\mathcal{SE}U^\dag\right].
\end{eqnarray}
The dynamical map in this problem is applicable only over a  compatibility domain of states, rather than over the complete state space of the system (see Sec. \ref{inicorr}).  The compatibility domain is the set of states of the system that are compatible with the correlations in $\rho^\mathcal{SE}$.  Formally, this problem defines an embedding map\footnote{The embedding map is also known as extension map, assignment map, or a preparation (not to be confused with the preparation map from Chap. \ref{chapprp}).} that relates the initial state of the system to the tota initial state, $\rho^\mathcal{SE}$.  The embedding map is linear but not necessarily a completely positive map \cite{modiext,pechukas94a, PhysRevLett.75.3020}.  How do we handle the initial correlations 
\index{compatibility domain}\index{embedding}\index{preparation! stochastic} experimentally in the case of stochastic preparation?

The stochastic preparation method provides a way for an experiment to be performed so that different input states can be prepared with a fixed environment state.  Consider the process equation
\index{process equation! linear} \ref{stocproceq} in section \ref{stochprep}.
\begin{eqnarray}
Q^{(m)} &=& \mbox{Tr}_\mathcal{E} \left[U\Omega^{(m)} \circ\Theta \rho^\mathcal{SE} U^\dagger\right]\nonumber \\
& =& \mbox{Tr}_\mathcal{S}\left[U P^{(m)} \otimes \rho^{\mathcal{E},\Theta} U^\dagger\right].\nonumber
\end{eqnarray}
Then the action of the process map on an arbitrary state of the system is then given by:
\begin{eqnarray}
\Lambda\rho^{\mathcal S}  &=& \mbox{Tr}_\mathcal{E} \left[ U \rho^{\mathcal S} \otimes \rho^{\mathcal{E},\Theta} U^\dagger\right].\nonumber
\end{eqnarray}
Therefore, in this context, the dynamical map is equivalent to the process map.  However, for consistency, we have to remember that the environment 
\index{process map! linear}
state is a constant to the problem, therefore the pin map $\Theta$ should also be a constant to the problem.

\section{Pseudo-pure states}
\index{pseudo-pure state}

Demanding pure input states in a quantum process tomography experiment guarantees initially uncorrelated state of the system.  Though if it is not possible to prepare pure states, we can still determine the process. 

Suppose the initial state is prepared by a pin map leading to
\begin{eqnarray}\label{pspure1}
\Theta\rho^\mathcal{SE}&=&
\left(pP^{(3,+)}+(1-p)P^{(3,-)}\right)\otimes\mathbb{I}\\
&=&\frac{1}{2}\{\mathbb{I}+p\sigma_3\}\otimes\mathbb{I},
\end{eqnarray}
where $0<<p<1$.  The rest of the input states are prepared by rotations.  Then the input states are
\begin{eqnarray}\label{psprj}
P^{(1,-)}=\frac{1}{2}\{\mathbb{I}-p\sigma_1\}&&
P^{(1,+)}=\frac{1}{2}\{\mathbb{I}+p\sigma_1\}\\
P^{(2,+)}=\frac{1}{2}\{\mathbb{I}+p\sigma_2\}&&
P^{(3,+)}=\frac{1}{2}\{\mathbb{I}+p\sigma_3\}.
\end{eqnarray}
For the unitary operator in Eq. \ref{unitarys}, the corresponding output states will be
\begin{eqnarray}
Q^{(1,-)}=\frac{1}{2}\{\mathbb{I}-p\cos^2(2\omega t)\sigma_1\},&&
Q^{(1,+)}=\frac{1}{2}\{\mathbb{I}+p\cos^2(2\omega t)\sigma_1\}\\
Q^{(2,+)}=\frac{1}{2}\{\mathbb{I}+p\cos^2(2\omega t)\sigma_2\},&&
Q^{(3,+)}=\frac{1}{2}\{\mathbb{I}+p\cos^2(2\omega t)\sigma_3\}.
\end{eqnarray}

The only change that we have to make to find the process map is to define a 
\index{negative maps! stochastic preparation}
\index{dual set! for a qubit}
dual proper set, in this case
\begin{eqnarray}\label{psdual}
\tilde{P}^{(1,-)} = \frac{1}{2p}\{p\mathbb{I} -\sigma_1-\sigma_2- \sigma_3\}, &&
\tilde{P}^{(1,+)} = \frac{1}{2p}\{p\mathbb{I} +\sigma_1-\sigma_2-\sigma_3 \}, \nonumber\\
\tilde{P}^{(2,+)} =\frac{1}{p}\sigma_2,&&
\tilde{P}^{(3,+)} =\frac{1}{p}\sigma_3. 
\end{eqnarray}
We can find the process map using Eq. \ref{sqptlin}.  The process map here case turns out to be the same as in Eq. \ref{sact}.  This is to expected; the process map for stochastic preparation is defined over set of all states, including the input states above.

The downside of course is that there is no way to distinguish the states in Eq. \ref{pspure1} from the following state
\begin{eqnarray}
\Theta\rho^\mathcal{SE}=\frac{1}{2}\{\mathbb{I}\otimes\mathbb{I}+p\sigma_3+c_{23}\sigma_2\otimes\sigma_3\}.
\end{eqnarray}
Unlike in Eq. \ref{psprj}, this is a correlated state. Yet, both total
\index{negative maps! due to correlations}
states have the same reduced state for the system part.

Even in this case we can find the process map properly if we use the correct dual set given in Eq. \ref{psdual}.  The process map in that case will be the same as the dynamical map given in Sec. \ref{inicorr} by Eq. \ref{dynamicalmap}.  In both of the examples above, had we assumed that the input states were close enough to the pure states we desired, and used the dual set give by Eq. \ref{dual}, then the process map would contain an error of magnitude $1-p$.

These two examples illustrate the contrary argument to what some have suggested, that only the completely positive process maps should be considered physically valid.  We showed in the second example, that one can obtain a not-completely positive process map in a consistent fashion.  The real issue is that the correlations with environment are not convenient for experimental purposes.  But that does not mean that one should fix a not-completely positive process map to a completely positive process map with numerical methods \cite{havel03,ziman}.

\section{Negative maps due to control errors}
\index{control errors}

There is one more possibility for non-linearity and negativity in a process map.  This has to do with poor control in the stochastic preparation procedure.  Suppose the initial state is prepared well using a pin map
\begin{eqnarray}\label{pspure2}
\Theta\rho^\mathcal{SE}&=&P^{(3,+)}\otimes\mathbb{I}.
\end{eqnarray}
After obtaining this state, the other input states are prepared by local rotations.  Let us consider the case where one of the rotation is not perfect.
\begin{eqnarray}
V^{(1,-)}\ket{1}
\rightarrow\frac{1}{\sqrt{2}} \left(
\sqrt{1-\epsilon}\ket{1}-\sqrt{1+\epsilon}\ket{0}\right),
\end{eqnarray}
where $\epsilon$ is taken to be a small positive real number.  We introduced a small error for the preparation of $P^{(1,-)}$, but we have kept the error simple by not giving it an additional phase, i.e. keeping $\epsilon$ to be real.  For simplicity we have also assumed that the error of this sort occurs in the preparation of only one state.

Let us now pretend that we are not aware of this error.  Then in reality we have the following set of input states
\begin{eqnarray}\label{psprj2}
P^{(1,-)}&=&\frac{1}{2}\{\mathbb{I}
+\epsilon\sigma_3
-\sqrt{1-\epsilon^2}\sigma_1\},\;\;
P^{(1,+)}=\frac{1}{2}\{\mathbb{I}+\sigma_1\},\\
P^{(2,+)}&=&\frac{1}{2}\{\mathbb{I}+\sigma_2\},\;\;
P^{(3,+)}=\frac{1}{2}\{\mathbb{I}+\sigma_3\}.
\end{eqnarray}
Let us use the same unitary evolution as before given in Eq. \ref{unitarys}. The output states corresponding to the last three input states in Eq. \ref{psprj2} are the same as before, given by Eq. \ref{stochoutputs}.  For the input state $P^{(1,-)}$, the corresponding output state is follows
\begin{eqnarray}\label{poorout}
Q^{(1,-)}&=&\frac{1}{2}\{\mathbb{I}
+(\epsilon\sigma_3
-\sqrt{1-\epsilon^2}\sigma_1)\cos^2(2\omega t)\},\nonumber\\
Q^{(1,+)}&=&\frac{1}{2}\{\mathbb{I}+\cos^2(2\omega t)\sigma_1\}\nonumber\\
Q^{(2,+)}&=&\frac{1}{2}\{\mathbb{I}+\cos^2(2\omega t)\sigma_2\},\nonumber\\
Q^{(3,+)}&=&\frac{1}{2}\{\mathbb{I}+\cos^2(2\omega t)\sigma_3\}.
\end{eqnarray}
Using these output states, the duals given by Eq. \ref{dual}, and Eq. \ref{sqptlin}, we can find the process map.  

\begin{center}
\begin{figure}[!ht]
\resizebox{12.8 cm}{7.8 cm}{\includegraphics{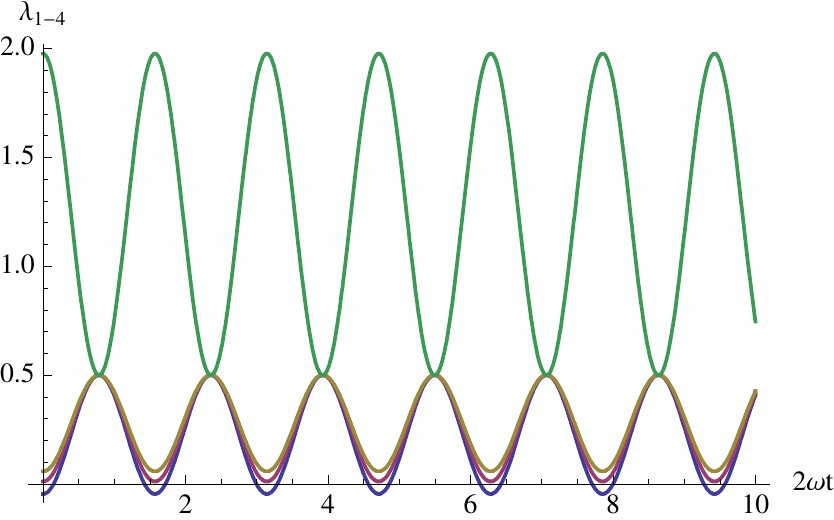}}
\caption{\label{negfig}
The eigenvalues of the process map found using the output states in Eq. \ref{poorout} and the duals in Eq. \ref{dual}.  One of the eigenvalue is negative for certain times; we have taken $\epsilon=0.1$.  The negativity is due to the errors in the unitary operation implemented to prepare one of the input states.}
\end{figure}
\end{center}

The process map turns out to be rather complicated, and its eigenvalues are even more complicated looking.  Therefore, we do not write them down, instead we have plotted the eigenvalues as function of $2\omega t$.  We 
\index{negative maps! control errors}
take the value for the error to be $\epsilon=0.1$ for the plot in Fig \ref{negfig}.  One of the eigenvalue in Fig. \ref{negfig} is negative for certain times.  This shows yet another cause for negative eigenvalues in a process map. The negative eigenvalues here have nothing to do with the initial correlations between the system and the environment.  The negative eigenvalues are attributed to poor control in the preparation procedure.

\section{Discussion}

Now we have discussed several scenarios that can lead to negative eigenvalues for a process map.  Two comments are in order at this point.  When the prepared input states are not pure and the process map has negative eigenvalues, 
\index{negative maps}
then one should be weary of initial correlations with the environment.  To further check this, the process map should be tested for linearity by sending several additional input states through the process.  If the process map predicts the output states properly, then one can be confident that the system is initially correlated with the environment.  If the process map does not predict the output states correctly (non-linear behavior), then there are additional problems with the experiment, including the possibility of poor control in the preparation procedure.  In the case where the input states are pure and the process map has negative eigenvalues, the negativity can only come from either inconsistencies in the preparation procedure or poor preparation control.



\chapter{Dynamical $\mathcal{M}$-map}\label{chapgMmap}
\index{Dynamical $\mathcal{M}$-map @\emph{Dynamical $\mathcal{M}$-map}}%

As we saw in Chap. \ref{chapqptex}, the additional step of preparation 
\index{$\mathcal{M}$-map|see{process map}}
leads to complications for quantum process tomography experiments.  At the same time, this step cannot be avoided for systems initially correlated with the environment.  Let us try to attack this problem directly.  Let us rewrite the process equation \ref{output}
\begin{eqnarray}
Q^{(m)}=\frac{1}{r^{(m)}}
\mbox{Tr}_{\mathcal E}[U\mathcal{P}^{(m)}\rho^\mathcal{SE} U^\dag].
\end{eqnarray}
In terms of matrix indices we have
\begin{eqnarray}
Q^{(m)}_{rs}=\frac{1}{r^{(m)}}\sum_{\epsilon}
U_{r\epsilon;r'\alpha}\mathcal{P}^{(m)}_{r'r'';s's''}\rho^\mathcal{SE}_{r''\alpha;s''\beta} U^*_{s\epsilon;s'\beta},
\end{eqnarray}
where the sum over $\epsilon$ denotes the trace with respect to the environment. We are interested in the reduced dynamics of the system states; not so much in the details of the preparations.  Since the preparation map only acts on the system, the trace with the environment has no effect on it. Thus, we can just pull the preparation map out of the trace,
\begin{eqnarray}
Q^{(m)}_{rs}
&=&\frac{1}{r^{(m)}}\nonumber
\mathcal{P}^{(m)}_{r'r'';s's''}\sum_{\epsilon}
U_{r\epsilon;r'\alpha}
\rho^\mathcal{SE}_{r''\alpha;s''\beta} 
U^*_{s\epsilon;s'\beta}\\
&=&\frac{1}{r^{(m)}}\label{RawProcessEquation}
\mathcal{M}^{(rs)}_{r'r'';s''s'}
\mathcal{P}^{(m)}_{r'r'';s's''}.
\end{eqnarray}
In the last equation, the matrix $\mathcal{M}$ is defined as:
\begin{eqnarray}\label{mmap}
\mathcal{M}^{(rs)}_{r'r''; s''s'} = \sum_{\epsilon} U_{r \epsilon,r' \alpha} 
{\rho^{\mathcal{SE}}}_{r''\alpha,s''\beta} U^*_{s \epsilon,s' \beta} .
\end{eqnarray}
Note that in Eq. \ref{RawProcessEquation} the superscript indices on $\mathcal{M}$ match the elements on the left hand side of the equation, 
\index{process equation! $\mathcal{M}$-map}
while the subscript indices are summed on the right hand side of the equation. 

The output state, $Q^{(m)}$, is given by the matrix $\mathcal{M}$ acting generally on the preparation of the input state, $\mathcal{P}^{(m)}$.  Therefore, matrix $\mathcal{M}$ fully describes the process before any preparation is made. $\mathcal{M}$ contains both $U$ and $\rho^{\mathcal{SE}}$; however knowing $\mathcal{M}$ is not sufficient to determine $U$ and $\rho^{\mathcal{SE}}$.  As expected, it should not be possible to determine $U$ and $\rho^{\mathcal{SE}}$ through measurements and preparations on the system alone without access to the environment.  Conversely, $\mathcal{M}$ contains all information necessary to fully determine the output state for any prepared state. We will call $\mathcal{M}$ \index{process map! $\mathcal{M}$-map} the \emph{dynamical $\mathcal{M}$-map} or just \emph{$\mathcal{M}$-map} for short.  The $\mathcal{M}$-map is due to Aik-Meng Kuah \cite{kuahdis}. Let us look at some basic properties of $\mathcal{M}$-map.

\noindent\emph{Trace of $\mathcal{M}$-map:}

Let us start with the trace of $\mathcal{M}$.  To take the trace, we equate 
\index{process map! $\mathcal{M}$-map! trace of}
indices $r$ with $s$, $r'$ with $s'$, and $r''$ with $s''$ to get: 
\begin{eqnarray*}
\mbox{Tr}[\mathcal{M}]&=&
\sum_{rr'r''}\delta_{rs}\delta_{r's'}\delta_{r''s''}
 M_{r'r'';s''s'}^{(r,s)}\\
&=&\sum_{ \epsilon}\sum_{rr'r''}
U_{r \epsilon ,r' \alpha} {\rho^{\mathcal{SE}}}_{r''\alpha ,r''\beta}U^*_{r \epsilon ,r' \beta}.
\end{eqnarray*}
Since $U^\dag U=\sum_{r\epsilon} U^*_{r\epsilon,r'\beta}U_{r\epsilon,r'\alpha}
=\mathbb{I}_{r' \alpha,r'\beta}$, then
\begin{eqnarray*}
\mbox{Tr}[\mathcal{M}]
=\sum_{\alpha,\beta r''r'} 
\mathbb{I}_{r'\alpha,r'\beta}
{\rho^{\mathcal{SE}}}_{r''\alpha,r''\beta}=d,
\end{eqnarray*}
where $d$ is the dimension of the system.

\noindent\emph{Hermiticity of $\mathcal{M}$-map:}

As with the case of linear process maps, matrix $\mathcal{M}$ is hermitian.  
\index{process map! $\mathcal{M}$-map! Hermiticity of}
This is easy to see by taking the complex conjugate of matrix $\mathcal{M}$,
\begin{eqnarray*}
\left(\mathcal{M}_{r'r'';s''s'}^{(r,s)}\right)^* &=&\left( \sum_{\epsilon} U_{r \epsilon,r' \alpha} 
{\rho^{\mathcal{SE}}}_{r''\alpha,s''\beta} U^*_{s \epsilon,s' \beta} \right)^*\\
&=&\sum_{\epsilon}  U_{s \epsilon,s' \beta}
{\rho^{\mathcal{SE}}}_{s''\beta ,r''\alpha}U^*_{r \epsilon,r' \alpha,} \\
&=&\mathcal{M}_{s's'';r''r'}^{(s,r)}.
\end{eqnarray*}
The complex conjugate of $\mathcal{M}$ is not only the transpose of $\mathcal{M}$, but each element of $\mathcal{M}$ is also transposed.  Hence $\mathcal{M}$ is a Hermitian matrix.

\noindent\emph{Positivity of $\mathcal{M}$-map:}

The $\mathcal{M}$-map is composed of a unitary matrix operating on a 
\index{process map! $\mathcal{M}$-map! positivity of}
density matrix.  Then we can take the square root of the density matrix to get
\begin{eqnarray*}
\mathcal{M}_{r'r'',s''s'}^{(r,s)}
&=&\sum_{\epsilon} 
U_{r \epsilon,r' \alpha} 
\left[\sqrt{\rho^{\mathcal{SE}}}\right]_{r''\alpha,\sigma\gamma} 
\left[\sqrt{\rho^{\mathcal{SE}}}\right]_{\sigma\gamma,s''\beta} 
U^*_{s \epsilon,s' \beta}\\
&=&\sum_{\epsilon} 
\left(\sum_{\alpha}U_{r \epsilon,r' \alpha} 
\left[\sqrt{\rho^{\mathcal{SE}}}\right]_{r''\alpha,\sigma\gamma} \right)
\left(\sum_{\beta}U_{s \epsilon,s' \beta}
\left[\sqrt{\rho^{\mathcal{SE}}}\right]_{s''\beta,\sigma\gamma}\right)^*\\
&=&\sum_{\mu}
M^{(\mu)}_{r;r'r''}
{M^{(\mu)}_{s's'';s}}^*,
\end{eqnarray*}
where $M=U \sqrt{\rho^{\mathcal{SE}}}$ and $\mu=\sigma\gamma\epsilon$.  We have written the $\mathcal{M}$-map in operator sum representation, hence it is completely positive.  This means, the $\mathcal{M}$-map acting on any completely positive preparation procedure will lead to a physical state.

The action of $\mathcal{M}$-map can now be written as 
\begin{eqnarray}
\mathcal{M}_{r'r'';s''s'}^{(rs)}\mathcal{P}_{r'r'';s's''}=
\sum_{\mu}
M^{(\mu)}_{r;r'r''}
\mathcal{P}_{r'r'';s's''}
M^{(\mu)}_{s's'';s}
\end{eqnarray}

The advantage of dealing with the $\mathcal{M}$-map is that we no longer need to worry about the preparation procedure.  Once we know the $\mathcal{M}$-map, we can act with it on any preparation procedure to find the corresponding output state.  Since we took out the preparation part out of the $\mathcal{M}$-map, it contains all of the dynamical information for the system.  In the next section we will try to extract some of this information from the $\mathcal{M}$-map.

\section{Uninterrupted dynamics}
\index{process map! $\mathcal{M}$-map! uninterrupted dynamics}

The $\mathcal{M}$-map contains the dynamics of the system before any preparation is made on the system.  Perhaps we can learn something about that dynamics. Let us start by looking at the definition of the $\mathcal{M}$-map.
\begin{eqnarray}\label{mmap1}
\mathcal{M}^{(rs)}_{r'r''; s''s'} = \sum_{\epsilon} U_{r \epsilon,r' \alpha} 
{\rho^{\mathcal{SE}}}_{r''\alpha,s''\beta} U^*_{s \epsilon,s' \beta} .
\end{eqnarray}

The system is labeled by indices $r''$ and $s''$. Then by tracing over everything else we can find the initial state of the system from $\mathcal{M}$-map.
\begin{eqnarray}\label{mmap2}
\sum_{rr'}\delta_{rs}\delta_{r's'}\mathcal{M}^{(rs)}_{r'r''; s''s'}
&=&\sum_{\epsilon}\sum_{rr'} 
U_{r \epsilon,r' \alpha} 
\rho^{\mathcal{SE}}_{r''\alpha;s''\beta} 
U^*_{r \epsilon,r' \beta}\\
&=&\sum_{r'} 
\left(\sum_{r\epsilon}U^*_{r \epsilon,r' \beta}
U_{r \epsilon,r' \alpha} \right)
\rho^{\mathcal{SE}}_{r''\alpha;s''\beta}\\
&=&\sum_{r'}\mathbb{I}_{r'r'}
\sum_{\alpha \beta} \mathbb{I}_{\alpha\beta}
\rho^{\mathcal{SE}}_{r''\alpha;s''\beta} \\
&=&d\;\rho^\mathcal{S}_{r''s''}.\label{initialstates}
\end{eqnarray}

Furthermore, if we apply an identity preparation, defined as
\begin{eqnarray}\label{iprep}
\mathcal{P}^{(\mathbb{I})}_{r'r'';s's''}
=\mathbb{I}_{r'r''}\mathbb{I}^{*}_{s''s'},
\end{eqnarray}
on the $\mathcal{M}$-map then we get the final state of the system.
\begin{eqnarray}\label{mmap3}
\mathcal{M}^{(rs)}_{r'r''; s'r''}\mathcal{P}^{(\mathbb{I})}_{r'r'';s's''}
&=&\mathbb{I}_{r'r''}
\mathcal{M}^{(rs)}_{r'r''; s's''}
\mathbb{I}^{*}_{s''s'}\\
&=&\sum_{\epsilon} 
U_{r \epsilon,r' \alpha} 
\mathbb{I}_{r'r''}
\rho^{\mathcal{SE}}_{r''\alpha;s''\beta} 
\mathbb{I}_{s''s'}
U^*_{s \epsilon,s' \beta}\\
&=&\sum_{\epsilon} 
U_{r \epsilon,r' \alpha} 
\rho^{\mathcal{SE}}_{r'\alpha;s'\beta} 
U^*_{s \epsilon,s' \beta}\\
&=&\rho^{\mathcal{S}}_{rs}(t).
\end{eqnarray}

Knowing the initial and the final states of the system (without any preparations) allows us to find the dynamics uninterrupted. From this information we can find the mapping from the initial to the final state.  In general this will not be the same as the dynamical map of Chap. \ref{chapopendyn}.  The dynamical map is defined over a set of initial states of the system.  Here we get a mapping from an initial state to a final state.

\section{Memory due to correlations}
\index{memory matrix}

When a system is initially correlated with the environment, the state of the system in the future will depend on these correlations.  This is the memory of a system, and it is a key feature of non-Markovian dynamics \cite{cesarnonmarkov}.  We will show here that we can find the dynamics of the memory from the $\mathcal{M}$-map.

The memory of the system is obtained from the $\mathcal{M}$-map by tracing over the system, indices $r''$ and $s''$.
\begin{eqnarray}\label{mmap4}
\sum_{r''}\delta_{r''s''}\mathcal{M}^{(rs)}_{r'r''; s's''}
=\sum_{\epsilon} U_{r \epsilon,r' \alpha} 
\rho^{\mathcal{E}}_{\alpha\beta} U^*_{s \epsilon,s' \beta}.
\end{eqnarray}
The last equation is the same as Eq. \ref{processmapsto} for the process map found by stochastic preparations. Thus,
\begin{eqnarray}\label{processfromM}
\Lambda_s=\sum_{r''}\delta_{r''s''}\mathcal{M}^{(rs)}_{r'r'';s's''}.
\end{eqnarray}
This means, even though the $\mathcal{M}$-map contains the correlated total state, we can find the process for the ideal stochastic preparation from it.  

Consider the following matrix composed of the matrices in Eqs. \ref{initialstates} and \ref{processfromM}
\begin{eqnarray}
\mathcal{L}^{(rs)}_{r'r'';s's''}&=&
\Lambda_{rr';ss'}\rho^\mathcal{S}_{r''s''}\\
&=&d\left(\sum_{r''}\delta_{r''s''}\mathcal{M}^{(rs)}_{r'r'';s's''}\right)
\left(\sum_{rr'}\delta_{rs}\delta_{r's'}\mathcal{M}^{(rr)}_{r'r''; s'r''}\right)\\
&=&\sum_{\epsilon} U_{r \epsilon,r' \alpha} 
\rho^\mathcal{S}_{r''s''}\rho^{\mathcal{E}}_{\alpha\beta} U^*_{s \epsilon,s' \beta}
\end{eqnarray}

The last equation similar to the expression for the $\mathcal{M}$-map, except the state of the system and the state of the environment are uncorrelated.  Let us now write the total correlated state as 
\begin{eqnarray}\label{corrmatrix}
\rho^{\mathcal{SE}}=\rho^{\mathcal{S}}\otimes\rho^{\mathcal{E}}+\chi,
\end{eqnarray}
where $\chi$ is the correlation matrix \cite{CarteretTernoZyczkowski05}.  
\index{correlation matrix}
Writing the $\mathcal{M}$-map in terms of the correlation matrix we get
\begin{eqnarray}
\mathcal{M}^{(rs)}_{r'r'';s's''}
=\sum_{\epsilon} U_{r \epsilon,r' \alpha} 
(\rho^{\mathcal{S}}_{r''s''}\rho^{\mathcal{E}}_{\alpha\beta}+\chi_{r''\alpha;s''\beta})
U^*_{s \epsilon,s' \beta}.
\end{eqnarray}

Now consider the following quantity:
\begin{eqnarray}
\mathcal{K}^{(rs)}_{r'r'';s's''}
&=&\mathcal{M}^{(rs)}_{r'r'';s's''}-
\mathcal{L}^{(rs)}_{r'r'';s's''}\\
&=&\sum_{\epsilon} U_{r \epsilon,r' \alpha}
(\rho^{\mathcal{S}}_{r''s''}\rho^{\mathcal{E}}_{\alpha\beta}+\chi_{r''\alpha;s''\beta}-\rho^\mathcal{S}_{r''s''}\rho^{\mathcal{E}}_{\alpha\beta}) U^*_{s \epsilon,s' \beta}\\
&=&\sum_{\epsilon} U_{r \epsilon,r' \alpha}
\chi_{r''\alpha;s''\beta} U^*_{s \epsilon,s' \beta}.\label{memoryeq}
\end{eqnarray}
The last equation gives us the reduced dynamics of the correlations between the system and the environment.  If the system and the environment were initially uncorrelated, then $\mathcal{K}$ will be zero, i.e. no memory is 
\index{memory matrix}
present.  We call $\mathcal{K}$ matrix the memory matrix.

This is an important result for studying non-Markovian systems.  It is important to keep in mind that, the memory matrix will be different for different preparations. But, if we want to study how the correlations act as memory in the uninterrupted system then we can just act with the identity preparation, defined in Eq. \ref{iprep} as,
\begin{eqnarray}
\chi^{\mathcal S}(t) &=&\mathcal{K}^{rs}_{r'r'';s's''}\mathcal{P}^{(\mathbb{I})}_{r'r'';s's''}\\
&=&\sum_{\epsilon}U_{r\epsilon,r'\alpha}\chi_{r'\alpha,s'\beta}U^*_{s\epsilon,s'\beta},
\end{eqnarray}
where $\chi^{\mathcal S}(t)=\mbox{Tr}_\mathcal{E}[U\chi U^\dag]$.  The quantity above is the memory for the uninterrupted system due to initial correlations with the environment.

\section{An example with $\mathcal{M}$-map and the memory matrix}
\index{example of! $\mathcal{M}$-map}
\index{example of! memory matrix}
Now, let us consider an example of the $\mathcal{M}$-map and the memory matrix. Let us use the same initial states (\ref{sep1}) and the unitary operator (\ref{unitarys}) as before. The $\mathcal{M}$-map is given by Eq. \ref{mmap} and the memory matrix is given by Eq. \ref{memoryeq}.

If we make the identity preparation on the $\mathcal{M}$-map then we get
\begin{eqnarray}
\mathcal{M}\circ\mathcal{P}^{(\mathbb{I})}
&=&\rho^{\mathcal S}(t)\\
&=&\frac{1}{2}\{ \mathbb{I} + \cos^2\left(2\omega t\right) a_j
\sigma_j -c_{23}\cos\left(2\omega t\right) \sin\left(2\omega
t\right)\sigma_1 \},
\end{eqnarray}
which is the same as the final state of the system in 
\index{preparation! identity}
Chap. \ref{chapopendyn} Eq. \ref{finalstate}.  While making the identity preparation on memory matrix we get
\begin{eqnarray}
\mathcal{K}\circ\mathcal{P}^{(\mathbb{I})}
&=&\chi^{\mathcal S}(t)
=\mbox{Tr}_{\mathcal E}[U\chi U^\dag]\\
&=&\frac{1}{2}\{-c_{23}\cos\left(2\omega t\right) 
\sin\left(2\omega t\right)\sigma_1 \},
\end{eqnarray}
where $\chi$ is the correlation matrix defined in Eq. \ref{corrmatrix} and $\chi^\mathcal{S}(t)$ is the reduced part of the time evolved correlation matrix.  This is precisely the contribution to $\rho^\mathcal{S}$ due to the correlations with the environment.

\begin{landscape}
The $\mathcal{M}$-map and the memory matrix are as follows:
\begin{center}
\begin{eqnarray*}
\mathcal{M}=\left(\begin{array}{llllllll}
\frac{a_3^+C^2_+}{2} &
0 & 0 &
a_3^+C^2  & 
 \frac{-i  c  S^2+a^-C^2_+}{2} & 
 0 & 0 &
 -c CS+a^- C^2 \\
 0 & 
 \frac{a_3^+C^2_-}{2} & 
 0 & 0 & 0 & 
 \frac{i  c + a^-}{2}S^2 & 
 0 & 0 \\
 0 & 0 & 
 \frac{a_3^+C^2_-}{2} & 
 0 & 0 & 0 &
 \frac{-i  c + a^-}{2}S^2 & 
 0 \\
a_3^+C^2 & 
0 & 0 & 
\frac{a_3^+C^2_+}{2} & 
c  CS+a^-C^2 & 
0 & 0 & 
\frac{i  c S^2+a^-C^2_+}{2} \\
\frac{i  c S^2 +a^+C^2_+}{2} & 
0 & 0 & 
c  CS+a^+C^2 & 
\frac{a_3^- C^2_+}{2}
& 0 & 0 & 
a_3^-C^2  \\
0 & 
\frac{-i c +a^+}{2}S^2 & 
0 & 0 & 0 & 
\frac{a_3^-C^2_-}{2} & 
0 & 0 \\
0 & 0 & 
\frac{i  c + a^+}{2}S^2 & 
0 & 0 & 0 & 
\frac{a_3^-C^2_-}{2} &
0 \\
- c CS+a^+C^2 & 
0 &  
\frac{-i c S^2+a^+C^2_+}{2} &
a_3^-C^2
& 0 & 0 & 0 &
\frac{a_3^-C^2_+}{2}
\end{array}\right),
\end{eqnarray*}
\begin{eqnarray}
\mathcal{K}=\left(
\begin{array}{cccccccc}
 1+C^2 & 0 & 0 & 2 C^2 & -i c_{23} S^2 & 0 & 0 & 2 c_{23}C S \\
 0 & 1-C^2 & 0 & 0 & 0 & i c_{23} S^2 & 0 & 0 \\
 0 & 0 & 1-C^2 & 0 & 0 & 0 & -i c_{23}S^2 & 0 \\
 2 C^2 & 0 & 0 & 1+C^2 & 2 c_{23} C S & 0 & 0 & i c_{23}S^2 \\
 i c_{23} S^2 & 0 & 0 & 2 c_{23}C S & 1+C^2 & 0 & 0 & 2 C^2 \\
 0 & -1 c_{23}S^2 & 0 & 0 & 0 & 1-C^2 & 0 & 0 \\
 0 & 0 & i c_{23}S^2 & 0 & 0 & 0 & 1-C^2 & 0 \\
 2 c_{23} C S & 0 & 0 & -i c_{23}S^2 & 2 C^2 & 0 & 0 & 1+C^2
\end{array}
\right)\nonumber,
\end{eqnarray}
\end{center}
where $C=\cos (2\omega t)$, $S=\sin(2\omega t)$, $C^2_\pm=1\pm C^2$, $a^\pm_3=1\pm a_3$, $a^+=a_1+ia_2$, $a^-=a_1-ia_2$, and $c=c_{23}$.
\end{landscape}


\section{Discussion}

The $\mathcal{M}$-map and the memory matrix are difficult to obtain in practice.   We have given a recipe to find  $\mathcal{M}$-map partially for a qubit in Apps. \ref{mqpt} and \ref{mrec}.  Our recipe is based on projective preparations only.  This issue needs to be further investigated  using techniques such as direct characterization of quantum dynamics \cite{mohseni:170501, mohseni:062331}.  Yet, the $\mathcal{M}$-map and memory matrix $\mathcal{K}$ are powerful theoretical tools in their own right. The $\mathcal{M}$-map allows us to determine various output states for any preparation procedure, while the memory matrix $\mathcal{K}$ provides a quantitative definition for the memory effect due to the initial correlations for any preparation procedures. This allows us to investigate and optimize preparation procedures.


%

\chapter{Analysis of experiments}\label{chapexp}
\index{Analysis of experiments @\emph{Analysis of experiments}}%

In this chapter we analyze two quantum process tomography experiments performed by Myrskog et al.\cite{myrskog:013615} and M. Howard et. al \cite{Howard05,Howard06}.  
\index{quantum process tomography}
Our critiques emphasizes the importance of having a consistent procedure of state preparation.

\section{Experiment by Myrskog et al.}

In this experiment, quantum process tomography of the motional states of  trapped $^{85}Rb$ atoms in the potential wells of a one dimensional optical 
\index{negative maps! experimental}
lattice is performed.  Only two bound bands are considered, which are labeled as states $\ket{0}$ and $\ket{1}$. The states are prepared stochastically.  An initial state of the system is the ground state $\ket{g}\bra{g}$, and from it states $\ket{r}\bra{r}$, $\ket{i}\bra{i}$, and the fully mixed state $\frac{1}{2}\mathbb{I}$ are prepared.  The states $\ket{r}$ and $\ket{i}$ stand for the real and imaginary coherence states, which are prepared by applying appropriate unitary transformations on the ground state.  This is achieved by displacing the lattice for the real coherence state and for the imaginary coherence state, a quarter-period delay is added after the displacement. The identity state is prepared by letting a superposition state decohere. The input states are allowed to evolve and the output states are determine by quantum state tomography.

The process map is found following the usual procedure laid out in Chap. \ref{chapqpt}.  Since there are particles lost to the neighboring cells, the map is not required to be trace preserving.  Based on this loss they also argue that the map can pick non-physical behavior (not-completely positive).  The map is forced to be ``physical" (completely positive) by using the maximum likelihood method \cite{Ziman06}.

In our terminology, the states prepared are $P^{(1,+)}$, $P^{(2,+)}$, $P^{(3,-)}$, and $\frac{1}{2}\mathbb{I}$.  The three projective states are prepared by a single consistent stochastic preparation, while the fully mixed state is prepared by letting the state $P^{(1,+)}$ decohere. Which means an additional pin map is applied to prepare one of the input states.  
\index{preparation! multiple stochastic}
As we saw in the example in Sec. \ref{exmultstoch}, this can lead to a not-completely positive and non-linear process map.

Furthermore, the input states have varying values for polarization. Their data is listed in the table below.
\begin{center}
\begin{tabular}{l | llll}
\hline
& $\rho_g$ & $\rho_{\mathbb I}$ & $\rho_r$ & $\rho_i$ 
\\ \hline 
$P^{(3,-)}$ & 0.90 & 0.60 & 0.69 & 0.69 \\
$P^{(3,+)}$ & 0.10 & 0.40 & 0.31 & 0.31 \\
$P^{(1,+)}$ & 0.82 & 0.59 & 0.85 & 0.63 \\
$P^{(2,+)}$ & 0.84 & 0.58 & 0.64 & 0.37 \\
\hline
\end{tabular}
\end{center}
where $\rho_j$ are the experimentally prepared  states projected onto projectors $P^{(m)}$. The polarization of the imaginary state in the $\sigma_2$ direction is very low.  This could mean that it is correlated with the environment, while the polarization of the ground state along the negative $\sigma_3$ direction is almost unity, meaning it is only weakly correlated at best.

Here, we have two potential causes for the process map to have negative eigenvalues.  The first problem is with the experimental procedure; applying multiple stochastic preparations that may affect the state of the environment differently.  The second problem may be unavoidable; since it is extremely difficult to prepare pure states in a setup like this.  Even if the negative eigenvalues are due to the initial correlations, we do not have a prescription to obtain a process map in a consistent fashion.  For the second example shown in Chap. \ref{appdvp}, we assumed that the initial correlations were constant throughout the process.  This does not seem to be the case here. Since the ground state is almost pure, while the imaginary state is clearly not pure.  Therefore the correlation with environment for these two inputs must be different.  This concludes our analysis of this experiment.  Let us now consider the second experiment.

\section{Experiment by Howard et al.}

In this experiment, the system that is studied is an electron configuration formed in a nitrogen vacancy defect in a diamond lattice.  The quantum state of the system is given by a spin triplet ($S=1$).  Again we will write the initial state of the system and environment as $\rho^\mathcal{SE}$.  

The system is prepared by optical pumping, which results in a strong spin polarization.  The state of the system is said to have a 70\% chance of being in a pure state $\ket{\phi}$.  Or more mathematically, the probability of obtaining $\ket{\phi}$ is $\mbox{Tr}[\ket{\phi}\bra{\phi} \rho^\mathcal{SE}] = 0.7$.  

Since the population probability is high, an assumption was made that the state of the system can be simply approximated as a pure state $\ket{\phi}\bra{\phi}$.  From this initial state, different input states can be prepared by suitably applying microwave pulses resonant with the transition levels.  After preparation, the system is allowed to evolve, and 
\index{preparation! stochastic}
the output states are determined by quantum state tomography.  With the knowledge of the input state and the measured output states, a linear process map is constructed.

It was found that the linear process map has negative eigenvalues, so the map was ``corrected" using a least squares fit between the experimentally determined map and a theoretical map based on Hermitian parametrization \cite{havel03}, while enforcing complete positivity. 

However, if we do not regard the negative eigenvalues of the map as aberrations, then we should consider the assumptions about the preparation of the system more carefully.  The assumption about the initial state of the system is:
\begin{eqnarray}
\rho^\mathcal{SE} \rightarrow \ket{\phi}\bra{\phi}\otimes\tau .
\end{eqnarray}
This is in effect a  pin map.  Along with the pin map, the stochastic transformations are applied on the initial state to prepare the various 
\index{pin map}
input states; this is identical to the stochastic preparation method discussed in Sec. \ref{stoprep}.

It is clear that the pure initial state assumption is unreasonable given our knowledge now of how the process is sensitive to the initial correlations between the system and the environment.  In effect the action of the pin map in this experiment is not perfect, and the pin map can be ignored.  Then the process equation is:
\begin{eqnarray}
Q^{(m)} = \mbox{Tr}_\mathbb{B} [U \Omega^{(m)} \rho^\mathcal{SE} U^\dagger] 
\end{eqnarray}
where $\Omega^{(m)}$ is the stochastic transformation that prepares the $m$th input state. In this experiment, $\Omega^{(m)}$ is nothing more than a unitary transformation $V^{(m)}$ satisfying $V^{(m)} \ket{\phi} = \ket{\psi^{(m)}}$, where $\ket{\psi^{(m)}}$ is the desired pure $m$th input state.

We can write the unitary transformation for a two-level system as:
\begin{eqnarray}
V^{(m)} = \ket{\psi^{(m)}}\bra{\phi}+ \ket{\psi^{(m)}_\perp }\bra{\phi_\perp}
\end{eqnarray}
where $\braket{\psi^{(m)}|\psi^{(m)}_\perp } = \braket{\phi | \phi_\perp } = 0$.  This defines $V^{(n)}$ as a transformation from the basis $\{\ket{\phi}\}$ to the basis $\{\ket{\psi^{(n)}_i}\}$.  The equation for the process becomes:
\begin{eqnarray*}
Q^{(m)} &=& 
\mbox{Tr}_\mathcal{E} \left[U \ket{\psi^{(m)}}\braket{\phi| 
\rho^\mathcal{SE} |\phi}\bra{\psi^{(m)}} U^\dagger\right]\\
&& +  \mbox{Tr}_\mathcal{E} \left[U \ket{\psi^{(m)}_\perp}\braket{\phi_\perp| \rho^\mathcal{SE} |\phi}\bra{\psi^{(m)}} U^\dagger\right]\\
&& +  \mbox{Tr}_\mathcal{E} \left[U \ket{\psi^{(m)}}\braket{\phi| \rho^\mathcal{SE} |\phi_\perp}\bra{\psi^{(m)}_\perp} U^\dagger\right]\\
&& +  \mbox{Tr}_\mathcal{E} \left[U \ket{\psi^{(m)}_\perp}\braket{\phi_\perp| \rho^\mathcal{SE} |\phi_\perp}\bra{\psi^{(m)}_\perp} U^\dagger\right].
\end{eqnarray*}
Therefore, since $\braket{\phi | \rho^\mathcal{SE} | \phi} = 0.7$ to first approximation, the process is a linear mapping on the states $\ket{\psi^{(m)}}\bra{\psi^{(m)}}$.  However, it is clear that if all terms are included, the process is not truly linear in the states 
\index{not-completely positive dynamics|see{ negative maps}}
\index{negative maps! experimental}
$\ket{\psi^{(m)}}\bra{\psi^{(m)}}$.  The  negative eigenvalues are therefore a result of fitting results into a linear map when the process is not truly represented by a linear map.


\chapter{Conclusion and future directions}\label{chapcon} 
\index{Conclusion and future directions
@\emph{Conclusion and future directions}}%

Our study of quantum process tomography started by noting that the dynamical map acting on a system can have negative eigenvalues.  The dynamical map has negative eigenvalues when the system is initially correlated with the environment.  In the course of our studies of quantum process tomography, we showed that the preparation procedure cannot be  neglected for any quantum system that interacts with an environment.  These are the two major themes discussed in this dissertation. Though, along the way,  we presented a method of quantum process tomography that is independent of the preparation procedure.  The map arising from this procedure lead us to an expression that quantifies the memory effect on the dynamics of the system due to the initial  correlations with the environment. Determining the memory effect is an important task in coherence control.  

\section{Future directions}

Though this dissertation has come to an end, the work goes unfinished.  There 
are many open problems that arise from our work.  We pointed out several causes 
for errors in quantum process tomography. We have not developed a complete 
method of distinguishing and correcting these errors.  This is an important problem
that has not been fully addressed.

Another problem that needs consideration is fully determining $\mathcal{M}$-map.
This is a challenging task, though there are some possibilities in 
resolving this problem by considering a scheme like the direct
characterization of quantum dynamics.  In this method a map is found with the 
aid of an ancillary system, which allows for direct estimation of the elements 
of a map without the aid of quantum state tomography.

Another avenue of research that may prove fruitful is studying various 
preparation procedures using $\mathcal{M}$-map.  Furthermore, the memory matrix can aid in better better understanding of the various preparation procedures.  
This will especially be the case when the system is multi-partite.

\def\thechapter{\Alph{chapter}}	
\appendix \index{Appendix @\emph{Appendix}}


\chapter{Quantum state tomography}\label{appqst}
\index{quantum state tomography 
@\emph{Quantum state tomography}}

Quantum state tomography \cite{Nielsen00a} is the procedure to experimentally determine the state of a quantum system. Suppose we want to determine the density matrix $Q$.  To fully determine the state, we need to measure the magnitude of the polarization along the principal directions.  For a qubit we need to make measurements along the $\sigma_x$, $\sigma_y$, and $\sigma_z$ directions.

However, a single experiment will not tell us the magnitude along any direction.  Therefore we make many measurements of the magnitude along each of these direction.  In general, the outcome of a set of the experiment along $j$th direction is written as 
$$
a_j=\mbox{Tr}[Q\sigma_j],
$$
which is the expectation value for the observable $\sigma_j$.  Where $\sigma_j$ are the generalized Pauli-Gell-Mann-Tilma \cite{TilmaSudarshan02} matrices.  Finally the total density matrix is given as follows
\begin{eqnarray}
Q&=&\frac{1}{d^\mathcal{S}}\left\{\mathbb{I}+\sum_j 
\mbox{Tr}[Q\sigma_j]\sigma_j\right\}\\
&=&\frac{1}{d^\mathcal{S}}\left\{\mathbb{I}+\sum_j 
a_j\sigma_j\right\}.
\end{eqnarray}

\chapter{$\mathcal{M}$-map process tomography}\label{mqpt}
\index{$\mathcal{M}$-map process tomography
@\emph{$\mathcal{M}$-map process tomography}}%


We now develop a tomography procedure to determine the $\mathcal{M}$-map. We will need to figure out a finite set of input states $P^{(m)}$ and the corresponding output states $Q^{(m)}$ that will allow us to determine the $\mathcal{M}$-map.  After-all, this is the objective of tomography; by performing measurements on a small number of select input states a complete description of a process can be obtained and predict the output state for \emph{any} input state. The process equation for projective preparations\footnote{Our procedure here only deals with projective preparations.} takes the following form
\begin{eqnarray}
Q^{(m)}&=&\frac{1}{r^{(m)}}
\mathcal{M}^{(rs)}_{r'r'';s's''}
\mathcal{P}^{(m)}_{r'r'';s's''}\\
&=&\frac{1}{r^{(m)}}
P^{(m)}_{r'r''}
\mathcal{M}^{(rs)}_{r'r'';s's''}
{P^{(m)}_{s's''}}^*.
\end{eqnarray}

The $\mathcal{M}$-map is a large ($d^3 \times d^3)$ matrix, to make it more
manageable, Eq. \ref{RawProcessEquation} can be interpreted as:
\begin{equation} 
\Brak{P^{(m)}}{\mathcal{M}}{P^{(m)}}= r^{(m)} Q^{(m)},
\label{SimplifiedProcessEquation} 
\end{equation} 
where $P^{(m)}$ is now
treated as a vector.  In this form, $\mathcal{M}$-map is a $d^2 \times d^2$ matrix,  where each element is a $d \times d$ matrix. 
\index{process equation! $\mathcal{M}$-map}
\index{process map! $\mathcal{M}$-map! matrix representation}
We will call Eq. \ref{SimplifiedProcessEquation} the `\emph{$\mathcal{M}$-map process equation}'. 

Since $\mathcal{M}$-map is a Hermitian matrix, it has $\frac{1}{2}(d^4+d^2)$ independent elements.  Therefore $\frac{1}{2}(d^4+d^2)$ independent equations in the form of Eq. \ref{SimplifiedProcessEquation} are necessary to fully determine $\mathcal{M}$-map.  It is clear that neither an orthonormal set of $d$ nor a linearly independent set of $d^2$ input states would provide sufficient equations to resolve the elements of the $\mathcal{M}$-map.

\section{The qubit case}

For a qubit system the projections, $P^{(m)}$, can be written in terms of three real parameter $a_j$ and Pauli spin matrices $\sigma_j$: 
\begin{equation} \label{PVectorDecomposition}
P^{(m)} =\frac{1}{2} \left(\mathbb{I} +a^{(m)}_j \sigma_j\right).
\end{equation} 
Since $P^{(m)}$ is a projection then there is have the additional constrain $\sum_j ({a^{(m)}_j})^2=1$. The matrices $\mathbb{I}$ and $\sigma_j$ together forms a vector basis for this space.  Therefore Eq. \ref{PVectorDecomposition} is simply a vector decomposition of the projections in a fixed basis. Taking this form for $P^{(m)}$ and substituting into Eq. \ref{SimplifiedProcessEquation} gives:
\begin{eqnarray} \label{OpenQubitProcessEquation}
4r^{(m)} Q^{(m)}
&=&4\bra{P^{(m)}}\mathcal{M}\ket{P^{(m)}}\nonumber\\
&=&\braket{\mathbb{I}|\mathcal{M}|\mathbb{I}} +a^{(m)}_j
\braket{\mathbb{I}|\mathcal{M}|\sigma_j} +a^{(m)}_k
\braket{\sigma_k|\mathcal{M}|\mathbb{I}}\nonumber\\ 
&&\hspace{3cm}+a^{(m)}_j a^{(m)}_k
\braket{\sigma_j|\mathcal{M}|\sigma_k}.
\end{eqnarray}
Observe that the terms $\braket{ \mathbb{I} | \mathcal{M} | \mathbb{I} }$,
$\braket{ \mathbb{I} | \mathcal{M} | \sigma_j }$, $\braket{ \sigma_j |
\mathcal{M} | \mathbb{I} }$ and $\braket{ \sigma_j | \mathcal{M} | \sigma_k
}$, are simply the matrix elements of $\mathcal{M}$-map in $\{ \mathbb{I},
\sigma_j\}$ basis.  We just need to find a set of projections $P^{(m)}$ that will allow us to solve for these matrix elements.

Consider the following specific projections defined as $P^{(j,\pm)}=\frac{1}{2}
(\mathbb{I} \pm \sigma_j)$ with $j=\{1,2,3\}$. 
\begin{eqnarray}
4r^{(j,\pm)}Q^{(j,\pm)} &=&4
\braket{P^{(j,\pm)}|\mathcal{M}|P^{(j,\pm)}}\nonumber\\
&=&\braket{\mathbb{I}|\mathcal{M}|\mathbb{I}}
+\braket{\sigma_j|\mathcal{M}|\sigma_j}
\pm\braket{\sigma_j|\mathcal{M}|\mathbb{I}}
+\braket{\mathbb{I}|\mathcal{M}|\sigma_j}. 
\end{eqnarray} 
Simultaneously solving the $(+)$ and the $(-)$ equations above to gives the following unknowns:
\begin{eqnarray*} 
\braket{\mathbb{I}|\mathcal{M}|\mathbb{I}} +
\braket{\sigma_j|\mathcal{M}|\sigma_j} &=&2\left(r^{(j,+)}Q^{(j,+)}
+r^{(j,-)}Q^{(j,-)}\right)\\ \braket{\mathbb{I}|\mathcal{M}|\sigma_j}
+\braket{\sigma_j|\mathcal{M}|\mathbb{I}} &=&2\left(r^{(j,+)}Q^{(j,+)}
-r^{(j,-)}Q^{(j,-)}\right). 
\end{eqnarray*}

To obtain the cross terms $\braket{ \sigma_j | \mathcal{M} | \sigma_k }$
consider projections such as $P^{\left(j+k+1,+\right)}=\frac{1}{2}
\left(\mathbb{I} + \frac{1}{\sqrt{2}} \sigma_j + \frac{1}{\sqrt{2}}
\sigma_k\right)$ for $k>j$ which give: 
\begin{eqnarray}
r^{(j+k+1,+)}Q^{(j+k+1,+)}&=&
\bra{P^{(j+k+1,+)}}M\ket{P^{(j+k+1,+)}}\nonumber\\ 
&=&\frac{1}{8}\left(
\braket{\mathbb{I}|\mathcal{M}|\mathbb{I}}
+\braket{\sigma_j|\mathcal{M}|\sigma_j} \right)\nonumber \\
&&+\frac{1}{8}\left( +\braket{\mathbb{I}|\mathcal{M}|\mathbb{I}}
+\braket{\sigma_k|\mathcal{M}|\sigma_k} \right)\nonumber \\
&&+\frac{1}{4\sqrt{2}}\left( \braket{\mathbb{I}|M|\sigma_j}
+\braket{\sigma_j|\mathcal{M}|\mathbb{I}} \right) \nonumber\\&&+
\frac{1}{4\sqrt{2}} \left( +\braket{\mathbb{I}|\mathcal{M}|\sigma_k}
+\braket{\sigma_k|\mathcal{M}|\mathbb{I}} \right)\nonumber\\
&&+\frac{1}{8}\left( \braket{\sigma_j|\mathcal{M}|\sigma_k}
+\braket{\sigma_j|\mathcal{M}|\sigma_k} \right). 
\end{eqnarray} 
Substitute the known terms and solve for the desired cross terms, \begin{eqnarray}
\braket{\sigma_j|\mathcal{M}|\sigma_k} +\braket{\sigma_k|\mathcal{M}|\sigma_j}
&=&-2\left(1+ {\sqrt{2}}\right)\left(r^{(j,+)}Q^{(j,+)} 
-2\left(1+\sqrt{2}\right) +r^{(k,+)}Q^{(k,+)}\right) \nonumber\\
&&-2\left(1-{\sqrt{2}}\right)\left(r^{(j,-)}Q^{(j,-)} 
-2\left(1-\sqrt{2}\right) +r^{(k,-)}Q^{(k,-)}\right) \nonumber\\
&&+8r^{(j+k+1,+)}Q^{(j+k+1,+)}. \nonumber
\end{eqnarray}

\index{input state! for $\mathcal{M}$-map tomography}
In summary using the following nine projections,
\begin{eqnarray}\label{choice1} 
P^{(j,+)}=\frac{1}{2} \left(\mathbb{I} +
\sigma_j\right),\;\; P^{(j,-)}=\frac{1}{2} 
\left(\mathbb{I} - \sigma_j\right),\;\; P^{(4,+)}=\frac{1}{2} 
\left(\mathbb{I}+\frac{1}{\sqrt{2}}\sigma_1+\frac{1}{\sqrt{2}}\sigma_2 \right),\nonumber \\ 
P^{(5,+)}=\frac{1}{2} 
\left(\mathbb{I}+\frac{1}{\sqrt{2}}\sigma_1+\frac{1}{\sqrt{2}} \sigma_3 \right),\;\;\; 
P^{(6,+)}=\frac{1}{2} 
\left(\mathbb{I}+\frac{1}{\sqrt{2}}\sigma_2+\frac{1}{\sqrt{2}}\sigma_3
\right),\nonumber 
\end{eqnarray} 
and solving them simultaneously yields all desired matrix elements: $\braket{ \mathbb{I} | \mathcal{M} | \mathbb{I} } +
\braket{ \sigma_j | \mathcal{M} | \sigma_j }$, $\braket{ \mathbb{I} |
\mathcal{M} | \sigma_j} + \braket{ \sigma_j | \mathcal{M} | \mathbb{I} }$, and
$\braket{ \sigma_j | \mathcal{M} | \sigma_k } + \braket{ \sigma_k |
\mathcal{M} | \sigma_j}.$

This is not enough to fully determine the $\mathcal{M}$-map, but these 
\index{process map! $\mathcal{M}-map$}
elements are sufficient to determine the output state for any input state. Using the property $\sum_j (a^{(m)}_j)^2= 1$ Eq.
\ref{OpenQubitProcessEquation} can be rewritten as: \begin{eqnarray}\label{M_Elements} 
4r^{(m)} Q^{(m)} &=&\sum_j
\left(a^{(m)}_j\right)^2 \left( \braket{\mathbb{I}|\mathcal{M}|\mathbb{I}}
+\braket{\sigma_j|\mathcal{M}|\sigma_j} \right) \nonumber\\ &&+\sum_j
a^{(m)}_j \left( \braket{\mathbb{I}| \mathcal{M} |\sigma_j}
+\braket{\sigma_j|\mathcal{M}|\mathbb{I}} \right) \\ &&+\sum_{k>j} a^{(m)}_j a^{(m)}_k \left( \braket{\sigma_j|\mathcal{M}|\sigma_k}
+\braket{\sigma_k|\mathcal{M}|\sigma_j} \right).\nonumber 
\end{eqnarray}

Observe that the sums of the cross terms $\braket{ \sigma_j | \mathcal{M} |
\sigma_k } + \braket{ \sigma_k | \mathcal{M} | \sigma_j }$ can appear together because the coefficients $a^{(m)}_j$ are real.  Also, the element
$\braket{\mathbb{I} | \mathcal{M} |\mathbb{I}}$ can always be paired with a
diagonal element $\braket{ \sigma_j|\mathcal{M}|\sigma_j}$ as long as the
state is a projection satisfying $\sum_j (a^{(m)}_j)^2= 1$.  The diagonal element $\braket{\mathbb{I}|\mathcal{M}|\mathbb{I}}$ only has to be known if the system can be prepared directly to a mixed state such that $\sum_j (a^{(m)}_j)^2< 1$.  This may be accomplished by a generalized preparations \cite{Kuah02}.  If generalized  preparations are allowed then just one more input state is needed, for example $\frac{1}{2}(\mathbb{I}+ \frac{1}{2} \sigma_1)$, which gives another independent equation that can be solved to obtain $\braket{\mathbb{I}|\mathcal{M}|\mathbb{I}}$.

Therefore the elements of the $\mathcal{M}$-map found in Eq. \ref{M_Elements} are all that are needed to describe the process. By measuring the outputs for the nine specified input states, the matrix $\mathcal{M}$-map can be calculated. We now have a good quantum process tomography procedure for an open qubit system.

\section{Beyond one qubit}

Note that the choice of these nine states used above is not unique.  The recipe which used to derive these nine states can be used in principle to derive other choices, and can also be partly generalized to $d$-level systems \cite{byrd:062322, TilmaSudarshan02}.  However, there are some non-trivial obstacles to overcome for the generalization to $d$-level systems.  In place of the Pauli matrices for two-level systems, the generalized Pauli-Gell--Mann-Tilma \cite{TilmaSudarshan02} Hermitian traceless $d \times d$ matrices can be used \cite{Greiner,TilmaSudarshan02} to decompose the $d \times d$ density matrix, and this decomposition will also have only real coefficients.  This trick eliminates certain degrees of freedom in the $\mathcal{M}$-map that is otherwise difficult to deal with.  Unfortunately, for $d > 2$, these real coefficients no longer satisfy just the simple constraint $\sum_j (a^{(m)}_j)^2= 1$.  The additional constraints on the coefficients complicate the task of constructing the projections needed to simultaneously span the matrix elements of $\mathcal{M}$-map.
\chapter{Experimental recipe for $\mathcal{M}$-map}\label{mrec} 
\index{Experimental recipe for $\mathcal{M}$-map 
@\emph{Experimental recipe for $\mathcal{M}$-map}}%


Although we have established the $\mathcal{M}$-map process tomography, let us make the ideas more concrete by developing a complete recipe for an experiment that can be used to determine whether a process is linear or given by the $\mathcal{M}$-map.  We will also show specifically how the  $\mathcal{M}$-map or linear map can be calculated from the measurement results.

For the $\mathcal{M}$-map process tomography nine input states are necessary.  For the nine states derived in App. \ref{mqpt} the first six states are three pairs of orthonormal projections, but the last three are not.  If we use projections to prepare the states, we will need use them as given by orthonormal pairs.  Therefore, let us instead use twelve
\index{input state! for $\mathcal{M}$-map tomography}
projections, nine from Eq. \ref{choice1} and three orthogonal to the last three in that equation: 
\begin{eqnarray} 
P^{(4,-)} = \frac{1}{2} \left(\mathbb{I} -
\frac{1}{\sqrt{2}} \sigma_1 - \frac{1}{\sqrt{2}} \sigma_2\right) , \nonumber\\
P^{(5,-)} = \frac{1}{2} \left(\mathbb{I} - \frac{1}{\sqrt{2}} \sigma_1 -
\frac{1}{\sqrt{2}} \sigma_3\right) , \\ 
P^{(6,-)} = \frac{1}{2} \left(\mathbb{I} -
\frac{1}{\sqrt{2}} \sigma_2 - \frac{1}{\sqrt{2}} \sigma_3\right) \nonumber.
\end{eqnarray}

These twelve projections are neatly grouped into six different sets of
orthonormal pairs.  If the states are prepared using projections,
these would correspond to measurements in the $\pm\sigma_1$, $\pm\sigma_2$,
$\pm\sigma_3$, $\sigma_1 \pm \sigma_2$, $\sigma_1 \pm \sigma_3$ and $\sigma_2 \pm \sigma_3$ directions.  Having twelve states is more than is necessary for the $\mathcal{M}$-map process tomography, but the extra states and the corresponding output states can serve as consistency checks.

After recording the corresponding output states for all twelve input states, we can verify if the process is linear by checking the following eight linear sum rules (that the input states satisfy) have to be satisfied by the corresponding output states:
\begin{eqnarray*} 
&& Q^{(j,+)} + Q^{j,-)} = Q^{(k,+)} +
Q^{(k,-)}\;\;\;(\mbox{for}\;  j,k=1,2,3)\\ Q^{(4,+)} 
&=& \left(\frac{1}{2} -
\frac{1}{\sqrt{2}}\right) \left(Q^{(1,+)} + Q^{(1,-)}\right) +
\frac{1}{\sqrt{2}} \left(Q^{(1,+)} + Q^{(2,+)}\right)\\ Q^{(5,+)} &=&
\left(\frac{1}{2} - \frac{1}{\sqrt{2}}\right) \left(Q^{(1,+)} +
Q^{(1,-)}\right) + \frac{1}{\sqrt{2}} \left(Q^{(1,+)} + Q^{(3,+)}\right)\\
Q^{(6,+)} &=& \left(\frac{1}{2} - \frac{1}{\sqrt{2}}\right) 
\left(Q^{(1,+)} +Q^{(1,-)}\right) + \frac{1}{\sqrt{2}} \left(Q^{(2,+)} + Q^{(3,+)}\right).
\end{eqnarray*}

If the eight sum rules are satisfied, then the process is not described by 
the $\mathcal{M}$-map, and we can be confident the process is described by a linear map.  The linear map $\Lambda$ can then be computed as follows: 
\begin{eqnarray} 
\Lambda\left(\frac{1}{2}(\mathbb{I}+p_j \sigma_j)\right) = 
\left(Q^{(1,+)}+ Q^{(1,-)}\right) + p_j \left(Q^{(j,+)} - Q^{(j,-)}\right) \end{eqnarray}

If the eight sum rules are not satisfied, then the process may be given by the $\mathcal{M}$-map.  We will attempt to verify that the process given by the $\mathcal{M}$-map and calculate it.

Note that the probabilities $r^{(m)} = \mbox{Tr}[\rho^{\mathcal{SE}}
P^{(m)}]$ associated with each preparation should be found experimentally. 
The probabilities should be complete for an orthonormal set of projections, in other words: $r^{(j,+)} + r^{(j,-)} = 1.$ Therefore the
probabilities can be calculated from the fraction of the $(+)$ states as
compared to the $(-)$ states, for all preparations made in the same direction.

To be certain that the process is given by the $\mathcal{M}$-map, we can check if the three additional states we included evolve in a way that is consistent with the $\mathcal{M}$-map derived from the other nine states.  The following equations are derived from this condition.  If these equations are satisfied, we can be confident that the process is given by the $\mathcal{M}$-map: 
\begin{eqnarray*} \sqrt{2}
\left(r^{(4,+)} Q^{(4,+)} -r^{(4,-)} Q^{(4,-)}\right)
&=&r^{(1,+)} Q^{(1,+)} -r^{(1,-)} Q^{(1,-)}\\ &&+r^{(2,+)}
Q^{(2,+)} -r^{(2,-)} Q^{(2,-)}\\ \sqrt{2} \left(r^{(5,+)} Q^{(5,+)}-
r^{(5,-)} Q^{(5,-)}\right) &=&r^{(1,+)} Q^{(1,+)} -r^{(1,-)}
Q^{(1,-)}\\ &&+r^{(3,+)} Q^{(3,+)} -r^{(3,-)} Q^{(3,-)}\\ \sqrt{2}
\left(r^{(6,+)} Q^{(6,+)}- r^{(6,-)} Q^{(6,-)}\right)
&=&r^{(2,+)} Q^{(2,+)} -r^{(2,-)} Q^{(2,-)}\\ &&+r^{(3,+)}
Q^{(3,+)} -r^{(3,-)} Q^{(3,-)}. 
\end{eqnarray*} 
If the conditions above are satisfied then the process is given by the $\mathcal{M}$-map, then the $\mathcal{M}$-map can be computed by following the recipe in App. \ref{mqpt}.

Once the matrix elements of $\mathcal{M}$-map are determined, the evolution of a generic state $X =\frac{1}{2}(\mathbb{I}+\sum_j x_j \sigma_j)$ is given by: 
\begin{eqnarray}
4\braket{X|\mathcal{M}|X} &=& r^{(X)} Q^{(X)}\\ &=& \sum_j x_j^2 \left(
\braket{\mathbb{I}|\mathcal{M}|\mathbb{I}}
+\braket{\sigma_j|\mathcal{M}|\sigma_j} \right)\\ &&+\sum_j x_j \left(
\braket{\mathbb{I}|\mathcal{M}|\sigma_i}
+\braket{\sigma_j|\mathcal{M}|\mathbb{I}} \right)\\ &&+\sum_{k>j} x_j x_k
\left( \braket{\sigma_j|\mathcal{M}|\sigma_k}
+\braket{\sigma_k|\mathcal{M}|\sigma_j} \right).
\end{eqnarray}

Note that since $Q$ is a normalized state, the normalization constant $r$
is the measurement probability $r = \mbox{Tr}[X \rho^{\mathcal{SE}}]$. 
Although we had not explicitly mentioned this before, the $\mathcal{M}$-map
contains all information about the measurement probabilities, that is why we needed the measurement probabilities $r^{(m)}$ to calculate the
$\mathcal{M}$-map.

Finally, note that if both the test for linearity and the $\mathcal{M}$-map fail, then the process cannot be consistently described by either a linear map or the $\mathcal{M}$-map.  The experiment then should be carefully analyzed for problems such as any non-linear dependence that may have been introduced if the input states are not accurately prepared or if there is some dependence of $\rho^{\mathcal{SE}}$ on the prepared state.


If the process is given by a linear process map, then the nine input states
(Eq. \ref{choice1}) are over complete; only four input states are needed to
determine a linear process map.  This discrepancy is summarized by the
following linear sum rules: 
\begin{eqnarray}\label{linrules} P^{(1,+)} &+&
P^{(1,-)} = P^{(2,+)} + P^{(2,-)} = P^{(3,+)} + P^{(3,-)} , \nonumber\\
P^{(4,+)} &=& \left(\frac{1}{2}-\frac{1}{\sqrt{2}}\right)\left(P^{(1,+)} +
P^{(1,-)}\right) \nonumber
+ \frac{1}{\sqrt{2}}
\left(P^{(1,+)} + P^{(2,+)}\right) , \nonumber\\ 
P^{(5,+)} &=& \left(\frac{1}{2}-\frac{1}{\sqrt{2}}\right)\left(P^{(1,+)}+ P^{(1,-)}\right)
+ \frac{1}{\sqrt{2}} \left(P^{(1,+)} + P^{(4,+)}\right) ,
\nonumber\\ 
P^{(6,+)}  &=&  \left(\frac{1}{2}-\frac{1}{\sqrt{2}}\right)\left(P^{(1,+)}+
P^{(1,-)}\right)\nonumber
+ \frac{1}{\sqrt{2}}
\left(P^{(2,+)} + P^{(4,+)}\right)\nonumber . 
\end{eqnarray}

If the process is linear, then the output states must satisfy the same sum
rules, which are obtained from the above equations by suitably writing $Q$ in place of $P$.  If these sum rules are not satisfied, then the process is not linear.  However satisfying the sum rules is necessary but not sufficient to determine if the process is linear; the $\mathcal{M}$-map can still be constructed from this set of input and output states without any contradictions.  Therefore an additional input state, distinct from the above nine input states should be tested.


\chapter{Notation}\label{appnot}
\index{notation @\emph{notation}}

\begin{eqnarray*}
\vec{a}&:&\mbox{Bloch vector}\\
 \mathcal{A}&:&\mbox{stochastic map}\\
 \mathcal{B}&:&\mbox{dynamical map}\\
c_{jk}&:&\mbox{correlation parameters}\\
\chi&:&\mbox{correlation matrix}\\
 H&:&\mbox{Hamiltonian}\\
 \mathbb{I}&:&\mbox{indentity matrix} \hspace{3cm} \\
 \mathcal{I}&:&\mbox{identity map}\\
 \mathcal{K}&:&\mbox{memory matrix}\\
 \Lambda&:&\mbox{linear process map}\\
 \mathcal{M}&:&\mbox{Kuah's dynamical $\mathcal{M}$-map}\\
 \Omega^{(m)}&:&m\mbox{th local rotaion}\\
 \mathcal{P}^{(m)}&:&m\mbox{th preparation map}\\
 P^{(m)}&:&m\mbox{th input states}\\
 Q^{(m)}&:&m\mbox{th output states}\\
 \rho^\mathcal{A}&:&\mbox{state of the ancillary system}\\
 \rho^\mathcal{E}&:&\mbox{state of the environment}\\
 \rho^\mathcal{S}&:&\mbox{state of the system}\\
 \rho^\mathcal{SE}&:&\mbox{state of the system plus environment}\\
 \sigma_j&:&\mbox{Pauli spin matrices}\\
 \Theta&:&\mbox{pin map}\\
 U&:&\mbox{unitary operators}\\
 X&:&\mbox{an arbitrary input state}\\
\end{eqnarray*}


\bibliography{dissertation} 										 %
\index{Bibliography@\emph{Bibliography}}%
\bibliographystyle{is-unsrt} 						 			     %

\printindex

\begin{vita}
\index{Vita @\emph{Vita}}
Kavan was born in Valsad, Gujarat, India on the 21st of October 1978 to Bhagavati and Kishore Modi. He grew up in India till the age of thirteen, then spent one year in Staten Island, New York. He spent his high school years in Parsippany, New Jersey, graduating from Parsippany High School in 1997. Upon graduation he enrolled in Embry-Riddle Aeronautical University in Daytona Beach, Florida.  He held summer internships at the Fermi National Accelerator Laboratory and NASA's Kennedy Space Center while attending Embry-Riddle Aeronautical University. In May 2001, he received his Bachelor of Science in Engineering Physics from Embry-Riddle Aeronautical University.  In the fall of 2001 he joined the University of Texas at Austin as a graduate student. He received his Master of Arts in physics from there in May 2004.  He has received the 2007 John Bardeen and the 2008 Rolfe Glover awards from the Forum in History of Physics of the American Physical Society. Beside physics, he enjoys a diverse social life and traveling.
\end{vita}

\end{document}